\documentclass[envcountsame]{article}

\newif\ifSpecNotes\SpecNotesfalse
\newif\ifImplNotes\ImplNotesfalse
\newif\ifPresentationNotes\PresentationNotesfalse

\usepackage[normalem]{ulem}

\usepackage{a4wide}
\usepackage{amsfonts}
\usepackage[fleqn]{amsmath}
\usepackage{amssymb}
\usepackage{amsthm}
\usepackage{authblk}
\usepackage{cite}
\usepackage[T1]{fontenc}
\usepackage{floatrow}
\newfloat{Sidebar}{tbp}{saux}\floatstyle{ruled}\restylefloat{Sidebar}
\usepackage{hyperref}
\usepackage{listings} 
\usepackage{mathtools}
\usepackage{pdflscape}
\usepackage{tcolorbox}\tcbset{boxrule=0.5pt,colback=red!3!white,beforeafter skip=\baselineskip}
\usepackage{tikz}\usetikzlibrary{automata,positioning,arrows}
\usepackage{verbatim}
\usepackage[nohyphen]{underscore}
\usepackage{url}

\newcommand{\TLSMessage}[1]{\texttt{#1}}

\newcommand{\ClientHello}{\TLSMessage{ClientHello}}
\newcommand{\ServerHello}{\TLSMessage{ServerHello}}
\newcommand{\HelloRetryRequest}{\TLSMessage{HelloRetryRequest}}

\newcommand{\EncryptedExtensions}{\TLSMessage{EncryptedExtensions}}

\newcommand{\CertificateRequest}{\TLSMessage{CertificateRequest}}
\newcommand{\Certificate}{\TLSMessage{Certificate}}
\newcommand{\CertificateVerify}{\TLSMessage{CertificateVerify}}

\newcommand{\Finished}{\TLSMessage{Finished}}

\newcommand{\EndOfEarlyData}{\TLSMessage{EndOfEarlyData}}

\newcommand{\KeyUpdate}{\TLSMessage{KeyUpdate}}
\newcommand{\NewSessionTicket}{\TLSMessage{NewSessionTicket}}

\newcommand{\TLSRecord}[1]{\texttt{#1}}

\newcommand{\TLSPlaintext}{\TLSRecord{TLSPlaintext}}
\newcommand{\TLSCiphertext}{\TLSRecord{TLSCiphertext}}

\newcommand{\TLSInnerPlaintext}{\TLSRecord{TLSInnerPlaintext}}

\newcommand{\TLSType}[1]{\textsf{#1}}

\newcommand{\TLSCipherSuite}{\TLSType{CipherSuite}}

\newcommand{\TLSField}[1]{\texttt{#1}}

\newcommand{\TLSalgorithm}{\TLSField{algorithm}}
\newcommand{\TLSapplicationLayerProtocolNegotiation}{\TLSField{application\_layer\_protocol\_negotiation}}
\newcommand{\TLScertificateAuthorities}{\TLSField{certificate\_authorities}}
\newcommand{\TLScertificateRequestContext}{\TLSField{certificate\_request\_context}}
\newcommand{\TLScipherSuite}{\TLSField{cipher\_suite}}
\newcommand{\TLScipherSuites}{\TLSField{cipher\_suites}}
\newcommand{\TLScertificateList}{\TLSField{certificate\_list}}
\newcommand{\TLSclientCertificateType}{\TLSField{client\_certificate\_type}}
\newcommand{\TLScookie}{\TLSField{cookie}}
\newcommand{\TLSearlyData}{\TLSField{early\_data}}
\newcommand{\TLSencryptedRecord}{\TLSField{encrypted\_record}}

\newcommand{\TLSextensions}{\TLSField{extensions}}
\newcommand{\TLSfragment}{\TLSField{fragment}}
\newcommand{\TLSheartbeat}{\TLSField{heartbeat}}
\newcommand{\TLSkeyShare}{\TLSField{key\_share}}
\newcommand{\TLSlegacyCompressionMode}{\TLSField{legacy\_compression\_method}}
\newcommand{\TLSlegacyCompressionModes}{\TLSField{legacy\_compression\_methods}}
\newcommand{\TLSlegacyRecordVersion}{\TLSField{legacy\_record\_version}}
\newcommand{\TLSlegacySessionId}{\TLSField{legacy\_session\_id}}
\newcommand{\TLSlegacySessionIdEcho}{\TLSField{legacy\_session\_id\_echo}}
\newcommand{\TLSlegacyVersion}{\TLSField{legacy\_version}}
\newcommand{\TLSlength}{\TLSField{length}}
\newcommand{\TLSmaxFragmentLength}{\TLSField{max\_fragment\_length}}
\newcommand{\TLSoidFilters}{\TLSField{oid\_filters}}
\newcommand{\TLSopaqueType}{\TLSField{opaque\_type}}
\newcommand{\TLSpadding}{\TLSField{padding}}
\newcommand{\TLSpostHandshakeAuth}{\TLSField{post\_handshake\_auth}}
\newcommand{\TLSpskModes}{\TLSField{psk\_key\_exchange\_modes}}
\newcommand{\TLSpsk}{\TLSField{pre\_shared\_key}}
\newcommand{\TLSrandom}{\TLSField{random}}
\newcommand{\TLSrequestUpdate}{\TLSField{request\_update}}
\newcommand{\TLSserverCertificateType}{\TLSField{server\_certificate\_type}}
\newcommand{\TLSserverName}{\TLSField{server\_name}}
\newcommand{\TLSsignature}{\TLSField{signature}}
\newcommand{\TLSsignatureAlgorithms}{\TLSField{signature\_algorithms}}
\newcommand{\TLSsignatureAlgorithmsCert}{\TLSField{signature\_algorithms\_cert}}
\newcommand{\TLSsignedCertificateTimestamp}{\TLSField{signed\_certificate\_timestamp}}
\newcommand{\TLSstatusRequest}{\TLSField{status\_request}}
\newcommand{\TLSsupportedGroups}{\TLSField{supported\_groups}}
\newcommand{\TLSsupportedVersions}{\TLSField{supported\_versions}}
\newcommand{\TLSsupportedSignatureAlgorithms}{\TLSField{supported\_signature\_algorithms}}
\newcommand{\TLSticket}{\TLSField{ticket}}
\newcommand{\TLSticketAgeAdd}{\TLSField{ticket\_age\_add}}
\newcommand{\TLSticketLifetime}{\TLSField{ticket\_lifetime}}
\newcommand{\TLSticketNonce}{\TLSField{ticket\_nonce}}
\newcommand{\TLStype}{\TLSField{type}}
\newcommand{\TLSuseSrtp}{\TLSField{use\_srtp}}
\newcommand{\TLSverifyData}{\TLSField{verify\_data}}
\newcommand{\TLSzeros}{\TLSField{zeros}}

\newcommand{\TLSConstant}[1]{\textsf{#1}}

\newcommand{\TLSapplicationData}{\TLSConstant{application\_data}}

\newcommand{\TLShandshake}{\TLSConstant{handshake}}

\newcommand{\TLSPskKe}{\TLSConstant{psk\_ke}}
\newcommand{\TLSPskDheKe}{\TLSConstant{psk\_dhe\_ke}}

\newcommand{\TLSmessageHash}{\TLSConstant{message\_hash}}

\newcommand{\TLSfinishedKey}{\TLSField{finished\_key}}

\newcommand{\TLSAlert}[1]{\textsf{#1}}

\newcommand{\TLSillegalParameter}{\TLSAlert{illegal\_parameter}}
\newcommand{\TLShandshakeFailure}{\TLSAlert{handshake\_failure}}
\newcommand{\TLSinsufficientSecurity}{\TLSAlert{insufficient\_security}}
\newcommand{\TLSmissingExtension}{\TLSAlert{missing\_extension}}
\newcommand{\TLSunexpectedMessage}{\TLSAlert{unexpected\_message}}
\newcommand{\TLSdecodeError}{\TLSAlert{decode\_error}}
\newcommand{\TLSbadCertificate}{\TLSAlert{bad\_certificate}}
\newcommand{\TLSdecryptError}{\TLSAlert{decrypt\_error}}
\newcommand{\TLScertificateRequired}{\TLSAlert{certificate\_required}}
\newcommand{\TLSunsupportedExtension}{\TLSAlert{unsupported\_extension}}

\newcommand{\TLSrecordOverflow}{\TLSAlert{record\_overflow}}
\newcommand{\TLSbadRecordMac}{\TLSAlert{bad\_record\_mac}}

\newcommand{\TLSFunction}[1]{\textsf{#1}}

\newcommand{\TranscriptHash}{\TLSFunction{Transcript-Hash}}
\newcommand{\Truncate}{\TLSFunction{Truncate}}
\newcommand{\Hash}{\TLSFunction{Hash}}
\newcommand{\HMACHash}{\TLSFunction{HMAC}}
\newcommand{\HashLength}{\Hash\TLSFunction{.length}}
\newcommand{\HKDFExpand}{\TLSFunction{HKDF-Expand}}
\newcommand{\HKDFExpandLabel}{\TLSFunction{HKDF-Expand-Label}}
\newcommand{\HKDFExtract}{\TLSFunction{HKDF-Extract}}
\newcommand{\DeriveSecret}{\TLSFunction{Derive-Secret}}

\newcommand{\AEADEnc}{\TLSFunction{AEAD-Encrypt}}
\newcommand{\AEADDec}{\TLSFunction{AEAD-Decrypt}}

\newcommand{\TLSEarlySecret}{\TLSField{Early Secret}}
\newcommand{\TLSHandshakeSecret}{\TLSField{Handshake Secret}}
\newcommand{\TLSMasterSecret}{\TLSField{Master Secret}}

\newcommand{\TLSbinderKey}{\TLSField{binder\_key}}
\newcommand{\TLSclientEarlyTrafficSecret}{\TLSField{client\_early\_traffic\_secret}}

\newcommand{\TLShandshakeTrafficSecret}[1][{[sender]}]{\TLSField{#1\_handshake\_traffic\_secret}}
\newcommand{\TLSclientHandshakeTrafficSecret}{\TLShandshakeTrafficSecret[client]}
\newcommand{\TLSserverHandshakeTrafficSecret}{\TLShandshakeTrafficSecret[server]}
\newcommand{\TLSapplicationTrafficSecret}[2][{[sender]}]{\TLSField{#1\_application\_traffic\_secret\_#2}}
\newcommand{\TLSapplicationTrafficSecretN}[1][N]{\TLSapplicationTrafficSecret{#1}}
\newcommand{\TLSclientApplicationTrafficSecret}[1][0]{\TLSapplicationTrafficSecret[client]{#1}}
\newcommand{\TLSserverApplicationTrafficSecret}[1][0]{\TLSapplicationTrafficSecret[server]{#1}}

\newcommand{\TLSresumptionMasterSecret}{\TLSField{resumption\_master\_secret}}  
\newcommand{\TLSwriteKey}[1][{[sender]}]{\TLSField{#1\_write\_key}}
\newcommand{\TLSclientWriteKey}{\TLSwriteKey[client]}
\newcommand{\TLSserverWriteKey}{\TLSwriteKey[server]}
\newcommand{\TLSwriteIV}[1][{[sender]}]{\TLSField{#1\_write\_iv}}
\newcommand{\TLSclientWriteIV}{\TLSwriteIV[client]}
\newcommand{\TLSserverWriteIV}{\TLSwriteIV[server]}

\newcommand{\TLSHandshakeMessage}{\textit{Transcript}} 
\newcommand{\TLSContext}{\textit{Context}}  
\newcommand{\TLSHKDFSalt}{\textit{Salt}}
\newcommand{\TLSHKDFSecret}{\textit{Secret}}
\newcommand{\TLSHKDFLabel}{\textit{Label}}
\newcommand{\TLSHKDFLength}{\textit{Length}}

\newcommand{\TLSHKDFExpLabel}{\textit{ExpLabel}}

\newcommand{\TLSHKDFLabelExt}{\textit{HkdfLabel}}

\newcommand{\zeros}{0\textrm{s}}

\makeatletter
\long\def\@ympar#1{%
  \@savemarbox\@marbox{\scriptsize #1}%
  \global\setbox\@currbox\copy\@marbox
  \@xympar}
\makeatother

\makeatletter
\def\lst@MSkipToFirst{%
    \global\advance\lst@lineno\@ne
    \ifnum \lst@lineno=\lst@firstline
        \def\lst@next{\lst@LeaveMode \global\lst@newlines\z@
        \lst@OnceAtEOL \global\let\lst@OnceAtEOL\@empty
        \lst@InitLstNumber % Added to work with modified \lsthk@PreInit.
        \lsthk@InitVarsBOL
        \c@lstnumber=\numexpr-1+\lst@lineno % this enforces the displayed line numbers to always be the input line numbers
        \lst@BOLGobble}%
        \expandafter\lst@next
    \fi}
\makeatother

\newlength{\rawgobble}
\newlength{\gobble}
\newlength{\gobblea}
\sbox0{\Huge\ttfamily \ }
\makeatletter
\def\sepstar#1*#2\relax{%
    \def\sepstarone{#1}%
    \def\sepstartwo{#2}%
}
\lst@Key{widthgobble}{0*0}{%
    \sepstar #1\relax
    \setlength{\gobble}{0.9\rawgobble}%
    \setlength{\gobble}{\sepstarone\gobble}%
    \setlength{\gobble}{\sepstartwo\gobble}%
    \setlength{\gobblea}{\gobble}%
    \addtolength{\gobblea}{10pt}%
    \def\lst@xleftmargin{\gobble}%
    \def\lst@framexleftmargin{\gobble}%
    \def\lst@numbersep{\gobblea}%
}
\makeatother

\makeatletter
\def\tinyskip{\vspace\tinyskipamount}
\newskip\tinyskipamount \tinyskipamount=2pt 

\def\lst@MProcessListing{%
    \lst@XPrintToken \lst@EOLUpdate \lsthk@InitVarsBOL
    \global\advance\lst@lineno\@ne
    \ifnum \lst@lineno>\lst@lastline
        \lst@ifdropinput \lst@LeaveMode \fi
        \ifx\lst@linerange\@empty
            \expandafter\expandafter\expandafter\lst@EndProcessListing
        \else
            \lst@interrange
            \lst@GetLineInterval
            \expandafter\expandafter\expandafter\lst@SkipToFirst{}\tinyskip
        \fi
    \else
        \expandafter\lst@BOLGobble
    \fi}
\makeatother

\tikzset{
  initial text={},
  every edge/.style={draw,->,>=stealth',semithick}
}

\lstdefinestyle{inlinestyle}{
  basicstyle=\lst@ifdisplaystyle\footnotesize\fi, %https://tex.stackexchange.com/a/161551/59197
  columns=fullflexible, %https://tex.stackexchange.com/questions/79952/unequal-letter-spacing-in-listings
  breaklines=true,
  literate={\\-}{}{0\discretionary{-}{}{}}
}

\newcommand{\code}{\lstinline[style=inlinestyle]} 
\title{TLS 1.3 for engineers: 
        An exploration of the TLS 1.3 specification 
        and OpenJDK's Java implementation %Thanks to Adam Petcher (email, 28 May 2020) for pointing out that discrepancy
}

\author{Ben Smyth}
\affil{\href{https://cryptostream.ltd/}{Crypto Stream} Ltd. 
          \& \href{https://x26.io/}{Ampersand} (0x26 Ltd.), UK}
\date{30 Sep 2020}

\begin{document}

\maketitle 
\thispagestyle{empty} 
%\ifPresentationNotes
%\input{abstract}
%\marginpar{Potential venues: computer science review, ACM Computing Surveys, Proceedings of the IEEE, and Springer's Information Security and Cryptography (ISC) series.}
%\fi
\newpage

\section*{Contribute}

This manuscript is far from perfect: 
Interesting aspects are omitted, because I %the authors
didn't have the time, knowledge, or expertise to add them. For instance, the specification 
hasn't been entirely covered, as is documented 
(\href{https://github.com/BenSmyth/tls-tutorial/blob/master/rfc8446-annotated.txt}{rfc8446-annotated.txt}); 
discussion of security guarantees are notably lacking; and an introduction to the underlying 
cryptography is absent. (E.g., some details on DHKE, AEAD, etc.\ would be grand.) 
Directions for further exploration are missing, hands-on teaching opportunities foregone. For 
instance, a Davies-style exploration of TLS on-the-wire, with notes on Wireshark and 
\code{SSLKEYLOGFILE} -- perhaps as dirty as readers can get, without bursting out soldering 
irons. (Cf. Davies's \emph{Implementing SSL/TLS: Using Cryptography and PKI}.)
Mistakes and issues are no doubt numerous. (Some are flagged by tokens \texttt{\textbackslash{}ifSpecNotes}, 
\texttt{\textbackslash{}ifImplNotes}, and \texttt{\textbackslash{}ifPresentationNotes}.)

%Publication is antiquated; manuscripts can continually evolve: 
\paragraph{Publication is antiquated; help evolve this manuscript:}
I %The authors
encourage \emph{you} to improve this manuscript. Fix a typo. Patch grammar.
Revise awkward, overcomplicated, or otherwise poorly-written passages. 
Contribute an entire section. Help evolve this manuscript:
\begin{center}
  \url{https://github.com/BenSmyth/tls-tutorial/}
\end{center}
(Perhaps get in touch prior to writing an entire section! We should probably reach consensus 
on what to add.) Contributions will be recognised through acknowledgements or co-authorship.

\section*{Change history}

\begin{tabular}{l|l}
Date        & Description \\ \hline
22 Jan 2019 & First draft of \S\ref{sec:intro}, \S\ref{sec:handshake}, \& \S\ref{sec:record}    \\
22 Feb 2019 & Added first draft of \S\ref{sec:JSSE}, plus minor revisions throughout            \\
27 May 2020 & Added first draft of \S\ref{sec:handshakeEarlyData}, plus minor additions and 
              revisions elsewhere                                                               \\
30 Sep 2020 & GitHub release, plus minor revisions
\end{tabular}

\section*{Acknowledgements}

IETF's TLS mailing list provided useful insights and I am particularly grateful 
to Eric Rescorla. Adam Petcher corrected misattribution of Java code to Oracle, 
rather than OpenJDK.

\newpage

\section*{Copyright and licensing}
%This manuscript contains source code by Oracle under license ``GNU General Public License v2, 
%with the Classpath Exception (GPLv2+CPE),'' which is available from the following URL: 
%\url{http://hg.openjdk.java.net/jdk/jdk11/file/1ddf9a99e4ad/LICENSE}.

Extracts of Oracle's code are subject to the following copyright and licensing notice:

\begin{quote}
Copyright (c) 1996, 2003, 2015, 2018, Oracle and/or its affiliates. All rights reserved.

This code is free software; you can redistribute it and/or modify it
under the terms of the GNU General Public License version 2 only, as
published by the Free Software Foundation.  Oracle designates this
particular file as subject to the ``Classpath'' exception as provided
by Oracle in the LICENSE file that accompanied this code.

This code is distributed in the hope that it will be useful, but WITHOUT
ANY WARRANTY; without even the implied warranty of MERCHANTABILITY or
FITNESS FOR A PARTICULAR PURPOSE.  See the GNU General Public License
version 2 for more details (a copy is included in the LICENSE file that
accompanied this code).

You should have received a copy of the GNU General Public License version
2 along with this work; if not, write to the Free Software Foundation,
Inc., 51 Franklin St, Fifth Floor, Boston, MA 02110-1301 USA.

Please contact Oracle, 500 Oracle Parkway, Redwood Shores, CA 94065 USA
or visit www.oracle.com if you need additional information or have any
questions.
\end{quote}

\noindent
The LICENSE file is available from the following URL: 
\url{http://hg.openjdk.java.net/jdk/jdk11/file/1ddf9a99e4ad/LICENSE}.

\newpage

\tableofcontents
%\listoffigures
\newpage

\lstlistoflistings
\newpage

\section{Introduction}\label{sec:intro}

%
%Sources suggest we 
We are nearing an all-encrypted Internet; 
yet, the underlying encryption technology %used to secure communication channels
is only understood by a select few. 
%% "a select few" is idiomatic
This manuscript broadens understanding by exploring TLS, an encryption technology 
used to protect application layer communication (including HTTP, FTP and SMTP traffic), 
and by examining OpenJDK's Java implementation. 
%We focus on the most recent version of which is defined by RFC8446,
We focus on the most recent TLS release, namely, version 1.3, which is defined by RFC~8446.
\ifPresentationNotes
\textcolor{red}{needs extending/revising or just plain rewriting}
\fi

\ifPresentationNotes
\marginpar{The history of TLS appears in many manuscripts on TLS. 
  It has been done; it can probably be omitted here. (If not, then push to a sidebar 
  or an appendix.)}
\fi

TLS is a protocol that establishes a channel between an initiating \emph{client} and a 
interlocutory \emph{server} (also known as \emph{endpoints} and \emph{peers}),
%. The protocol is designed to enable:
for the purpose of enabling:

\begin{description}

\item Authentication. 
  %The client's belief of the server's identity
  %is correct, and similarly for the server's belief.
  An endpoint's belief of their peer's identity is correct.

\item Confidentiality. Communication over an established channel is only
  visible to endpoints.

\item Integrity. Communication over an established channel is received-as-sent, 
  or tampering is detected.

\end{description}

\noindent
These properties should hold even in the presence of an adversary that has 
complete control of the underlying network, i.e., an adversary that may read, 
modify, drop, and inject messages. 

\ifPresentationNotes
\marginpar{\emph{cryptographic primitives} vs. \emph{cryptographic schemes} vs. ...}
\fi

The TLS protocol commences with a \emph{handshake}, wherein cryptographic primitives 
and parameters are negotiated, and shared (traffic) keys are established. 
Moreover, the handshake %\sout{typically} 
includes unilateral authentication of the server. (Mutual authentication of both 
the client and the server is also possible.) %\sout{, as-is unauthenticated communication}.) 
The handshake results in a channel which uses the negotiated cryptography and 
parameters, along with a shared key, to protect communication.

%\marginpar{I'd like to use \emph{key} only for traffic keys, and \emph{secrets} 
%  for the inputs used to derive them. But, the term \emph{pre-shared keys} makes
%  that impossible and (EC)DHE key share \& (EC)DHE key makes it awkward.}

\begin{comment}
Shared keys are established using one of the three supported key exchange 
modes: Ephemeral Diffie-Hellman over finite fields (DHE) or elliptic curves (ECDHE), 
pre-shared key (PSK), or PSK with (EC)DHE. (EC)DHE key exchange requires no prior knowledge, whereas PSK-based key exchanges requires knowledge of a pre-shared key, 
which may have been established out-of-band or during a previous connection. Such a 
pre-shared key also serves to authenticate endpoints, whereas (EC)DHE-only key 
exchange is reliant on asymmetric cryptography for authentication.
\end{comment}
%
% The above implicitly assumes knowledge of key exchange, let's try to focus
% on the high-level functional requirement instead:
%
%\begin{comment}
The handshake does not require any prior knowledge: A shared key may be derived 
from secrets established using Diffie-Hellman key exchange over finite fields (DHE) 
or elliptic curves (ECDHE). Alternatively, %endpoints may derive a shared (traffic) key 
such a shared key may be derived from a secret pre-shared key (PSK), which %they 
endpoints
establish %out-of-band 
externally or during a previous connection. (Shared keys 
%are derived from underlying secrets 
are combined with nonces to ensure they are always unique, regardless of whether
secrets have been previously used.)
The former achieves \emph{forward secrecy} -- i.e., confidentiality is 
preserved even if long-term keying material is compromised after the handshake, 
as long as (EC)DHE secrets are erased  -- whereas the latter does not.
The two key exchange modes can be combined, using PSK with (EC)DHE key exchange,
to achieve forward secrecy with pre-shared keys. 
%\sout{Pre-shared keys serve to authenticate endpoints, whereas (EC)DHE-only key 
%exchange is reliant on asymmetric cryptography for authentication. }
%\end{comment}
%
% If using the above, mention PSK with (EC)DHE
%

The handshake is itself a protocol (summarised in Figure~\ref{fig:handshake}). 
It is commenced by the client sending a 
\emph{\ClientHello} message, comprising: a nonce; offered protocol versions,  
symmetric ciphers, and hash functions; offered Diffie-Hellman key 
shares, pre-shared key labels, or both; 
and %any additional extensions.
details of any extended functionality.
The protocol proceeds with the server receiving the client's 
message, establishing mutually acceptable cryptographic primitives and parameters, 
and responding with a \emph{\ServerHello} message, containing: a nonce; 
selected protocol version, symmetric cipher, and hash function; and
a Diffie-Hellman key share, a selected pre-shared key label, or both. (The server 
may respond with a \emph{\HelloRetryRequest} message, if the offered key shares
are unsuitable.) Once the client receives the server's message, a shared
(handshake traffic) key can be established to enable confidentiality and integrity for the 
remainder of the handshake protocol. In particular, that shared key
is used to protect an \emph{\EncryptedExtensions} message, sent by the server to the 
client, which may detail extended functionality.

The handshake protocol concludes with unilateral authentication of the 
server. (Client authentication is also possible.) For (EC)DHE-only key 
exchange, after sending the \EncryptedExtensions\ message, the server sends 
a \emph{\Certificate} message, containing a certificate (or some other 
suitable material corresponding to the server's long-term, private key), 
and a \emph{\CertificateVerify} message, containing a signature 
(using the private key corresponding to the public key in the certificate)
over a hash of the handshake protocol's \emph{transcript} (i.e., a 
concatenation of each handshake message, e.g., \ClientHello, \ServerHello,
\EncryptedExtensions, and \Certificate, in this instance).
Finally, the server 
sends a \emph{\Finished} message, containing a Message Authentication Code (MAC) 
over the protocol's transcript, which provides key confirmation, binds 
the server's identity to the exchanged keys, and, for PSK-based
key exchange, authenticates the handshake. Moreover,
the client responds with a \Finished\ message of its own. A shared (application
traffic) key can then be established to protect communication of application data. 

\begin{figure}
\caption[Handshake protocol]{
  A client initiates the handshake protocol by sending a \ClientHello\ (CH) message. 
  After sending that message, the client waits for a \ServerHello\ (SH) message 
  followed by an \EncryptedExtensions\ (EE) message, or a \HelloRetryRequest\ (HRR) 
  message. An \EncryptedExtensions\ message might be followed by a \CertificateRequest\
  (CR) message (requesting client authentication). Moreover, for certificate-based
  server authentication, the client waits for a \Certificate\ (CT) message followed 
  by a \CertificateVerify\ (CV) message. The handshake protocol concludes upon 
  an exchange of  \Finished\ (FIN) messages from each of the client and server. 
  (We omit the client's \Finished\ message for brevity.) The client's \Finished\ 
  message may be preceded by client generated \Certificate\ and \CertificateVerify\ 
  messages, when client authentication is requested. (We omit those messages for
  brevity.) 
}
\label{fig:handshake}
\begin{tikzpicture}
  \node[state, initial] (1) {Begin};
  \node[state, right=of 1] (2) {Wait};
  \node[state, right=of 2] (3) {Wait};
  \node[state, right=of 3] (4) {Wait};
  \node[state, right=of 4] (5) {Wait};
  \node[state, right=of 5] (6) {Wait};
  \node[state, right=of 6, accepting] (7) {End};
  
  \draw (1) edge[bend left, above] node{CH} (2)
        (2) edge[bend left, below] node{HRR} (1)
        (2) edge[above] node{SH} (3)
        (3) edge[above] node{EE} (4)
        (4) edge[looseness=4,above] node{CR} (4)
        (4) edge[bend right, above] node{FIN} (7)  
        (4) edge[above] node{CT} (5)  
        (5) edge[above] node{CV} (6)
        (6) edge[above] node{FIN} (7);
\end{tikzpicture}
\end{figure}

Beyond the handshake protocol, TLS defines a \emph{record protocol} which 
writes handshake protocol messages (and application data, as well as 
error messages) to the transport layer, after adding headers and, where
necessary, protecting messages. 

\paragraph{Contribution and structure.}

We explore the TLS handshake (\S\ref{sec:handshake}) and record (\S\ref{sec:record})
protocols, as defined by RFC~8446,\footnote{\url{https://tools.ietf.org/html/rfc8446}.}
moreover, we examine OpenJDK's Java 
implementation,\footnote{\url{http://hg.openjdk.java.net/jdk/jdk11/file/1ddf9a99e4ad/}.} 
namely, JDK~11 package \code{sun.security.ssl}.

\section{Handshake protocol}\label{sec:handshake}

A client initiates the handshake protocol from an initial context detailing the client's 
expectations, e.g., willingness to use particular cryptographic primitives and parameters.
%Similarly, a server participates in the handshake protocol with an initial context too. 
A server participates with a similar initial context.
Those contexts evolve during the handshake protocol, to reach agreement on cryptographic
primitives and parameters, along with shared keys.
(The protocol may abort if the endpoints cannot reach agreement.)

\ifPresentationNotes\marginpar{Styling of \texttt{tcolorbox}s could be better}\fi

\begin{tcolorbox}
Client and server contexts are implemented by classes \code{ClientHandshakeContext} and
\code{ServerHandshakeContext}, respectively, that share parent
\code{HandshakeContext} (which implements empty interface \code{ConnectionContext}).
Those classes are both parameterised by instances of classes \code{SSLContextImpl} (with parent
\code{SSLContext}) and \code{TransportContext} (which also implements empty 
interface \code{ConnectionContext}), which define initial contexts.
\end{tcolorbox}

  \subsection{\ClientHello}\label{sec:handshakeCH}\label{sec:CH}

\ifPresentationNotes
\marginpar{Maybe customise use of \texttt{lstinline}, which is wrapped inside 
  command \texttt{code} (defined in \texttt{main-tls-intro.tex}), to improve inline code listings.\\
  \textbf{Dear \LaTeX\ guru}, Please fix this, it will make reading this manuscript
  far easier for you (and everyone else). Please also fix other presentational/formatting issues.}
\fi

The handshake protocol is initiated by a \ClientHello\ message, %(specified in Appendix~\ref{sec:specification})
comprising the following fields:

\begin{description}

\item \TLSlegacyVersion: Constant 0x0303.
%
%\footnote{Previous versions of TLS used this field for the highest 
%  offered protocol version of the client, but experience has shown that
%  servers implement version negotiation rather poorly, in particular, 
%  some servers reject \ClientHello\ messages with a version number
%  higher than it supports. In TLS 1.3, the client offers protocol 
%  versions in an extension.}
%
(Previous versions of TLS used this field for the client's highest offered 
  protocol version. In TLS 1.3, protocol versions are offered in an 
  extension, as explained below.)
  % The field remains for backwards compatibility.)

\item \TLSrandom: A 32 byte nonce.

\item \TLSlegacySessionId: A zero-length vector, except to 
  resume an earlier pre-TLS 1.3 session or for ``compatibility 
  mode.'' (Previous versions of TLS used this field for 
  ``session resumption.'' In TLS 1.3, that feature has been 
  merged with pre-shared keys.)
  %Again, the field remains for backwards compatibility.)

\item \TLScipherSuites: A list of offered \emph{symmetric cipher suites} in
  descending order of client preference, where a suite defines a value identifying an Authenticated Encryption 
  with Associated Data (AEAD) algorithm and a hash function (Table~\ref{table:suites}).\footnote{
    Support for cipher suite TLS\_AES\_128\_GCM\_SHA256 is mandatory (unless an implementation 
    explicitly opts out), and cipher suites TLS\_AES\_256\_GCM\_SHA384 and 
    TLS\_CHACHA20\_POLY1305\_SHA256 should also be supported.
  }

\begin{table}[tbp]
\caption{Symmetric cipher suites defined by a value identifying
  an AEAD algorithm and a hash function. Suites are named in the 
  format TLS\_AEAD\_HASH, where AEAD and HASH are replaced by the 
  corresponding algorithm and function names.}
\label{table:suites}

\centering

\begin{tabular}{l|l}
  Name                             & Value  \\ \hline
  TLS\_AES\_128\_GCM\_SHA256       & 0x1301 \\
  TLS\_AES\_256\_GCM\_SHA384       & 0x1302 \\
  TLS\_CHACHA20\_POLY1305\_SHA256  & 0x1303 \\
  TLS\_AES\_128\_CCM\_SHA256       & 0x1304\\
  TLS\_AES\_128\_CCM\_8\_SHA256    & 0x1305
\end{tabular}
\end{table}

\item \TLSlegacyCompressionModes: Constant 0x00. %byte set to zero. 
  (Previous versions of TLS used this field to list supported compression methods.
  In TLS 1.3, this feature has been removed.)
  % The field remains for backwards compatibility.)

\ifPresentationNotes\marginpar{Maybe mention that hex values are used in place of (extension) names}\fi
\item \TLSextensions: A list of \emph{extensions}, where an extension comprises 
  a name along with associated data. The list must contain at least extension
  \TLSsupportedVersions\ %(0x002B) \marginpar{drop hex names?}
  associated with a list of offered protocol versions 
  in descending order of client preference, minimally including constant 0x0304,
  denoting TLS 1.3.

\end{description}

\noindent
Legacy fields \TLSlegacyVersion, \TLSlegacySessionId, and \TLSlegacyCompressionModes\
are included for backwards compatibility.

\ifPresentationNotes
\marginpar{Perhaps elaborate on the produce/consume design}

\marginpar{I've overridden \texttt{listings.sty} (in \texttt{main-tls-intro.tex}) 
  to add vertical space (controlled by \texttt{tinyskip}) at the end of a 
  \texttt{linerange}, e.g., see Listing~\ref{lst:ClientHelloMessage}, between 
  Lines~71~\&~74, for example. I'm unsure whether it helps.}
\fi

\begin{tcolorbox}
The \ClientHello\ message is implemented by class 
\code{ClientHello.ClientHelloMessage} 
%(Listings~\ref{lst:ClientHelloMessage} \&~\ref{lst:ClientHelloMessageB}). 
(Listing~\ref{lst:ClientHelloMessage}).
Instances of that class are produced by class 
\code{ClientHello.Client\-Hello\-Kickstart\-Producer} 
(Listing~\ref{lst:ClientHelloKickstartProducer}), %\footnotemark\ 
which is instantiated as static constant \code{ClientHello.kickstart\-Producer}. That constant is used by method 
\code{SSLHandshake.kickstart}. 
\begin{comment}
Class \code{ClientHello} is reliant on classes 
\code{SSLExtensions} %(Listings~\ref{lst:SSLExtensions} \& \ref{lst:SSLExtensionsB}) 
and \code{SSLExtension} (Appendix~\ref{sec:extensions}) 
for extensions.
\end{comment}
\end{tcolorbox}

\begin{comment}
\footnotetext{Class \code{ClientHello.ClientHelloKickstartProducer} implements
  interface \code{SSLProducer}, which defines a single method --
    namely, \code{byte[] produce(ConnectionContext context) throws IOException}
  -- common to \ClientHello\ messages originating from the client, %\HelloRequest\ and 
  \NewSessionTicket\ messages originating from the server, and 
  \KeyUpdate\ messages originating from either endpoint. Parameter \code{context} defines the active context and may 
  be cast to children \code{ClientHandshakeContext} and \code{ServerHandshakeContext}, 
  as seen in Line~398 of Listing~\ref{lst:ClientHelloKickstartProducer}, for instance.
  %That context is updated during production of \ClientHello\ %, \HelloRequest,
  %and \NewSessionTicket\ messages, whereas production of \KeyUpdate\ messages
  %concludes by returning \textcolor{red}{``the encoded producing'' (reword)}.
  That context is updated during production of \ClientHello, \KeyUpdate\
  and \NewSessionTicket\ messages.
  \textcolor{red}{Are these details worth knowing?}
}
\end{comment}

\begin{lstlisting}[
  float=tbp,
  linerange={
    71-71,%class declaration
    74-76,79-83,%variable declaration
    84-88,90-93,100-100,102-102,105-107,%constructor
    160-162,165-167,185-190,192-192,194-200,%constructor
    312-316,318-322,326-329,%ClientHello output stream
    382-383%closing brace
  },
  label=lst:ClientHelloMessage,
  caption={[\code{ClientHello.ClientHelloMessage} defines \ClientHello]
  Class \code{ClientHello.ClientHelloMessage} defines the six fields 
  of a \ClientHello\ message (Lines~74--81) and constructors to instantiate them 
  from parameters (Lines~85--106) or an input buffer (Lines~160--200). The former 
  constructor does not populate the extensions field (and a call to method \code{SSLExtensions.produce},
  Listing~\ref{lst:SSLExtensions}, is required), whereas the latter may (Line~195--196).
  Method \code{send} (Lines~312--316) writes those fields to an output stream, using method 
  \code{sendCore} (Lines~318--328) to write all fields except the extensions 
  field, which is written by method \code{SSLExtensions.send} 
  (Listing~\ref{lst:SSLExtensions}, Lines~293--307). 
}]
    static final class ClientHelloMessage extends HandshakeMessage {
\end{lstlisting}

\begin{comment}
\lstinputlisting[
  float=tbp,
  linerange={
%    281-289,%getEncodedCipherSuites()
    312-322,326-329,%ClientHello output stream
    382-383%closing brace
  },
  label=lst:ClientHelloMessageB,
  caption={Class \code{ClientHello.ClientHelloMessage} (continued from 
  Listing~\ref{lst:ClientHelloMessage}) defines method \code{send} (Lines~312--316) 
  to write \ClientHello\ message fields to an output stream, using method 
  \code{sendCore} (Lines~318--327) to write all fields except the extensions 
  field, which is written by method \code{SSLExtensions}.\code{send} 
  (Listing~\ref{lst:SSLExtensions}, Lines~293--307).
}]{listings/ClientHello.java}
\end{comment}

\begin{lstlisting}[
  float=tbp,
  linerange={
    50-51,%kickstartProducer
    387-388,%class declaration
    396-396,%method declaration
    398-398,%clientHandshakeContext
%    404-404,
    407-407,%sessionId
    410-410,%cipherSuites
%%%%%%%
    415-419,
%    420-426,
%    428-433, %chc.reservedServerCerts
%    435-436,
%    442-443,
    445-447, %sessionSuite
%    455-456,
%    458-462,
%    468-469,
%    513-513,
 %   522-529,
%    534-534,
    542-542,
    553-555,
%%%%%%
%    607-608,
    615-615,%clientVersion
%    617-617,
    618-621,%Construct ClientHelloMessage
    622-630,%cache clientRandom & clientHelloVersion
    635-638,%output
    639-646,%manage clientHandshakeContext
    653-657%closing braces etc.
  },
  label=lst:ClientHelloKickstartProducer,
  caption={[\code{ClientHello.ClientHelloKickstartProducer} produces \ClientHello]
  Class \code{ClientHello.ClientHelloKickstartProducer} defines 
  method \code{produce} which instantiates a \ClientHello\ message 
  (Lines~619--621), populates the extension field for the active context 
  (Lines~628--630), writes the \ClientHello\ message to an output 
  stream (Lines~637--638), and prepares the client's active context for the server's 
  response (Lines~624--625, 642, \& 645--646).
  The \ClientHello\ message parameterises \TLSlegacySessionId\ as a zero-length 
  byte array (Line~407); \ifImplNotes\textcolor{red}{truthify the following (EC)DHE
  vs. PSK}\fi \TLScipherSuites\ as the list of available cipher suites, for 
  (EC)DHE-only key exchange (Line~410), or as a list containing the cipher suite 
  %used by the previous session,
  associated with the pre-shared key, 
  for PSK-based key exchange (Line~542); and \TLSlegacyVersion\ as constant 0x0303 (Line~615). 
  (Prior versions of TLS are supported by the class and constants other than 
  0x0303 may be assigned to \TLSlegacyVersion. We omit those details for brevity.)
  The output stream is written-to using method \code{ClientHello.ClientHelloMessage.write}, 
  defined by parent class \code{SSLHandshake.HandshakeMessage}, which in 
  turn uses method \code{ClientHello.ClientHelloMessage.send}
  (Listing~\ref{lst:ClientHelloMessage}).
}]
    static final SSLProducer kickstartProducer =
        new ClientHelloKickstartProducer();
\end{lstlisting}

The primary goal of the handshake protocol is to establish a channel that protects 
communication using one of the symmetric cipher suites offered by the client
and a key shared between the endpoints. That key is derived
from (secret) client and server key shares for 
%Ephemeral Diffie-Hellman key exchange over finite fields (DHE) or elliptic curves (ECDHE), 
(EC)DHE key exchange, from a (secret) pre-shared key for PSK-only key exchange, or by a combination of 
key shares and a pre-shared key for PSK with (EC)DHE key exchange. The 
desired key exchange mode determines which extensions to include: 
  For (EC)DHE, extensions \TLSsupportedGroups\ and \TLSkeyShare\ are included;
  for PSK-only, extensions \TLSpsk\ and \TLSpskModes\ must be included, and
  extensions \TLSsupportedGroups\ and \TLSkeyShare\ may be included to allow 
  the server to decline resumption and fall back to a full handshake; and 
  for PSK with (EC)DHE, all four of the aforementioned extensions are included.
Those extensions are associated with data:

\begin{description}

\item \TLSsupportedGroups\ and \TLSkeyShare: A list of offered %\sout{ephemeral}
  Diffie-Hellman groups for key exchange (\TLSsupportedGroups) 
  and key shares for some or all of those groups (\TLSkeyShare), in  descending 
  order of client preference. Groups may be selected over finite fields or
  elliptic curves.\footnotemark\ A key share for a particular group must 
  be listed in the same order that the group is listed. However, a key share 
  for a particular group may be omitted, even when a key share for a less
  preferred group is present. This situation could arise when a group is 
  new or lacking support, making key shares for such groups redundant
  and wasteful. An empty vector of key shares can be used to request 
  group selection from the server. (Servers respond with \HelloRetryRequest\
  messages when no key share is offered for the server selected group.)

\ifSpecNotes
\textcolor{red}{Why include both extensions? They overlap. In particular, \TLSsupportedGroups\
  is associated with a list of type \texttt{NamedGroup} and \TLSkeyShare\ is associated with
  list of pairs, where the first element of each pair is also of type \texttt{NamedGroup}
  (the second element contains the key share and a zero-length key share could be used
  when a key share isn't offered for the group), hence, the group is repeated.}
\fi

\footnotetext{Supported groups include: 
  Finite field groups defined in RFC~7919, namely, 
    \TLSConstant{ffdhe2048} (0x0100), 
    \TLSConstant{ffdhe3072} (0x0101), 
    \TLSConstant{ffdhe4096} (0x0102),
    \TLSConstant{ffdhe6144} (0x0103), and
    \TLSConstant{ffdhe8192} (0x0104), and
  elliptic curve groups defined in either FIPS 186-4 
  or RFC~7748, namely, 
    \TLSConstant{secp256r1} (0x0017), 
    \TLSConstant{secp384r1} (0x0018), 
    \TLSConstant{secp521r1} (0x0019),
    \TLSConstant{x25519} (0x001D), and
    \TLSConstant{x448} (0x001E).
  Supporting group \TLSConstant{secp256r1} is mandatory (unless an implementation explicitly 
  opts out), and group \TLSConstant{x25519} should also be supported.
}

\item \TLSpsk\ and \TLSpskModes: A list of offered pre-shared
  key identifiers (\TLSpsk) and a key exchange mode for each 
  (\TLSpskModes). (Further details on extension \TLSpsk\
  appear in Section~\ref{sec:NST}, after \NewSessionTicket\ messages -- which 
  establish pre-shared keys for subsequent connections -- are introduced.)
  At least one offered cipher suite should define a hash function associated 
  with at least one of the identifiers. 
  \ifSpecNotes
  \textcolor{red}{A stronger requirement seems desirable here, e.g., a hash 
  function associated with each identifier, but that's not what's required by 
  the spec (p28)}
  \fi
  Key exchange modes include PSK-only (\TLSPskKe) and PSK with (EC)DHE (\TLSPskDheKe).
  Extension \TLSpsk\ must be the last extension in the \ClientHello\ message.
  (Other extensions may appear in any order.)

\end{description}

\noindent
A further goal of the handshake protocol is unilateral authentication
of the server, which for (EC)DHE key exchange mode is achieved by inclusion of extensions 
\TLSsignatureAlgorithms\ and \TLSsignatureAlgorithmsCert\ (for PSK-only and
PSK with (EC)DHE, authentication is derived from the \Finished\ message), 
and associated data:

\begin{description}

\begin{sloppypar}
\item \TLSsignatureAlgorithms\ and \TLSsignatureAlgorithmsCert:
  A list of accepted signature algorithms in descending order of client 
  preference for \CertificateVerify\ messages (\TLSsignatureAlgorithms)
  and \Certificate\ messages (\TLSsignatureAlgorithmsCert).\footnotemark\
  (Extension \TLSsignatureAlgorithmsCert\ may be omitted in favour of extension 
  \TLSsignatureAlgorithms, when accepted algorithms for \Certificate\ and 
  \CertificateVerify\ messages coincide. In such cases,
  algorithms listed by extension \TLSsignatureAlgorithms\ apply to certificates too.)
\end{sloppypar}

\footnotetext{Supported signature algorithms include: 
  RSASSA-PKCS1-v1_5 (RFC8017) or RSASSA-PSS (RFC8017) with 
  a corresponding hash function, namely, SHA256, SHA384, or 
  SHA512;
  ECDSA (American National Standards Institute, 2005) with 
  a corresponding curve \& hash function, namely, 
  secp256r1 \& SHA256, secp384r1 \& SHA384, or 
  secp521r1 \& SHA512; and
  EdDSA (RFC8032). (RSASSA-PKCS1-v1_5 is only supported for 
  \Certificate\ messages.) 
  Supporting RSA-based signatures with SHA256 (for certificates)
  and ECDSA signatures with secp256r1 \& SHA256 is mandatory (unless an implementation 
  explicitly opts out). (RSASSA-PSS must also be supported for \CertificateVerify\ messages.)
}

%\item certificate\_authorities (47)

\end{description}

\noindent
Additional extensions exist and may be included in \ClientHello\ messages.
(Appendix~\ref{sec:extensions} lists all extensions.)

A \ClientHello\ message is consumed by the server: The 
server first checks that the message is a TLS 1.3 \ClientHello\ 
message, which is achieved by checking that extension \TLSsupportedVersions\
is present and that constant 0x0304 is the first listed preference.
(The \ClientHello\ message format is backward compatible with 
previous versions of TLS, hence, the message might need to be processed
by a prior version of TLS. Those details are beyond the scope of this 
manuscript.) The server may 
also check that field \TLSlegacyVersion\
is set to constant 0x0303 and field \TLSlegacySessionId\ is set to a zero-length vector.
Moreover, the server checks field \TLSlegacyCompressionModes\ is set to constant 0x00
and aborts with an \TLSillegalParameter\ alert if the check fails.\footnote{%
  RFC 8446 does not explicitly require servers to check fields 
  \TLSlegacyVersion\ and \TLSlegacySessionId, it merely requires clients to
  set those fields correctly. Accordingly, we assume servers \emph{may} 
  perform these checks, rather than mandating them. By comparison,
  RFC 8446 explicitly requires field \TLSlegacyCompressionModes\ to be
  correctly set.
}
Next,\label{comp:CH:cons:cipher} the server selects an acceptable cipher suite from field \TLScipherSuites, 
disregarding suites that are not recognised, unsupported, or otherwise unacceptable, 
and aborting with a \TLShandshakeFailure\ or an \TLSinsufficientSecurity\ alert if no 
mutually acceptable cipher suite exists. Finally, the server processes any remaining
extensions:

\begin{description}

\item \TLSsupportedGroups\ and \TLSkeyShare: The server selects an acceptable group 
  from the list; aborting with a \TLSmissingExtension\ alert if extension \TLSsupportedGroups\ 
  is present and extension \TLSkeyShare\ is absent, or vice versa; aborting with a 
  \TLShandshakeFailure\ or an \TLSinsufficientSecurity\ alert if no mutually acceptable 
  group exists; and \label{comp:CH:cons:HRR} responding with a \HelloRetryRequest\ 
  message if extension \TLSkeyShare\ does not offer a key share for the selected group.

\item \TLSpsk\ and \TLSpskModes: The server selects an acceptable key identifier
  from the list (\ifSpecNotes\textcolor{red}{the spec requires that identifier to be compatible 
  with the server-selected cipher suite. That's rather vague. I think compatibility varies 
  between out-of-band and NST provisioned keys. I think the following captures the meaning 
  of compatibility:}\fi{}that identifier must be associated with a hash function, 
  AEAD algorithm, or both, which are defined by the server-selected cipher suite), 
  disregarding unknown identifiers, aborting with an \TLSillegalParameter\ 
  alert if extension \TLSpsk\ is not the last extension in the \ClientHello\ message,
  and aborting if extension \TLSpsk\ is present without \TLSpskModes.
  The server also selects a key exchange mode.\label{comp:CH:cons:psk}
  If no mutually acceptable key identifier exists and extensions \TLSsupportedGroups\ 
  and \TLSkeyShare\ are present, then the server should perform a non-PSK handshake.

\item \TLSsignatureAlgorithms\ and \TLSsignatureAlgorithmsCert:
  The server selects acceptable signature algorithms for \CertificateVerify\ 
  and \Certificate\ messages.

\end{description}

\noindent
Any unrecognised extensions are ignored and the server aborts with a 
\TLSmissingExtension\ alert if extension \TLSpsk\ is absent as-is
either extension \TLSsupportedGroups, \TLSsignatureAlgorithms, or 
both. (Alerts are formally defined by RFC 8446, as 
discussed in Appendix~\ref{sec:alerts}.)

\begin{tcolorbox}
Consumption is implemented by class 
\code{ClientHello.ClientHelloConsumer} (Listing~\ref{lst:ClientHelloConsumer}).
That class checks the presence of extension \TLSsupportedVersions, to determine
whether the message is a TLS 1.3 \ClientHello\ message, and the remainder of the 
message is processed by class \code{ClientHello.T13ClientHelloConsumer} 
(Listings~\ref{lst:T13ClientHelloConsumer} \&~\ref{lst:T13ClientHelloConsumerC}), 
if it is a TLS 1.3 message. %\footnotemark\
\begin{comment}
which is reliant on classes \code{SSLExtensions} 
%(Listings~\ref{lst:SSLExtensions} \&~\ref{lst:SSLExtensionsB}) 
and \code{SSLExtension} 
%(Listing~\ref{lst:SSLExtension}).\footnotemark\
(Appendix~\ref{sec:extensions}).
\end{comment}
%The latter is in turn reliant on a variable \code{onLoadConsumer} of (interface) type 
%\code{SSLExtension.ExtensionConsumer} which is instantiated by a constant in 
%the form \code{ThisNameExtension.chOnLoadConsumer}, where \code{ThisName}
%corresponds to extension \TLSField{this\_name}. 
Consumption of the \ClientHello\ message may result in the server aborting or 
responding with either a \ServerHello\ or \HelloRetryRequest\ message.
\end{tcolorbox}

\begin{comment}
\footnotetext{Class \code{ClientHello.ClientHelloConsumer} implements
  interface \code{SSLConsumer}, which defines a single method, namely, 
  \code{void consume(ConnectionContext context, ByteBuffer message) throws IOException}.
  Similarly, class \code{ClientHello.T13ClientHelloConsumer} implements
  interface \code{HandshakeConsumer}, which defines method
  \code{void consume(ConnectionContext context, HandshakeMessage message) throws IOException}.
  Parameter \code{context} defines the active context in both interfaces 
  and may be cast to children \code{ClientHandshakeContext} and \code{ServerHandshakeContext}, 
  as seen in Line~770 of Listing~\ref{lst:ClientHelloConsumer}, for instance.
  \textcolor{red}{Are these details worth knowing?}
}
\end{comment}

\begin{lstlisting}[
  float=tbp,
  linerange={
    52-53,%handshakeConsumer
    760-760,%ClientHelloConsumer class declaration
    767-768,%consume method delaration
    770-770,
    781-783,
    785-786,
    791-793,
    795-796,
%    800-816,
    800-808,
    %810-811,
    809-815,
    816-816,
    %831-831,
    830-834,
    836-836,
%    840-842,%negotiateProtocol according to ClientHello.client_version
%    844-846,
%    852-854,
%    857-859,
%    868-868,
    873-875,
    877-880,
    884-884,
    888-892,
    902-903
  },
  label=lst:ClientHelloConsumer,
  caption={[\code{ClientHello.ClientHelloConsumer} consumes generic \ClientHello]
  Class \code{ClientHello.ClientHelloConsumer} defines method \code{consume} to 
  instantiate a (generic) \ClientHello\ message from an input buffer (Lines~785--786); update the 
  server's active context to include the client's offered versions (Lines~800--803), 
  indirectly using method 
  \code{SupportedVersionsExtension.CHSupportedVersionsConsumer.consume}, which calls
  \code{context.handshakeExtensions.put(CH\_SUPPORTED\_VERSIONS, spec)}, where 
  parameter \code{spec} is a byte array encoding of extension \TLSsupportedVersions;
  %(variable \code{context} is named \code{shc} in the method);
  select the first server preferred version that the client offered (Lines~810--811 \& 880--892); 
  and update the active context to include that selected version preference as the negotiated 
  protocol (Line~816). Further processing is deferred (Line 831) to class 
  \code{ClientHello.T13ClientHelloConsumer} (Listing~\ref{lst:T13ClientHelloConsumer}).
  \ifPresentationNotes\textcolor{red}{Maybe mention consumption of TLS 1.2 \ClientHello\ messages}\fi
}]
    static final SSLConsumer handshakeConsumer =
        new ClientHelloConsumer();
\end{lstlisting}

\begin{lstlisting}[
  float=tbp,
  linerange={
    59-60,
    1075-1076,
    1083-1084,
    1086-1087,
    1097-1119,
    1120-1127
  },
  label=lst:T13ClientHelloConsumer,
  caption={[\code{ClientHello.T13ClientHelloConsumer} consumes \ClientHello]
  Class \code{ClientHello.T13ClientHelloConsumer} defines method \code{consume}
  to process incoming (TLS 1.3) \ClientHello\ messages (further to processing 
  shown in Listing~\ref{lst:ClientHelloConsumer}). The method updates the server's 
  active context to include
  any pre-shared key identifiers and key exchange modes offered by the client (Lines~1101--1105),
  indirectly using the \code{consume} method of classes 
  \code{PskKeyExchangeModesExtension.PskKeyExchangeModesConsumer} 
  and \code{PreSharedKeyExtension.CHPreSharedKeyConsumer}; updates the active 
  context to include any further (enabled) extensions 
  (Lines~1113--1119), excluding those that have already been added to the active context,
  namely, extensions \TLSsupportedVersions, \TLSpsk, and \TLSpskModes; and proceeds by 
  producing either a \HelloRetryRequest\ message if extension \TLSkeyShare\ does not
  offer a key share for the server selected group (method  
  \code{KeyShareExtension.CHKeyShareConsumer.consume} may add a producer 
  for \HelloRetryRequest\ messages which ensures \code{!shc.handshakeProducers.isEmpty()} 
  holds) or a \ServerHello\ message otherwise (Lines~1121--1126).
  \ifImplNotes
  \textcolor{blue}{Contrary to comments (Lines~1107--1109) extension key\_share doesn't 
  appear to be ignored in Lines 1113--1119, extensions \TLSpskModes, \TLSpsk, and 
  \TLSsupportedVersions\ are. (If it were ignored, then \HelloRetryRequest\ messages 
  would never be produced.) }
  \fi
}]
    private static final HandshakeConsumer t13HandshakeConsumer =
            new T13ClientHelloConsumer();
    private static final
            class T13ClientHelloConsumer implements HandshakeConsumer {
        public void consume(ConnectionContext context,
                HandshakeMessage message) throws IOException {
            ServerHandshakeContext shc = (ServerHandshakeContext)context;
            ClientHelloMessage clientHello = (ClientHelloMessage)message;
            // Check and launch the "psk_key_exchange_modes" and
            // "pre_shared_key" extensions first, which will reset the
            // resuming session, no matter the extensions present or not.
            shc.isResumption = true;
            SSLExtension[] extTypes = new SSLExtension[] {
                    SSLExtension.PSK_KEY_EXCHANGE_MODES,
                    SSLExtension.CH_PRE_SHARED_KEY
                };
            clientHello.extensions.consumeOnLoad(shc, extTypes);

            // Check and launch ClientHello extensions other than
            // "psk_key_exchange_modes", "pre_shared_key", "protocol_version"
            // and "key_share" extensions.
            //
            // These extensions may discard session resumption, or ask for
            // hello retry.
            extTypes = shc.sslConfig.getExclusiveExtensions(
                    SSLHandshake.CLIENT_HELLO,
                    Arrays.asList(
                            SSLExtension.PSK_KEY_EXCHANGE_MODES,
                            SSLExtension.CH_PRE_SHARED_KEY,
                            SSLExtension.CH_SUPPORTED_VERSIONS));
            clientHello.extensions.consumeOnLoad(shc, extTypes);

            if (!shc.handshakeProducers.isEmpty()) {
                // Should be HelloRetryRequest producer.
                goHelloRetryRequest(shc, clientHello);
            } else {
                goServerHello(shc, clientHello);
            }
        }
\end{lstlisting}

\begin{lstlisting}[
  float=tbp,
  linerange={
    1129-1133,
    1135-1135,
    1147-1147,
    1149-1150,
    1154-1154,
    1159-1180,
    1185-1193
  },
  label=lst:T13ClientHelloConsumerC,
  caption={[\code{ClientHello.T13ClientHelloConsumer} consumes \ClientHello\ (cont.)]
  Class \code{ClientHello.T13ClientHelloConsumer} (continued from 
  Listing~\ref{lst:T13ClientHelloConsumer}) defines methods \code{goHelloRetryRequest} 
  to produce a \HelloRetryRequest\ message and \code{goServerHello} 
  to produce a \ServerHello\ message. The latter method prepares the server's active 
  context for the client's response (Lines~1154 \& 1159--1162); updates the active context to 
  include a producer for \ServerHello\ messages (Lines~1168--1169); constructs
  an array of producers that servers might use during the handshake protocol, 
  namely, producers for messages \ServerHello, \EncryptedExtensions, 
  \CertificateRequest, \Certificate, \CertificateVerify, and \Finished,
  in the order that they might be used (Lines~1171--1180); and uses those producers 
  to produce messages when the active context includes the producer
  (Lines~1185--1191). Since a \ServerHello\ message producer is 
  included, a \ServerHello\ message is always produced, using 
  method \code{ServerHello.T13ServerHelloProducer.produce} 
  (Listing~\ref{lst:T13ServerHelloProducer}). That method 
  adds producers for \EncryptedExtensions\ and \Finished\ messages
  (Listing~\ref{lst:T13ServerHelloProducer}, Lines~560--563), 
  since those messages must be sent. Other producers 
  may also be added.
}]
        private void goHelloRetryRequest(ServerHandshakeContext shc,
                ClientHelloMessage clientHello) throws IOException {
            HandshakeProducer handshakeProducer =
                    shc.handshakeProducers.remove(
                            SSLHandshake.HELLO_RETRY_REQUEST.id);
                    handshakeProducer.produce(shc, clientHello);
        }
        private void goServerHello(ServerHandshakeContext shc,
                ClientHelloMessage clientHello) throws IOException {
            shc.clientHelloRandom = clientHello.clientRandom;
            if (!shc.conContext.isNegotiated) {
                shc.conContext.protocolVersion = shc.negotiatedProtocol;
                shc.conContext.outputRecord.setVersion(shc.negotiatedProtocol);
            }

            // update the responders
            //
            // Only ServerHello/HelloRetryRequest producer, which adds
            // more responders later.
            shc.handshakeProducers.put(SSLHandshake.SERVER_HELLO.id,
                SSLHandshake.SERVER_HELLO);

            SSLHandshake[] probableHandshakeMessages = new SSLHandshake[] {
                SSLHandshake.SERVER_HELLO,

                // full handshake messages
                SSLHandshake.ENCRYPTED_EXTENSIONS,
                SSLHandshake.CERTIFICATE_REQUEST,
                SSLHandshake.CERTIFICATE,
                SSLHandshake.CERTIFICATE_VERIFY,
                SSLHandshake.FINISHED
            };
            for (SSLHandshake hs : probableHandshakeMessages) {
                HandshakeProducer handshakeProducer =
                        shc.handshakeProducers.remove(hs.id);
                if (handshakeProducer != null) {
                    handshakeProducer.produce(shc, clientHello);
                }
            }
        }
    }
\end{lstlisting}

\begin{comment}

%Pushed to Listing~\ref{lst:SSLExtensions}
\begin{lstlisting}[
  float=tbp,
  linerange={
    132-134,
    163-164,
    169-170    
  },
  label=lst:SSLExtensionsB,
  caption={Class \code{SSLExtensions}}?
}]
    void consumeOnLoad(HandshakeContext context,
            SSLExtension[] extensions) throws IOException {
        for (SSLExtension extension : extensions) {
            ByteBuffer m = ByteBuffer.wrap(extMap.get(extension));
            extension.consumeOnLoad(context, handshakeMessage, m);
        }
    }
\end{lstlisting}

%Pushed to Listing~\ref{lst:SSLExtension}
\begin{lstlisting}[
  float=tbp,
  linerange={
    539-547
  },
  label=lst:SSLExtensionB,
  caption={Class \code{SSLExtension} 
}]
    public void consumeOnLoad(ConnectionContext context,
            HandshakeMessage message, ByteBuffer buffer) throws IOException {
        if (onLoadConsumer != null) {
            onLoadConsumer.consume(context, message, buffer);
        } else {
            throw new UnsupportedOperationException(
                    "Not yet supported extension loading.");
        }
    }
\end{lstlisting}

\end{comment}

  \subsection{\ServerHello}\label{sec:handshakeSH}\label{sec:SH}

A server that is able to successfully consume a \ClientHello\ message responds
with a \ServerHello\ message, %(specified in Appendix~\ref{sec:specification}), 
comprising fields \TLSlegacyVersion, \TLSrandom, and \TLSextensions\ as per the 
\ClientHello\ message and the following fields:

\begin{description}

\item \TLSlegacySessionIdEcho: The contents of \ClientHello.\TLSlegacySessionId.

\item \TLScipherSuite:  The cipher suite selected by the server from \ClientHello.\TLScipherSuites.

\item \TLSlegacyCompressionMode: Constant 0x00.

\end{description}

\noindent
Legacy fields are included for backwards compatibility.

\begin{tcolorbox}
The \ServerHello\ message is implemented by class \code{ServerHello.ServerHelloMessage}
(Listings~\ref{lst:ServerHelloMessage} \&~\ref{lst:ServerHelloMessageB}). Instances 
of that class are produced by class \code{Server\-Hello.T13\-Server\-Hello\-Producer}
(Listings~\ref{lst:T13ServerHelloProducer} \& \ref{lst:T13ServerHelloProducerB}), %\footnotemark\
which is instantiated as static constant \code{Server\-Hello.t13\-Handshake\-Producer}. That 
constant is used indirectly -- via class \code{SSL\-Handshake.SERVER_HELLO} -- to produce
\ServerHello\ messages in class \code{Client\-Hello.T13\-Client\-Hello\-Consumer}
(Listing~\ref{lst:T13ClientHelloConsumerC}).
\end{tcolorbox}

\begin{comment}
\footnotetext{Class \code{ServerHello.T13ServerHelloProducer} implements
  interface \code{HandshakeProducer}, which defines a single method, namely, 
  \code{byte[] produce(ConnectionContext context, HandshakeMessage message) throws IOException}, 
  where parameter \code{context} defines the active context in both interfaces 
  and may be cast to children \code{ClientHandshakeContext} and \code{ServerHandshakeContext}.
  That context may be updated during production of messages or production may conclude
  by returning \textcolor{red}{``the encoded producing'' (reword)}.
  \textcolor{red}{Are these details worth knowing?}
}
\end{comment}

\begin{lstlisting}[
  float=tbp,
  linerange={
    85-122
  },
  label=lst:ServerHelloMessage,
  caption={[\code{ServerHello.ServerHelloMessage} defines \ServerHello/\HelloRetryRequest]
  Class \code{ServerHello.ServerHelloMessage} defines the six fields 
  of a \ServerHello\ message (Lines~86--91), two additional fields 
  for production and consumption of a \HelloRetryRequest\ message (Lines~96 \&~100), 
  and constructors to instantiate those fields from parameters (Lines~102--122) 
  or an input buffer (Listing~\ref{lst:ServerHelloMessageB}). The former constructor does 
  not populate the extensions field,
  %\sout{ (and a call to method \code{SSLExtensions.produce}, Listing~\ref{lst:SSLExtensions}, is required)}
  whereas the latter may (Listing~\ref{lst:ServerHelloMessageB}, Line~173--174).
}]
    static final class ServerHelloMessage extends HandshakeMessage {
        final ProtocolVersion           serverVersion;      // TLS 1.3 legacy
        final RandomCookie              serverRandom;
        final SessionId                 sessionId;          // TLS 1.3 legacy
        final CipherSuite               cipherSuite;
        final byte                      compressionMethod;  // TLS 1.3 legacy
        final SSLExtensions             extensions;

        // The HelloRetryRequest producer needs to use the ClientHello message
        // for cookie generation.  Please don't use this field for other
        // purpose unless it is really necessary.
        final ClientHelloMessage        clientHello;

        // Reserved for HelloRetryRequest consumer.  Please don't use this
        // field for other purpose unless it is really necessary.
        final ByteBuffer                handshakeRecord;

        ServerHelloMessage(HandshakeContext context,
                ProtocolVersion serverVersion, SessionId sessionId,
                CipherSuite cipherSuite, RandomCookie serverRandom,
                ClientHelloMessage clientHello) {
            super(context);

            this.serverVersion = serverVersion;
            this.serverRandom = serverRandom;
            this.sessionId = sessionId;
            this.cipherSuite = cipherSuite;
            this.compressionMethod = 0x00;      // Don't support compression.
            this.extensions = new SSLExtensions(this);

            // Reserve the ClientHello message for cookie generation.
            this.clientHello = clientHello;

            // The handshakeRecord field is used for HelloRetryRequest consumer
            // only.  It's fine to set it to null for generating side of the
            // ServerHello/HelloRetryRequest message.
            this.handshakeRecord = null;
        }
\end{lstlisting}

\begin{lstlisting}[
  float=tbp,
  linerange={
    124-126,
    128-129,
    131-133,
    141-142,
    149-155,
    157-157,
    163-170,
    172-177,
    179-183,
    205-216,
    246-246
  },
  label=lst:ServerHelloMessageB,
  caption={[\code{ServerHello.ServerHelloMessage} defines \ServerHello/\HelloRetryRequest\ 
    (cont.)]
  Class \code{ServerHello.ServerHelloMessage} (continued from 
  Listing~\ref{lst:ServerHelloMessage}) defines a constructor 
  which instantiates \ServerHello\ or \HelloRetryRequest\ messages
  from an input buffer (Lines~124-183), checking that the server-selected cipher suite 
  (Lines~149--150) is amongst those offered by the client (Lines~151--155), and method 
  \code{send} to write %\ServerHello\ or \HelloRetryRequest\ fields 
  such messages 
  to an output stream, using  
  method \code{SSLExtensions.send} to write the extensions field
  (Listing~\ref{lst:SSLExtensions}, Lines~293--307).
}]
        ServerHelloMessage(HandshakeContext context,
                ByteBuffer m) throws IOException {
            super(context);
            // Reserve for HelloRetryRequest consumer if needed.
            this.handshakeRecord = m.duplicate();
            byte major = m.get();
            byte minor = m.get();
            this.serverVersion = ProtocolVersion.valueOf(major, minor);
            this.serverRandom = new RandomCookie(m);
            this.sessionId = new SessionId(Record.getBytes8(m));
            int cipherSuiteId = Record.getInt16(m);
            this.cipherSuite = CipherSuite.valueOf(cipherSuiteId);
            if (cipherSuite == null || !context.isNegotiable(cipherSuite)) {
                context.conContext.fatal(Alert.ILLEGAL_PARAMETER,
                    "Server selected improper ciphersuite " +
                    CipherSuite.nameOf(cipherSuiteId));
            }
            this.compressionMethod = m.get();
            SSLExtension[] supportedExtensions;
            if (serverRandom.isHelloRetryRequest()) {
                supportedExtensions = context.sslConfig.getEnabledExtensions(
                            SSLHandshake.HELLO_RETRY_REQUEST);
            } else {
                supportedExtensions = context.sslConfig.getEnabledExtensions(
                            SSLHandshake.SERVER_HELLO);
            }
            if (m.hasRemaining()) {
                this.extensions =
                    new SSLExtensions(this, m, supportedExtensions);
            } else {
                this.extensions = new SSLExtensions(this);
            }
            // The clientHello field is used for HelloRetryRequest producer
            // only.  It's fine to set it to null for receiving side of
            // ServerHello/HelloRetryRequest message.
            this.clientHello = null;        // not used, let it be null;
        }
        public void send(HandshakeOutStream hos) throws IOException {
            hos.putInt8(serverVersion.major);
            hos.putInt8(serverVersion.minor);
            hos.write(serverRandom.randomBytes);
            hos.putBytes8(sessionId.getId());
            hos.putInt8((cipherSuite.id >> 8) & 0xFF);
            hos.putInt8(cipherSuite.id & 0xff);
            hos.putInt8(compressionMethod);

            extensions.send(hos);           // In TLS 1.3, use of certain
                                            // extensions is mandatory.
        }
    }
\end{lstlisting}

\begin{lstlisting}[
  float=tbp,
  linerange={
    59-60,
    484-485,
    492-493,
    495-501,
%    495-496,
%    501-501,
    513-514,
    516-516,
    518-525,
%    519-522,
%    525-525,
    531-532,
    535-536,
    538-542,
%    539-542,
    546-546,
    549-551,
%    550-551,
    557-563
  },
  label=lst:T13ServerHelloProducer,
  caption={[\code{ServerHello.T13ServerHelloProducer} produces \ServerHello]
  Class \code{ServerHello.T13ServerHelloProducer} defines method 
  \code{produce} to write a \ServerHello\ message to an output stream. 
  Prior to instantiating such a message, the server's active context is updated to 
  include extensions -- in particular,
  \TLSsignatureAlgorithms, \TLSsignatureAlgorithmsCert, and \TLSpsk\ -- that 
  may impact the \ServerHello\ message (Lines~519--522 or~539--542). Moreover, 
  the active context is updated to include a producer for \EncryptedExtensions\ 
  and \Finished\ messages (Lines~560--563). Code for writing the \ServerHello\ 
  message appears in Listing~\ref{lst:T13ServerHelloProducerB}.
}]
    static final HandshakeProducer t13HandshakeProducer =
        new T13ServerHelloProducer();
    private static final
            class T13ServerHelloProducer implements HandshakeProducer {
        public byte[] produce(ConnectionContext context,
                HandshakeMessage message) throws IOException {
            ServerHandshakeContext shc = (ServerHandshakeContext)context;
            ClientHelloMessage clientHello = (ClientHelloMessage)message;

            // If client hasn't specified a session we can resume, start a
            // new one and choose its cipher suite and compression options,
            // unless new session creation is disabled for this connection!
            if (!shc.isResumption || shc.resumingSession == null) {
\end{lstlisting}

\begin{lstlisting}[
  float=tbp,
  linerange={
    565-578,
    582-585,
    677-692,
    694-694,
    697-701,
    713-714,
    725-727
  },
  label=lst:T13ServerHelloProducerB,
  caption={[\code{ServerHello.T13ServerHelloProducer} produces \ServerHello\
    (cont.)]
  Class \code{ServerHello.T13ServerHelloProducer} defines method 
  \code{produce} (continued from Listing~\ref{lst:T13ServerHelloProducer}) 
  which instantiates a \ServerHello\ message~(Lines 566--571), populates 
  the extension field for the server's active context (Lines~575--578), and writes 
  the \ServerHello\ message to an output stream (Lines~584--585). The 
  \ServerHello\ message parameterises \TLSlegacyVersion\ as constant 0x0303 
  (Line~567), \TLSlegacySessionIdEcho\ as \ClientHello.\TLScipherSuites\ 
  (Line~568), and \TLScipherSuite\ as the negotiated cipher suite (Line~569), 
  which is the server selected cipher suite for (EC)DHE-only key exchange 
  (Lines~525 \&~531, Listing~\ref{lst:T13ServerHelloProducer}), selected using 
  method \code{chooseCipherSuite}, or the cipher suite %used by the previous session 
  associated with the pre-shared key for PSK-based key exchange (Line~546, Listing~\ref{lst:T13ServerHelloProducer}). 
  \ifImplNotes\textcolor{red}{Truthify regarding DHE/PSK/both}\fi
  The output stream is written to using method \code{ServerHello.ServerHelloMessage.write}, 
  defined by parent class \code{SSLHandshake.HandshakeMessage}, which in 
  turn uses method \code{ServerHello.ServerHelloMessage.send} (Listing~\ref{lst:ServerHelloMessageB}). (After outputting the message, the server 
  updates the active context to include new keying material in preparation 
  for the server's response, Section~\ref{sec:hkdf}.)
  Method \code{chooseCipherSuite} instantiates lists of 
  preferred and proposed cipher suites as the list of available 
  cipher suites and the list of offered cipher suites, respectively,
  or vice-versa, depending on the active context (Lines~684--692),
  and returns the first preferred cipher suite that is amongst those 
  proposed (Lines~694--714) or \code{null} if no such suite exists
  (Line~725).
}]
            // Generate the ServerHello handshake message.
            ServerHelloMessage shm = new ServerHelloMessage(shc,
                    ProtocolVersion.TLS12,      // use legacy version
                    clientHello.sessionId,      // echo back
                    shc.negotiatedCipherSuite,
                    new RandomCookie(shc),
                    clientHello);
            shc.serverHelloRandom = shm.serverRandom;

            // Produce extensions for ServerHello handshake message.
            SSLExtension[] serverHelloExtensions =
                    shc.sslConfig.getEnabledExtensions(
                        SSLHandshake.SERVER_HELLO, shc.negotiatedProtocol);
            shm.extensions.produce(shc, serverHelloExtensions);

            // Output the handshake message.
            shm.write(shc.handshakeOutput);
            shc.handshakeOutput.flush();
            // The handshake message has been delivered.
            return null;
        }

        private static CipherSuite chooseCipherSuite(
                ServerHandshakeContext shc,
                ClientHelloMessage clientHello) throws IOException {
            List<CipherSuite> preferred;
            List<CipherSuite> proposed;
            if (shc.sslConfig.preferLocalCipherSuites) {
                preferred = shc.activeCipherSuites;
                proposed = clientHello.cipherSuites;
            } else {
                preferred = clientHello.cipherSuites;
                proposed = shc.activeCipherSuites;
            }
            CipherSuite legacySuite = null;
            for (CipherSuite cs : preferred) {
                if (!HandshakeContext.isNegotiable(
                        proposed, shc.negotiatedProtocol, cs)) {
                    continue;
                }
                return cs;
            }
            return null;
        }
    }
\end{lstlisting}

In addition to mandatory extension \TLSsupportedVersions, message \ServerHello\ 
must include additional extensions depending on the key exchange mode: For ECDHE/DHE,
\label{comp:SH:prof:keyShare} extension
\TLSkeyShare\ is included in association with the server's key share, which must 
be in the group selected by the server from \ClientHello.\TLSsupportedGroups; 
for PSK-only, extension \TLSpsk\ is included in association with 
the pre-shared key identifier selected by the server from \ClientHello.\TLSpsk; 
and for PSK with (EC)DHE, both of those extensions are included.
Additional extensions are sent separately 
in the \EncryptedExtensions\ message.

\begin{sloppypar}
A \ServerHello\ message is consumed by the client: The client first checks that
the message is a TLS 1.3 \ServerHello\ message, which is achieved by checking 
that extension \TLSsupportedVersions\ is present and that constant 0x0304 is
the first listed preference. 
\ifSpecNotes
\begin{color}{red}
The specification seems to contain conflicting requirements:\marginpar{conflict reported 1 May 2020}
\begin{quote}
  Clients MUST check for this [supported_versions] extension \emph{prior to
  processing} the rest of the ServerHello (although they will have to           
  parse the ServerHello in order to read the extension)
\end{quote}
and 
\begin{quote}
  Upon receiving a message with type server_hello, implementations MUST      
  \emph{first examine} the Random value
\end{quote}
which conflict on the emphasised text. Presumably, these must be the 
first two checks, but their ordering is unclear.
\end{color}
\fi
Next, the client checks whether the server's nonce 
(\TLSrandom) is a special value (defined by constant 
\code{RandomCookie.hrrRandomBytes}) indicating that the \ServerHello\ message
is a \HelloRetryRequest\ message and should be processed as such 
(\S\ref{sec:HelloRetryRequest}). \ifPresentationNotes\marginpar{Maybe drop inline reference to \code{RandomCookie}
  if including OpenSSL, or include a reference to OpenSSL's handling.}\fi
The client also checks whether \label{comp:SH:cons:version} the server selected protocol version
(\TLSsupportedVersions) is amongst those offered (\ClientHello.\TLSsupportedVersions)
and is at least version 1.3, whether the server selected cipher suite
(\TLScipherSuite) is amongst those offered (\ClientHello.\TLScipherSuites),
and whether field \TLSlegacySessionIdEcho\ matches 
\ClientHello.\TLSlegacySessionId, aborting with an \TLSillegalParameter\
alert if any check fails. Finally, the client processes any remaining 
extensions:
\end{sloppypar}

\begin{description}
\ifSpecNotes
\item   \textcolor{red}{\TLSkeyShare: 
    I expected the client to check whether the server's key share is 
    in a group [selected by the server] from \ClientHello.\TLSsupportedGroups, 
    but that doesn't seem to be required.}
\fi

\item \TLSpsk: The client checks whether the server-selected key 
  identifier is amongst those offered by the client, 
  the server-selected cipher suite defines a hash function associated 
  with that identifier, and 
  extension \TLSkeyShare\ is present if the offered key exchange 
  mode for that identifier is PSK with (EC)DHE, aborting with an 
  \TLSillegalParameter\ alert if either check fails.
\end{description}

%\noindent 
%\textcolor{red}{...XXX...}

\begin{tcolorbox}
Consumption is implemented by class \code{ServerHello.ServerHelloConsumer}
(Listing~\ref{lst:ServerHelloConsumer}). That class checks the presence of extension 
\TLSsupportedVersions, to determine whether the message is a TLS~1.3 \ServerHello\ message, and 
the remainder of the message is processed by \code{ServerHello.T13ServerHelloConsumer}
(Listing~\ref{lst:T13ServerHelloConsumer}), if it is a TLS 1.3 message. 
\end{tcolorbox}

\begin{lstlisting}[
  float=tbp,
  linerange={
    55-56,%handshakeConsumer
    839-840,
    847-848,
    850-850,
    864-864,
    869-874,
    928-929,
    933-936,
%    937-960, svs != null for TLS 1.3
    937-941,
    943-944,
    948-954,
    956-956,
%    957-960,
    984-984,
    993-994
  },
  label=lst:ServerHelloConsumer,
  caption={[\code{ServerHello.ServerHelloConsumer} consumes generic 
            \ServerHello/\HelloRetryRequest]
  Class \code{ServerHello.ServerHelloConsumer} defines method \code{consume} to 
  instantiate a (generic) \ServerHello\ message from an input buffer (Line~864) and processes the message
  as a \HelloRetryRequest\ (Line~870) or a \ServerHello\ message (Line~872). The latter
  updates the client's active context to include the server's selected version (Lines~933--936), 
  using method \code{SupportedVersionsExtension.SHSupportedVersionsConsumer.consume}, which 
  calls \code{chc.handshakeExtensions.put(SH\_SUPPORTED\_VERSIONS, spec)},
  and checks whether that version was offered by the client (Lines~949), aborting if it
  was not (Lines~950--953) and, otherwise, updating the active context to include that version as the   
  negotiated protocol (Lines~956--960). 
  (Variable \code{serverVersion}, Lines~943--944, cannot be null for (TLS 1.3) \HelloRetryRequest\ 
  nor \ServerHello\ messages.)
  Further processing is deferred (Line~984) to class \code{ServerHello.T13ServerHelloConsumer}
  (Listing~\ref{lst:T13ServerHelloConsumer}).
  %\textcolor{red}{Maybe explicitly mention that Lines~958 \&~959 update \code{TransportContext}.}
}]
    static final SSLConsumer handshakeConsumer =
        new ServerHelloConsumer();
\end{lstlisting}

\begin{lstlisting}[
  float=tbp,
  linerange={
    69-70,
    1172-1173,
    1180-1181,
    1183-1184,
    1189-1190,
    1193-1193,
    1200-1203,
    1214-1216,
    1219-1219,
    1221-1222,
    1228-1228,
    1230-1231,
%    1236-1236, %Calls MaxFragExtension.shOnTradeConsumer (uninteresting?)
    1330-1330,
    1339-1356,
    1362-1363
  },
  label=lst:T13ServerHelloConsumer,
  caption={[\code{ServerHello.T13ServerHelloConsumer} consumes \ServerHello/\HelloRetryRequest]
  Class \code{ServerHello.T13ServerHelloConsumer} defines method \code{consume}
  to process incoming (TLS 1.3) \ServerHello\ or \HelloRetryRequest\ messages
  (further to processing shown in Listing~\ref{lst:ServerHelloConsumer}). The method
  updates the client's active context to include the server's selected cipher suite as the 
  negotiated suite (Line~1190), extensions, including \TLSpsk\ (Lines~1200--1202), and 
  additional session information  (Lines~1203--1231). (The client also updates the active 
  context to include new keying material, Section~\ref{sec:hkdf}.) Moreover, the  
  active context is made ready to received further server messages (Lines~1330--1356).
}]
    private static final HandshakeConsumer t13HandshakeConsumer =
        new T13ServerHelloConsumer();
    private static final
            class T13ServerHelloConsumer implements HandshakeConsumer {
        public void consume(ConnectionContext context,
                HandshakeMessage message) throws IOException {
            ClientHandshakeContext chc = (ClientHandshakeContext)context;
            ServerHelloMessage serverHello = (ServerHelloMessage)message;

            chc.negotiatedCipherSuite = serverHello.cipherSuite;
            chc.serverHelloRandom = serverHello.serverRandom;
            SSLExtension[] extTypes = chc.sslConfig.getEnabledExtensions(
                    SSLHandshake.SERVER_HELLO);
            serverHello.extensions.consumeOnLoad(chc, extTypes);
            if (!chc.isResumption) {
                chc.handshakeSession = new SSLSessionImpl(chc,
                        chc.negotiatedCipherSuite,
                        serverHello.sessionId);
            } else {
                Optional<SecretKey> psk =
                        chc.resumingSession.consumePreSharedKey();
                chc.handshakeSession = chc.resumingSession;
                setUpPskKD(chc, psk.get());
            }
\end{lstlisting}

  \subsubsection{\HelloRetryRequest}\label{sec:HelloRetryRequest}\label{sec:HRR}

A server that consumes a \ClientHello\ message, without a share for the 
server-selected group, responds with a \HelloRetryRequest\ message.
That message is an instance of a \ServerHello\ message, with field \TLSrandom\
set to a special constant value.\footnote{
  For convenience, \HelloRetryRequest\ and \ServerHello\ messages are distinctly named 
  (in the specification), despite \HelloRetryRequest\ messages being instances 
  of \ServerHello\ messages. It follows that a \ServerHello\ message might be 
  confused for a \HelloRetryRequest\ message, but this only occurs with probability 
  $\frac{1}{2^{128}}$, hence, confusion will not occur in practice.
}
In addition to mandatory extension \TLSsupportedVersions, message \HelloRetryRequest\ 
\ifSpecNotes
\marginpar{The specification states ``[a \HelloRetryRequest] SHOULD 
contain the minimal set of extensions necessary for the client to 
generate a correct ClientHello pair," which I presume means the server
should include \TLSkeyShare, but surely that is mandatory?} 
\fi
should include extension \TLSkeyShare\ to indicate the server-selected 
group.\footnote{Extension \TLSkeyShare\ is associated with key shares
  for \ClientHello\ messages and a single key share for \ServerHello\
  messages, whereas the extension is associated with the server-selected 
  group for \HelloRetryRequest\ messages. Hence, data structures associated
  with extension \TLSkeyShare\ vary between messages.}
(The server should defer producing a key share for this group
until the client's response is received.) 
The server may also include extension \TLScookie\ associated 
with some data:

\begin{description}

\item \TLScookie: Some server-specific data for purposes including, but not limited to,
  first, offloading state (required to construct transcripts) to the client, by 
  storing the hash of the \ClientHello\ message in the cookie (with suitable integrity
  protection); and, secondly, DoS protection, by forcing the client to demonstrate
  reachability of their network address.

\end{description}

A \HelloRetryRequest\ message is consumed by the client, which performs
the checks specified for \ServerHello\ messages (above), 
additionally aborting with an \TLSillegalParameter\ alert if  
the server-selected group is not amongst those offered (\ClientHello.\TLSsupportedGroups)
or a key share for that group was already offered,
or aborting with an an \TLSunexpectedMessage\ alert if a \HelloRetryRequest\
message was already received in the same connection.
%
\begin{comment}
Additional abort [due to \emph{a key share for that group was already offered}] inferred 
(possibly incorrectly) from: 
\begin{quote}
  Clients MUST abort the handshake with an "illegal_parameter" 
  alert if the HelloRetryRequest would not result in any change 
  in the ClientHello.
\end{quote}
A change is required if key shares (plural) were offered in the Client Hello Message. (Those key 
shares must be replaced by a single key share.) I suspect the spec meant  "would not result in a
meaningful change," e.g., when a key share for the server-selected group was already offered.

Actually, this is made explicit later in the specification:
\begin{quote}
  the client MUST verify that...the selected_group field does not
  correspond to a group which was provided in the "key_share" extension
  in the original ClientHello. If [the check] fails, then the client MUST 
  abort the handshake with an "illegal_parameter" alert.
\end{quote}
\end{comment}
%
\ifSpecNotes
\begin{color}{red}\\
The specification seems to contain conflicting requirements: \marginpar{conflict reported 1 May 2020}
\begin{quote}
  Clients MUST check for this [supported_versions] extension \emph{prior to
  processing} the rest of the ServerHello (although they will have to           
  parse the ServerHello in order to read the extension)
\end{quote}
and
\begin{quote}
   the client \emph{MUST check} the
   legacy_version, legacy_session_id_echo, cipher_suite, and
   legacy_compression_method as specified in Section 4.1.3 and \emph{then}
   process \emph{the extensions}, starting with determining the version using          
   "supported_versions".
\end{quote}
which conflict on the emphasised text. 
\end{color}
\fi
A client that is able to successfully consume a \HelloRetryRequest\ message
responds with their original \ClientHello\ message, replacing the key shares 
in extension \TLSkeyShare\ with a single key share from the server-selected group,
removing extension \TLSearlyData\ if present,
including a copy \label{comp:HRR:cons:cookie} of extension \TLScookie\ and 
associated data if the extension appeared in the 
\HelloRetryRequest\ message, and updating extension \TLSpsk\ by recomputing its 
obfuscated age and binder values (\S\ref{sec:NST}).
Moreover, the client should remove any pre-shared
key identifiers that are incompatible with the server-selected cipher suite
\ifSpecNotes
\textcolor{red}{similarly to earlier issue regarding ``compatibility,''
  in relation to \ClientHello, the spec is rather vague}
\fi
(i.e., remove identifiers associated with a hash function, AEAD algorithm, 
  or both that are not defined by the server-selected cipher suite).
That \ClientHello\ message is consumed by the server (\S\ref{sec:CH}) and 
the server responds with a \ServerHello\ message, which must contain the
previously selected cipher suite, namely, \HelloRetryRequest.\TLSCipherSuite.
The \ServerHello\ message is consumed by the client as described above, 
additionally aborting with an \TLSillegalParameter\ alert if the server-selected
cipher suite differs from the previous server-selected cipher suite 
(\HelloRetryRequest.\TLSCipherSuite), if extension \TLSsupportedVersions\
is associated with a list of offered protocol versions that differ from 
the previous list (\HelloRetryRequest.\TLSsupportedVersions), or if the 
server's key share does not belong to the previous server-selected group 
(\HelloRetryRequest.\TLSkeyShare).

Beyond the above two instances of \ClientHello\ messages, a server that 
receives a \ClientHello\ message at any other time must abort with an 
\TLSunexpectedMessage\ alert.

\begin{tcolorbox}
The \HelloRetryRequest\ message is implemented by class \code{Server\-Hello.Server\-Hello\-Message}
(Listings~\ref{lst:ServerHelloMessage}--\ref{lst:ServerHelloMessageB}). Instances 
of that class are produced by class \code{Server\-Hello.T13\-Server\-Hello\-Producer} 
(Listings~\ref{lst:T13HelloRetryRequestProducer}), which is instantiated as static 
constant \code{Server\-Hello.t13\-Handshake\-Producer}. That constant is used indirectly
-- via class \code{SSL\-Handshake.HELLO_RETRY_REQUEST} -- to produce \HelloRetryRequest\
messages in class \code{Client\-Hello.T13\-Client\-Hello\-Consumer}
(Listing~\ref{lst:T13ClientHelloConsumerC}). Consumption is implemented by 
class \code{Server\-Hello.Server\-Hello\-Consumer} (Listing~\ref{lst:ServerHelloConsumer}
\&~\ref{lst:ServerHelloConsumerB}). That class checks the presence of extension
\TLSsupportedVersions, to determine whether the message is a TLS 1.3 \HelloRetryRequest\
message, and the remainder of the message is processed by class
\code{Server\-Hello.T13\-Hello\-Retry\-Request\-Consumer} 
(Listing~\ref{lst:T13HelloRetryRequestConsumer}), if it is a TLS 1.3 message. 
Successful consumption 
results in transmission of a further \ClientHello\ message (which is 
consumed by class \code{Client\-Hello.T13\-Client\-Hello\-Consumer}, 
Listings~\ref{lst:T13ClientHelloConsumer} \&~\ref{lst:T13ClientHelloConsumerC}), 
with any \TLScookie\ extension being indirectly processed -- via class \code{CookieExtension}
-- by class \code{HelloCookieManager}.
\end{tcolorbox}

\begin{lstlisting}[
  float=tbp,
  linerange={
    61-62,
    732-733,
    740-743,
    745-747,
    754-760,
    762-762,
    766-770,
    776-778,
    780-782,
    784-786,
    788-791  
  },
  label=lst:T13HelloRetryRequestProducer,
  caption={[\code{ServerHello.T13HelloRetryRequestProducer} produces \HelloRetryRequest]
  Class \code{ServerHello.T13HelloRetryRequestProducer} defines method 
  \code{produce} to instantiate a \HelloRetryRequest\ message, i.e., 
  a \ServerHello\ message with field \TLSrandom\ set to a special constant
  value~(Lines 754--760), populate 
  the extension field for the active context (Lines~767--770), write 
  the %\ServerHello\ 
  message to an output stream (Lines~777--778),
  and prepare the server's active context for the client's response (Lines~762 \& 785--786).  
}]
    static final HandshakeProducer hrrHandshakeProducer =
        new T13HelloRetryRequestProducer();
    private static final
            class T13HelloRetryRequestProducer implements HandshakeProducer {
        public byte[] produce(ConnectionContext context,
                HandshakeMessage message) throws IOException {
            ServerHandshakeContext shc = (ServerHandshakeContext) context;
            ClientHelloMessage clientHello = (ClientHelloMessage) message;
            // negotiate the cipher suite.
            CipherSuite cipherSuite =
                    T13ServerHelloProducer.chooseCipherSuite(shc, clientHello);
            ServerHelloMessage hhrm = new ServerHelloMessage(shc,
                    ProtocolVersion.TLS12,      // use legacy version
                    clientHello.sessionId,      //  echo back
                    cipherSuite,
                    RandomCookie.hrrRandom,
                    clientHello
            );
            shc.negotiatedCipherSuite = cipherSuite;
            // Produce extensions for HelloRetryRequest handshake message.
            SSLExtension[] serverHelloExtensions =
                shc.sslConfig.getEnabledExtensions(
                    SSLHandshake.HELLO_RETRY_REQUEST, shc.negotiatedProtocol);
            hhrm.extensions.produce(shc, serverHelloExtensions);
            // Output the handshake message.
            hhrm.write(shc.handshakeOutput);
            shc.handshakeOutput.flush();
            // Stateless, shall we clean up the handshake context as well?
            shc.handshakeHash.finish();     // forgot about the handshake hash
            shc.handshakeExtensions.clear();
            // What's the expected response?
            shc.handshakeConsumers.put(
                    SSLHandshake.CLIENT_HELLO.id, SSLHandshake.CLIENT_HELLO);
            // The handshake message has been delivered.
            return null;
        }
    }
\end{lstlisting}

\begin{lstlisting}[
  float=tbp,
  linerange={
    876-877,
    881-889,
    891-892,
    897-902,
    909-909,
%    914-919,
    924-924,
    926-926
  },
  label=lst:ServerHelloConsumerB,
  caption={[\code{ServerHello.ServerHelloConsumer} consumes generic \HelloRetryRequest\ (cont.)]
  Class \code{ServerHello.ServerHelloConsumer} (omitted from 
  Listing~\ref{lst:ServerHelloConsumer}) defines method 
  \code{onHelloRetryRequest} to consume a (generic) \HelloRetryRequest\ message.
  Similarly to method \code{ServerHello.ServerHelloConsumer.onHelloServer}
  (Listing~\ref{lst:ServerHelloConsumer}), the client's active context is updated to include
  the server's selected version (Lines~886--892), aborting if that version was not
  offered by the client (Lines~897--902). 
  Further processing is deferred (Line~924) to class \code{ServerHello.T13HelloRetryRequestConsumer}
  (Listing~\ref{lst:T13HelloRetryRequestConsumer}).
}]
        private void onHelloRetryRequest(ClientHandshakeContext chc,
                ServerHelloMessage helloRetryRequest) throws IOException {
            SSLExtension[] extTypes = new SSLExtension[] {
                    SSLExtension.HRR_SUPPORTED_VERSIONS
                };
            helloRetryRequest.extensions.consumeOnLoad(chc, extTypes);

            ProtocolVersion serverVersion;
            SHSupportedVersionsSpec svs =
                    (SHSupportedVersionsSpec)chc.handshakeExtensions.get(
                            SSLExtension.HRR_SUPPORTED_VERSIONS);
                serverVersion =            // could be null
                        ProtocolVersion.valueOf(svs.selectedVersion);
            if (!chc.activeProtocols.contains(serverVersion)) {
                chc.conContext.fatal(Alert.PROTOCOL_VERSION,
                    "The server selected protocol version " + serverVersion +
                    " is not accepted by client preferences " +
                    chc.activeProtocols);
            }
            chc.negotiatedProtocol = serverVersion;
\end{lstlisting}

\begin{lstlisting}[
  float=tbp,
  linerange={
    77-78,
    1373-1374,  
    1376-1377,
    1383-1383,
    1389-1392,
    1397-1397,
%    1448-1454,
    1459-1461
  },
  label=lst:T13HelloRetryRequestConsumer,
  caption={[\code{ServerHello.T13HelloRetryRequestConsumer} consumes \HelloRetryRequest]
  Class \code{ServerHello.T13HelloRetryRequestConsumer} defines method 
  \code{consume} to process incoming (TLS 1.3) \HelloRetryRequest\ messages (further to 
  processing shown in Listing~\ref{lst:ServerHelloConsumerB}). The method 
  updates the active context to include the server's selected
  cipher suite (Line~1383) and extensions (Line~1390--1397), and produces a 
  \ClientHello\ message (Line~1459).
}]
    private static final HandshakeConsumer t13HrrHandshakeConsumer =
        new T13HelloRetryRequestConsumer();
        public void consume(ConnectionContext context,
                HandshakeMessage message) throws IOException {
            ClientHandshakeContext chc = (ClientHandshakeContext)context;
            ServerHelloMessage helloRetryRequest = (ServerHelloMessage)message;
            chc.negotiatedCipherSuite = helloRetryRequest.cipherSuite;
            // Check and launch ClientHello extensions.
            SSLExtension[] extTypes = chc.sslConfig.getEnabledExtensions(
                    SSLHandshake.HELLO_RETRY_REQUEST);
            helloRetryRequest.extensions.consumeOnLoad(chc, extTypes);
            helloRetryRequest.extensions.consumeOnTrade(chc, extTypes);
\end{lstlisting}

  \subsection{Key establishment}\label{sec:hkdf}

Once a \ServerHello\ message has been sent, a shared (handshake traffic) key can be 
established, and that key can be used to enable confidentiality 
and integrity for the remainder of the handshake protocol, which includes the 
subsequent \EncryptedExtensions\ message (\S\ref{sec:EE}). The initial shared key 
is derived by application of a key derivation function, 
known as a \emph{HMAC-based Extract-and-Expand Key Derivation Function} (HKDF),
which applies the negotiated hash function to the handshake protocol's 
transcript and either the negotiated pre-shared key, the negotiated (EC)DHE key,
%(Appendix~\ref{sec:dheKey}), 
or both. Further shared (application traffic) keys can be established 
similarly, to protect additional data, including application data.
Since transcripts include client- and server-generated nonces, shared (traffic) keys
are always unique, regardless of whether the pre-shared key (or for that matter
(EC)DHE key shares) are used for multiple connections.

\subsubsection{Transcript hash}\label{sec:hkdf:transcript}

A protocol's transcript concatenates each of the protocol's 
messages, in the order that they were sent, including message headers (namely, type 
and length fields, as introduced in Section~\ref{sec:record}), 
but excluding record-layer headers. The concatenation of messages
starts with \ClientHello, optionally followed by \HelloRetryRequest\ 
and \ClientHello\ if present, and proceeded by \ServerHello.
That transcript is used in computing transcript traffic keys 
(which protect the remaining handshake messages). Thereafter, 
the concatenation of messages is extended with \EncryptedExtensions\
and optionally \CertificateRequest, \Certificate, and 
\CertificateVerify\ if sent. A MAC over that transcript 
is included in a server's \Finished\ message, and a signature
over the transcript (excluding message \CertificateVerify) is 
included in any \CertificateVerify\ message. Once extended with
that \Finished\ message, the transcript is used in computing
%
\begin{comment}
the application traffic key. 
\end{comment}
%
%% Clarified in email from Eric Rescorla, dated 11 May 2020: 
%
the application traffic keys (which protect application traffic).
Finally, for a client's \Finished\ message, the transcript
is further extended with their \EndOfEarlyData, \Certificate,  
and \CertificateVerify\ messages (as relevant), before computing a 
MAC, wherein any \CertificateVerify\ message includes a signature over
that transcript (excluding itself). 
%
\begin{comment}
Finally, the client's application
traffic key is computed after extending the transcript with the 
client's \Finished\ message.
\end{comment}

To capture a transcript hash (i.e., a hash of a transcript), 
we introduce function \TranscriptHash\ such that
\[
  \TranscriptHash(M_1,\dots,M_n) = \Hash(M_1 \parallel \dots \parallel M_n)
\]
for handshake protocol messages $M_1,\dots,M_n$ (sent in that order), where \Hash\ is the 
negotiated hash function and $\parallel$ denotes concatenation, except when 
messages $M_1$ and $M_2$ are \ClientHello\ and \HelloRetryRequest\ messages, 
respectively. In that case, $M_1$ is replaced by $M'_1$ in the hash, i.e., 
\[
  \TranscriptHash(M_1,\dots,M_n) = \Hash(M'_1 \parallel M_2 \parallel \dots \parallel M_n),
\]
where $M'_1$ is the following special, synthetic handshake message, 
namely, \ifSpecNotes\marginpar{Truthify, is the 0x0000 padding or something else}\fi
\begin{align*}
  M'_1 = 
    & \mathbin{\phantom{\parallel}}  0\textrm{x}FE &\textrm{/* header type \TLSmessageHash */}     \\
    & \parallel                      0\textrm{x}0000 \parallel \HashLength &\textrm{/* (padded) header length */} \\
    & \parallel                      \Hash(M_1) &\textrm{/* hash of \ClientHello\ message */}
\end{align*}
where \HashLength\ is the output length in bytes of negotiated hash function \Hash.
This special case enables servers to construct transcripts without maintaining 
state, in particular, they need not store an initial \ClientHello\ message,
since it can be stored in extension \TLScookie\ (\S\ref{sec:HRR}).\footnote{%
  \HelloRetryRequest\ messages need not be maintained by the server either, since
  they can be reconstructed from \ClientHello\ messages and the special constant
  value that is used by field \HelloRetryRequest\code{.}\TLSrandom.}

\begin{tcolorbox}
Transcript hashing is implemented by class \code{HandshakeHash} (Listing~\ref{lst:HandshakeHash}).
Instances of that class form part of the active client and server contexts (instantiated by classes 
\code{SSL\-Engine\-Impl} and \code{SSL\-Socket\-Impl}), which are updated by classes 
\code{SSL\-Engine\-Input\-Record} and \code{SSL\-Engine\-Output\-Record}, respectively classes 
\code{SSL\-Socket\-Input\-Record} and \code{SSL\-Socket\-Output\-Record}. Moreover, in the case 
of a \HelloRetryRequest\ message, it is updated by class 
\code{Server\-Hello.T13\-Hello\-Retry\-Request\-Consumer} (Listing~\ref{lst:T13HelloRetryRequestConsumerB})
and by class \code{HelloCookieManager} (Listing~\ref{lst:T13HelloCookieManager}), when consuming any 
\TLScookie\ extension associated with a corresponding \ClientHello\ message.
\end{tcolorbox}

\begin{lstlisting}[
  float=tbp,
  widthgobble=0*0,
  linerange={
    37-39,
    %40-40    %%Omit field hasBeenUsed and use of that field. It used by SSLMasterKeyDerivation and HandshakeContext
    41-44,
    %45-45,
    46-46,
    85-87,
    106-109,
    116-119,
    164-168,
    %169-169,
    170-170,
    172-176,
    178-180,
    %181-181,
    182-182,
    645-645    
  },
  label=lst:HandshakeHash,
  caption={[\code{HandshakeHash} supports transcript hashes]
    Class \code{HandshakeHash} defines field \code{reserves} to maintain
    a list of protocol messages (Line~39), which can be extended (e.g., with incoming 
    messages) using method \code{receive} (Lines~85--87), moreover, the class defines 
    field \code{transcriptHash} 
    as a message digest algorithm (Line~38, see also Lines~58 \& 551--644), whose digest 
    can be updated to include the aforementioned messages using methods \code{deliver}
    (Lines~116--119) and \code{update} (Lines~164--170). (The former method is used when 
    the digest should also include an additional message, e.g., an outgoing message, whereas 
    the latter only updates the digest with messages listed by field \code{reserves}.) 
    Furthermore, method \code{digest} returns the hash over the current digest (Lines~172--176) 
    and method \code{finish} resets all fields (Lines~178--182).
}]
final class HandshakeHash {
    private TranscriptHash transcriptHash;
    private LinkedList<byte[]> reserves;    // one handshake message per entry
\end{lstlisting}

\begin{lstlisting}[
  float=tbp,
  linerange={
    1401-1401,
    1403-1408,
    1411-1412,
    1415-1446
  },
  label=lst:T13HelloRetryRequestConsumerB,
  caption={[\code{ServerHello.T13HelloRetryRequestConsumer} modifies transcript hashes]
    Class \code{ServerHello.T13HelloRetryRequestConsumer} (omitted from 
    Listing~\ref{lst:T13HelloRetryRequestConsumer}) modifies the transcript hash 
    in the special case of \HelloRetryRequest\ messages: A hash of the \ClientHello\
    message is computed (Lines 1401--1415), using variable \code{chc.initialClientHelloMsg}
    that was initialised in Listing~\ref{lst:ClientHelloKickstartProducer}; a special,
    synthetic handshake message $M'_1$ is computed, as the concatenation of $0\textrm{x}FE$
    (Line~1426), $0\textrm{x}0000$ (Lines~1427--1428), \HashLength\ (Line~1429), and 
    the hashed \ClientHello\ message (Lines~1430--1431); the transcript hash's digest
    is reset and the special message is added (Line~1433--1434); a further message is computed 
    as the concatenation of $0\textrm{x}02$ (Line~1438), the \HelloRetryRequest\ message 
    length (Lines~1439--1441), and the \HelloRetryRequest\ (Lines~1443--1444); and that 
    message is added to the transcript hash's digest too (Line~1446). Thus, the client's
    active context includes the expected digest.
}]
            chc.handshakeHash.finish();     // reset the handshake hash
            // calculate the transcript hash of the 1st ClientHello message
            HandshakeOutStream hos = new HandshakeOutStream(null);
            try {
                chc.initialClientHelloMsg.write(hos);
            } catch (IOException ioe) {
                // unlikely
            }
            chc.handshakeHash.deliver(hos.toByteArray());
            byte[] clientHelloHash = chc.handshakeHash.digest();

            // calculate the message_hash
            //
            // Transcript-Hash(ClientHello1, HelloRetryRequest, ... Mn) =
            //   Hash(message_hash ||    /* Handshake type */
            //     00 00 Hash.length ||  /* Handshake message length (bytes) */
            //     Hash(ClientHello1) || /* Hash of ClientHello1 */
            //     HelloRetryRequest || ... || Mn)
            int hashLen = chc.negotiatedCipherSuite.hashAlg.hashLength;
            byte[] hashedClientHello = new byte[4 + hashLen];
            hashedClientHello[0] = SSLHandshake.MESSAGE_HASH.id;
            hashedClientHello[1] = (byte)0x00;
            hashedClientHello[2] = (byte)0x00;
            hashedClientHello[3] = (byte)(hashLen & 0xFF);
            System.arraycopy(clientHelloHash, 0,
                    hashedClientHello, 4, hashLen);

            chc.handshakeHash.finish();     // reset the handshake hash
            chc.handshakeHash.deliver(hashedClientHello);

            int hrrBodyLen = helloRetryRequest.handshakeRecord.remaining();
            byte[] hrrMessage = new byte[4 + hrrBodyLen];
            hrrMessage[0] = SSLHandshake.HELLO_RETRY_REQUEST.id;
            hrrMessage[1] = (byte)((hrrBodyLen >> 16) & 0xFF);
            hrrMessage[2] = (byte)((hrrBodyLen >> 8) & 0xFF);
            hrrMessage[3] = (byte)(hrrBodyLen & 0xFF);

            ByteBuffer hrrBody = helloRetryRequest.handshakeRecord.duplicate();
            hrrBody.get(hrrMessage, 4, hrrBodyLen);

            chc.handshakeHash.receive(hrrMessage);
\end{lstlisting}

\begin{lstlisting}[
  float=tbp,
  linerange={
    200-201,
    %202-216,
    271-272,
    278-279,
    284-284,
    291-292,
    312-335,
    338-339
  },
  label=lst:T13HelloCookieManager,
  caption={[\code{HelloCookieManager.T13HelloCookieManager} modifies transcript hashes]
    Class \code{HelloCookieManager.T13HelloCookieManager} processes cookies, in particular, 
    method \code{isCookieValid} tests the validity of cookies. That method also updates
    the transcript hash in the special case of \HelloRetryRequest\ messages: A
    \HelloRetryRequest\ message is reconstructed and added to the front of the 
    transcript hash's digest (Lines~322--324). Moreover, a special, synthetic handshake 
    message $M'_1$ is computed as the concatenation of $0\textrm{x}FE0000 \parallel \HashLength$ 
    and the hash (of a \ClientHello\ message) stored in the cookie, and message 
    $M'_1$ is added to the front of the transcript hash's digest (Lines 327--335).
    \ifImplNotes\textcolor{red}{I don't understand why \code{HandshakeHash.push} is needed, it
    seems that the transcript hash could be reset and the digest reconstructed, similarly
    to Listing~\ref{lst:T13HelloRetryRequestConsumerB}. Presumably there's a side-case
    when this doesn't work (probably because the digest already contains something useful, hence, 
    cannot be reset).}\fi
}]
    private static final
            class T13HelloCookieManager extends HelloCookieManager {
\end{lstlisting}

\subsubsection{Key derivation}\label{sec:secrets}

\begin{figure}
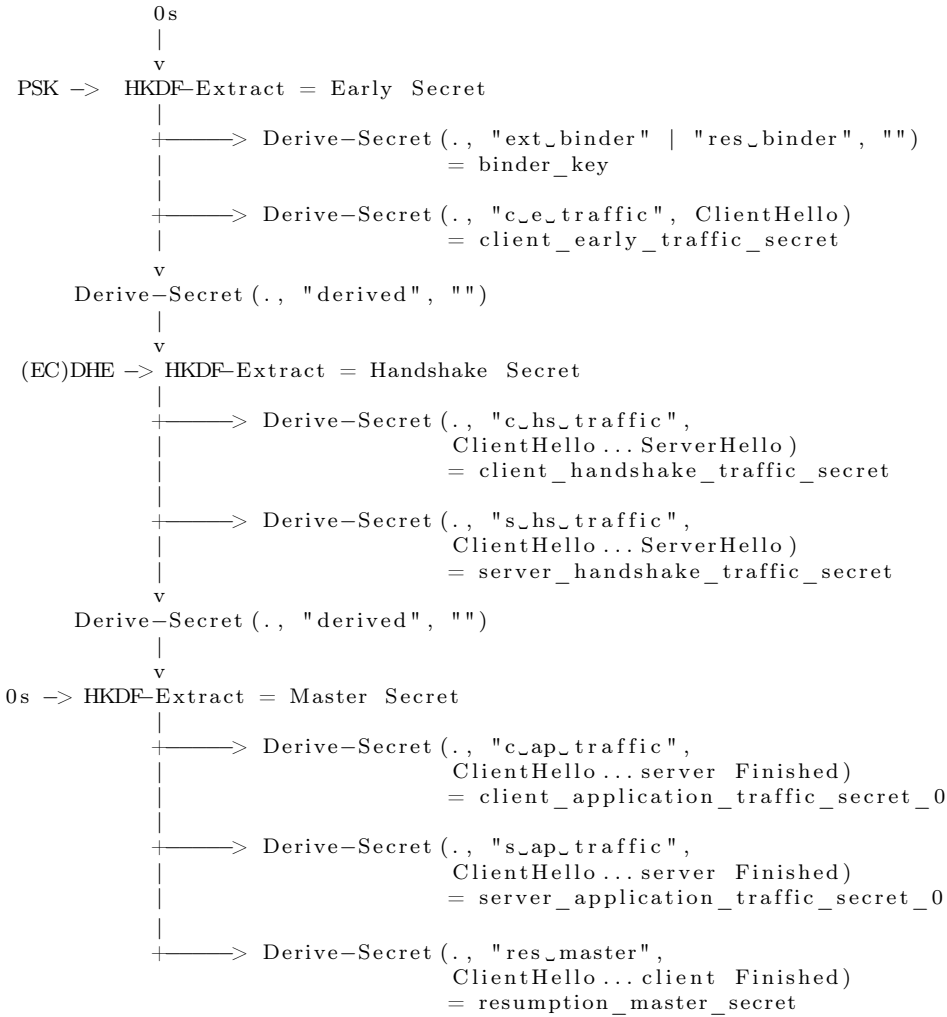

%%
%% Lines 5159 & 5189 of listings/rfc8446.txt is 0, whereas we're using 0s
%% 
\begin{lstlisting}[belowskip=0pt,numbers=none,frame=none,nolol=true]
             0s
\end{lstlisting}
\lstinputlisting[
  aboveskip=0pt,
  belowskip=0pt,
  numbers=none,
  frame=none,
  nolol=true,
  linerange={
    %5159-5162,
    %5166-5168,
    %5159-5168,
    5160-5168,
    5172-5188
  }
]{listings/rfc8446.txt}
\begin{lstlisting}[aboveskip=0pt,belowskip=0pt,numbers=none,frame=none,nolol=true]
  0s -> HKDF-Extract = Master Secret
\end{lstlisting}
\begin{lstlisting}[
  aboveskip=0pt,
  numbers=none,
  frame=none,
  nolol=true,
  linerange={
    5190-5197,
    5202-5205
  }
]
             |
             +-----> Derive-Secret(., "c ap traffic",
             |                     ClientHello...server Finished)
             |                     = client_application_traffic_secret_0
             |
             +-----> Derive-Secret(., "s ap traffic",
             |                     ClientHello...server Finished)
             |                     = server_application_traffic_secret_0
             |
             +-----> Derive-Secret(., "res master",
                                   ClientHello...client Finished)
                                   = resumption_master_secret
\end{lstlisting}

\caption[Key derivation process]{
  Key derivation process, showing application of functions \HKDFExtract\ and \DeriveSecret\
  to derive working keys. Function \HKDFExtract\ is shown inputting salt from 
  the top and secrets from the left, and outputs to the bottom, where the output's name 
  is shown to the right. Moreover, function \DeriveSecret\ is shown inputting secrets from the 
  incoming arrow and the remaining inputs appear inline, and some outputs are named below 
  the function's application (e.g., \TLSEarlySecret\ is input as the secret to generate
  \textrm{client\_early\_traffic\_secret}) and others serve as salt for subsequent 
  applications of the former (e.g., \TLSEarlySecret\ is input as the secret to 
  generate salt for \TLSHandshakeSecret). Output \TLSbinderKey\ is derived
  by application of function \DeriveSecret\ to 
  $``\textrm{ext binder}" \mid ``\textrm{res binder}"$ which denotes either 
  ``\textrm{ext binder}'' or ``\textrm{res binder}.'' The former is used for external 
  PSKs (those established independently of TLS) and the latter is used for resumption 
  PSKs (those established by \NewSessionTicket\ messages, using the resumption master 
  secret of a previous handshake), hence, one type of PSK cannot be substituted for the 
  other.\\
  \emph{Source: This figure is excerpted from RFC~8446.}
}
\label{fig:keyDerivation}
\end{figure}

The key derivation process combines the negotiated pre-shared key, the 
(EC)DHE key, or both, with the protocol's transcript. The process 
uses function \HKDFExtract, which is defined by RFC~5869
such that 
\[
  \HKDFExtract(\TLSHKDFSalt,\TLSHKDFSecret) = 
  \left\{
    \begin{matrix*}[l]
      \HMACHash(\zeros,\TLSHKDFSecret)             & \textrm{if \TLSHKDFSalt\ is null}\\
      \HMACHash(\TLSHKDFSalt,\TLSHKDFSecret)  & \textrm{otherwise},
    \end{matrix*}
  \right.
\]
%
%parameterised by salt value $\TLSHKDFSalt$ 
%%(and is a \HashLength-length string of zeros if not provided) 
%and secret $\TLSHKDFSecret$, 
where $\zeros$ denotes a \HashLength-length string of zeros
\ifSpecNotes
\textcolor{red}{Rather than \zeros, The spec uses 0, which isn't particularly intuitive, 
since it can be confused for, well, integer 0. We deviate for clarity.}
\fi
(and function \HMACHash\ is specified by RFC~2104 over keys and messages, hence, the above
definition treats $\TLSHKDFSalt$ as a key and $\TLSHKDFSecret$ 
as a message when applying \HMACHash).
In the context of key derivation, 
the salt is initially $\zeros$ 
and the secret is initially the pre-shared key or $\zeros$ if no such 
key was negotiated. The function's first output is known 
as \TLSEarlySecret. It follows that
\[
  \TLSEarlySecret = \HKDFExtract(\zeros,\textrm{PSK}),
\]
where PSK is $\zeros$ for (EC)DHE-only key exchange and
otherwise the pre-shared key, which provides raw 
entropy without context.
Context is added using function \DeriveSecret, defined 
such that 
\[
  \DeriveSecret(\TLSHKDFSecret,\TLSHKDFLabel,\TLSHandshakeMessage) 
    %= \HKDFExpandLabel(\TLSHKDFSecret, \TLSHKDFLabel, \TranscriptHash(\TLSHandshakeMessage), \HashLength)
      %= \HKDFExpand(\TLSHKDFSecret,\TLSHKDFLabelExt,\HashLength)
        %= \HMACHash(\TLSHKDFSecret, ``" \parallel \TLSHKDFLabelExt \parallel 0\textrm{x}01), \marginpar{Does $``" \parallel M = M$? If so, simplify left}
          = \HMACHash\left(\TLSHKDFSecret, \TLSHKDFLabelExt \parallel 0\textrm{x}01\right),
\]
%
%parameterised by secret \TLSHKDFSecret\ and label \TLSHKDFLabel, 
where \TLSHandshakeMessage\ is a concatenation of the protocol's messages (\S\ref{sec:hkdf:transcript})
and $\TLSHKDFLabelExt$ is defined as the following message,\footnote{
  RFC 8446 defines function \DeriveSecret\ in terms of functions \HKDFExpandLabel\
  and \HKDFExpand, namely,
  $\DeriveSecret(\allowbreak\TLSHKDFSecret,\allowbreak\TLSHKDFLabel,\allowbreak\TLSHandshakeMessage) 
    = \HKDFExpandLabel(\allowbreak\TLSHKDFSecret,\allowbreak \TLSHKDFLabel,\allowbreak \TranscriptHash(\TLSHandshakeMessage),\allowbreak \HashLength)
      = \HKDFExpand(\allowbreak\TLSHKDFSecret,\allowbreak \TLSHKDFLabelExt,\allowbreak \HashLength)
        = \HMACHash(\allowbreak\TLSHKDFSecret,\allowbreak %``" \parallel 
                                \TLSHKDFLabelExt \parallel 0\textrm{x}01)$, 
  where $\TLSHKDFLabelExt = \HashLength \parallel ``\textrm{tls13\textvisiblespace}" \parallel \TLSHKDFLabel \parallel \TranscriptHash(\TLSHandshakeMessage)$.
  By comparison, we define function \DeriveSecret\ more directly and defer the 
  additional functions to Section~\ref{sec:hkdf:trafficKeys}, where we consider
  functions \HKDFExpandLabel\ and \HKDFExpand\ more generally (in particular,
  the former may omit transcript hashes in favour of strings and the latter
  may consider lengths other than \HashLength).
}
namely,
\[
  \TLSHKDFLabelExt = \HashLength \parallel ``\textrm{tls13\textvisiblespace}" \parallel \TLSHKDFLabel \parallel \TranscriptHash(\TLSHandshakeMessage).
\]
Function \DeriveSecret\ is used (with the empty context) to derive salt for subsequent
applications of \HKDFExtract. Indeed, we have
\[
  \TLSHandshakeSecret = \HKDFExtract(\DeriveSecret(\TLSEarlySecret, ``\textrm{derived}", ``"),\textrm{K}),
\]
where \textrm{K} is \zeros\ for PSK-only key exchange and
otherwise the (EC)DHE key, moreover, 
\[
  \TLSMasterSecret = \HKDFExtract(\DeriveSecret(\TLSHandshakeSecret, ``\textrm{derived}", ``"),\zeros),
\]
noting that \TranscriptHash(``'') = \Hash(``''), that is, the hash of the empty string
(null ASCII character 0x00).
Traffic secrets are derived from \TLSEarlySecret, \TLSHandshakeSecret, and \TLSMasterSecret, 
as shown in Figure~\ref{fig:keyDerivation}, by adding context. Those secrets are used to 
derive traffic keys (\S\ref{sec:hkdf:trafficKeys}) to protect the data summarised in the 
following table:

\ifPresentationNotes
\marginpar{Table nor Listing are complete.}
\fi

\begin{table}[H]
\caption{Traffic secrets that underlie traffic keys used to protect data}
\label{table:secrets}
\centering
\begin{tabular}{l|l}
Underlying traffic secret & Protected data \\ \hline
%\TLSbinderKey & \textcolor{red}{to do}\\
\TLSclientEarlyTrafficSecret & 0-RTT \\
%\TLSearlyExporterMasterSecret & \textcolor{red}{omit exporterts?} \\
\TLShandshakeTrafficSecret & Handshake extensions %\EncryptedExtensions,  
  %\CertificateRequest, \Certificate, \CertificateVerify, and \Finished. 
  \\
\TLSapplicationTrafficSecretN & Application traffic %\\
%\TLSexporterMasterSecret & \textcolor{red}{omit exporterts?} \\
%\TLSresumptionMasterSecret& \textcolor{red}{to do}
\end{tabular}
\end{table}

\noindent
where \TLSField{[sender]} is either \TLSField{client} or \TLSField{server}, and
\TLSapplicationTrafficSecretN[N+1] is defined as follows when $N>0$, namely,
\begin{multline*}
  \TLSapplicationTrafficSecretN[N+1] \\
    = \DeriveSecret(\TLSapplicationTrafficSecretN, ``\textrm{traffic upd}", ``"),
\end{multline*}
which is used to update application-traffic secrets.

\begin{tcolorbox}
Function \HKDFExtract\ is implemented by class \code{HKDF} (Listing~\ref{lst:HKDF})
and function \DeriveSecret\ is implemented by class \code{SSL\-Secret\-Derivation} 
(Listing~\ref{lst:SSLSecretDerivation} \&~\ref{lst:SSLSecretDerivationB}).
\ifImplNotes
  \textcolor{red}{Note: Implementation of function \DeriveSecret\ 
  is reliant on the implementation of \HKDFExpand, 
  which we haven't introduced yet (beyond a footnote reference), since a more 
  direct definition of function \DeriveSecret\ suffices for this section
  (as per the footnote). This doesn't seem worth mentioning.}
\fi
Application of the former %function 
is dependent on the negotiated pre-shared 
key to derive \TLSEarlySecret, which is computed by static method 
\code{Server\-Hello.set\-Up\-Psk\-KD} (Listing~\ref{lst:setUpPskKD}), except
for (EC)DHE-only key exchange, which derives $\TLSEarlySecret$ as $\HKDFExtract(\zeros,\zeros)$
and is computed by class \code{DH\-Key\-Exchange.DHEKA\-Generator.DHEKA\-Key\-Derivation} or
class \code{ECDH\-Key\-Exchange.ECDHEKA\-Key\-Derivation}, which also compute 
\TLSHandshakeSecret. (PSK-only key exchange is unsupported and \TLSHandshakeSecret\ 
is only computed with an (EC)DHE key.) Those classes
are identical up to constructor names, strings \code{"DiffieHellman"}
and \code{"ECDH"}, and whitespace. (Refactoring has replaced those 
classes with \code{KAKeyDerivation} in JDK-13.) So, for brevity, we only present the 
former class (Listing~\ref{lst:DHEKAKeyDerivation}).
\end{tcolorbox}

\begin{lstlisting}[
  float=tbp,
  widthgobble=0*0,
  linerange={
    46-49,
    61-61,
    64-67,
    86-95,
    114-120,
    185-185
  },
  label=lst:HKDF,
  caption={[\code{HKDF} implements function \HKDFExtract]
    Class \code{HKDF} defines method \code{extract} to implement 
    function \HKDFExtract\ (RFC~5869), 
    over salt values of type \code{SecretKey} (Lines~86--95) and 
    \code{byte[]} (Lines~114--120), using a \HMACHash\ function 
    derived from the negotiated hash function (Lines~64--65), 
    where \code{JsseJce.getMac(hmacAlg)} computes \code{Mac.getInstance(hmacAlg)}
    or \code{Mac.getInstance(hmacAlg, cryptoProvider)}, depending 
    on whether \code{sun.security.ssl.SunJSSE.cryptoProvider} is 
    null.
    Both implementations allow the 
    salt to be \code{null} and will instantiate salt as a zero-filled 
    byte array of the same length as \HashLength\ (Lines~88--90 \& 116--118). 
    An HMAC is initialised with the salt as a key (Line~91) and a secret 
    as the message (Line~93), the resulting HMAC is returned as 
    a key of type \code{javax.crypto.spec.SecretKeySpec}.
}]
final class HKDF {
    private final String hmacAlg;
    private final Mac hmacObj;
    private final int hmacLen;
    HKDF(String hashAlg) throws NoSuchAlgorithmException {
        hmacAlg = "Hmac" + hashAlg.replace("-", "");
        hmacObj = JsseJce.getMac(hmacAlg);
        hmacLen = hmacObj.getMacLength();
    }
    SecretKey extract(SecretKey salt, SecretKey inputKey, String keyAlg)
            throws InvalidKeyException {
        if (salt == null) {
            salt = new SecretKeySpec(new byte[hmacLen], "HKDF-Salt");
        }
        hmacObj.init(salt);

        return new SecretKeySpec(hmacObj.doFinal(inputKey.getEncoded()),
                keyAlg);
    }
    SecretKey extract(byte[] salt, SecretKey inputKey, String keyAlg)
            throws InvalidKeyException {
        if (salt == null) {
            salt = new byte[hmacLen];
        }
        return extract(new SecretKeySpec(salt, "HKDF-Salt"), inputKey, keyAlg);
    }
}
\end{lstlisting}

\begin{lstlisting}[
  float=tbp,
  widthgobble=0*0,
  linerange={
    36-36,
    63-63,
    %64-64, %hkdfAlg
    65-67,
    69-73,
    %74-75, %init hkdfAlg (constructor)
    76-78,
    85-96,
    101-114
%    116-129,
  },
  label=lst:SSLSecretDerivation,
  caption={[\code{SSLSecretDerivation} implements function \DeriveSecret]
    Class \code{SSLSecretDerivation} implements function \DeriveSecret. The 
    class defines a constructor (Lines~69--78) that instantiates fields
    \code{context}, \code{transcriptHash} and \code{secret} 
    with data including the transcript hash, the hash of the corresponding digest 
    and a secret, respectively. Method \code{deriveKey} (Lines~85--114) is 
    instantiated with a string that references a label and returns an HMAC computed 
    by application of method \code{HKDF.expand} (Line~109) to inputs including 
    \TLSHKDFLabelExt, which is computed (Lines~105--106) over the 
    negotiated hash function's output length, the label prepended with 
    $``\textrm{tls13\textvisiblespace}"$, and a hash of 
    either the transcript's digest (when the resulting output will be used as a secret)
    or the empty digest (when the resulting output will be used as salt, i.e., when 
    \code{ks == SecretSchedule.TlsSaltSecret}), using static method 
    \code{SSLSecretDerivation.createHkdfInfo}  to handle concatenation.
    \ifImplNotes
    \textcolor{blue}{\code{SSLBasicKeyDerivation.createHkdfInfo} can be dropped in 
    favour of \code{SSLSecretDerivation.createHkdfInfo} (noting a slight
    difference regarding runtime exceptions). As can 
    \code{SSLTrafficKeyDerivation.T13TrafficKeyDerivation.createHkdfInfo}
    (noting the need to supply 0x00 as the context).}
    \textcolor{red}{Reported to security-dev@openjdk.java.net on 27 May 2020
    (\url{https://mail.openjdk.java.net/pipermail/security-dev/2020-May/021928.html}), 
    bug report added (\url{https://bugs.openjdk.java.net/browse/JDK-8245983})}
    \fi
}]
final class SSLSecretDerivation implements SSLKeyDerivation {
    private final HandshakeContext context;
\end{lstlisting}

\begin{lstlisting}[
  float=tbp,
  widthgobble=0*0,
  linerange={
    132-151,
    152-152    
  },
  label=lst:SSLSecretDerivationB,
  caption={[\code{SSLSecretDerivation.SecretSchedule} implements function \DeriveSecret\ (cont.)]
    Enum \code{SSLSecretDerivation.SecretSchedule} (omitted from Listing~\ref{lst:SSLSecretDerivation})
    maps strings to labels used by function \DeriveSecret, and prepends labels with 
    $``\textrm{tls13\textvisiblespace}"$.
}]
    private enum SecretSchedule {
        // Note that we use enum name as the key/secret name.
        TlsSaltSecret                       ("derived"),
        TlsExtBinderKey                     ("ext binder"),
        TlsResBinderKey                     ("res binder"),
        TlsClientEarlyTrafficSecret         ("c e traffic"),
        TlsEarlyExporterMasterSecret        ("e exp master"),
        TlsClientHandshakeTrafficSecret     ("c hs traffic"),
        TlsServerHandshakeTrafficSecret     ("s hs traffic"),
        TlsClientAppTrafficSecret           ("c ap traffic"),
        TlsServerAppTrafficSecret           ("s ap traffic"),
        TlsExporterMasterSecret             ("exp master"),
        TlsResumptionMasterSecret           ("res master");

        private final byte[] label;

        private SecretSchedule(String label) {
            this.label = ("tls13 " + label).getBytes();
        }
    }
}
\end{lstlisting}

\begin{comment}
\begin{lstlisting}[
  float=tbp,
  widthgobble=0*0,
  linerange={
    116-130    
  },
  label=lst:SSLSecretDerivationB,
  caption={[Class \code{SSLSecretDerivation} implements function \ExpandLabel]
    Class \code{SSLSecretDerivation} (omitted from Listing~\ref{lst:SSLSecretDerivationB})
    implements function \ExpandLabel...
}]
    public static byte[] createHkdfInfo(
            byte[] label, byte[] context, int length) {
        byte[] info = new byte[4 + label.length + context.length];
        ByteBuffer m = ByteBuffer.wrap(info);
        try {
            Record.putInt16(m, length);
            Record.putBytes8(m, label);
            Record.putBytes8(m, context);
        } catch (IOException ioe) {
            // unlikely
            throw new RuntimeException("Unexpected exception", ioe);
        }

        return info;
    }
\end{lstlisting}
\end{comment}

\begin{lstlisting}[
  float=tbp,
  widthgobble=0*0,
  linerange={
    1152-1153,
    1159-1170
  },
  label=lst:setUpPskKD,
  caption={[\code{ServerHello.setUpPskKD} derives \TLSEarlySecret\ over a pre-shared key]
    Static method \code{ServerHello.setUpPskKD} (omitted from Listings~\ref{lst:T13ServerHelloProducer}
    \&~\ref{lst:T13ServerHelloConsumer}) derives \TLSEarlySecret\ over a negotiated 
    pre-shared key. 
}]
    private static void setUpPskKD(HandshakeContext hc,
            SecretKey psk) throws SSLHandshakeException {
        try {
            CipherSuite.HashAlg hashAlg = hc.negotiatedCipherSuite.hashAlg;
            HKDF hkdf = new HKDF(hashAlg.name);
            byte[] zeros = new byte[hashAlg.hashLength];
            SecretKey earlySecret = hkdf.extract(zeros, psk, "TlsEarlySecret");
            hc.handshakeKeyDerivation =
                    new SSLSecretDerivation(hc, earlySecret);
        } catch  (GeneralSecurityException gse) {
            throw (SSLHandshakeException) new SSLHandshakeException(
                "Could not generate secret").initCause(gse);
        }
    }
\end{lstlisting}

\begin{lstlisting}[
  float=tbp,
  linerange={
    449-453,
    %454-463, %constructor
    464-465,    
    469-469,
    471-471,
    499-532
  },
  label=lst:DHEKAKeyDerivation,
  caption={[\code{DHKeyExchange.DHEKAGenerator.DHEKAKeyDerivation} derives keys]
  Class \code{DHKeyExchange.DHEKAGenerator.DHEKAKeyDerivation} defines method \code{t13DeriveKey}
  to derive the negotiated key (Lines~502--506); compute \TLSEarlySecret, for (EC)-DHE-only
  key exchange (Lines~511--520), i.e., when production or consumption of a \ServerHello\ 
  message did not call method \code{ServerHello.setUpPskKD}, which instantiates
  \code{context.handshakeKeyDerivation}; applies \DeriveSecret\ to \TLSEarlySecret\
  and label ``\textrm{derived}'' (Line~523); and uses the resulting output as salt when 
  applying \HKDFExtract\ to the negotiated key (Line~526), which produces 
  \TLSHandshakeSecret.
}]
        private static final
                class DHEKAKeyDerivation implements SSLKeyDerivation {
            private final HandshakeContext context;
            private final PrivateKey localPrivateKey;
            private final PublicKey peerPublicKey;
\end{lstlisting}

\subsubsection{Traffic keys}\label{sec:hkdf:trafficKeys}

Traffic keys are derived from traffic secrets listed in Table~\ref{table:secrets}, using 
function \HKDFExpandLabel, defined such that
\begin{multline*}
  \HKDFExpandLabel(\TLSHKDFSecret, \TLSHKDFLabel, \TLSContext, \TLSHKDFLength) \\
    = \HKDFExpand(\TLSHKDFSecret, \TLSHKDFLabelExt, \TLSHKDFLength),
\end{multline*}
where $\TLSHKDFLabelExt = \TLSHKDFLength \parallel ``\textrm{tls13\textvisiblespace}" \parallel \TLSHKDFLabel \parallel \TLSContext$
and function \HKDFExpand\ is defined by RFC 5869 such that
$
  \HKDFExpand(\TLSHKDFSecret,\allowbreak \TLSHKDFExpLabel,\allowbreak \TLSHKDFLength)
$
outputs the first \TLSHKDFLength-bytes of $T_1 \parallel \dots \parallel T_n$, where
$n = \lceil \frac{\TLSHKDFLength}{\HashLength} \rceil$ and 
\begin{align*}
  T_0 &= ``" \\
  T_1 &= \HMACHash(\TLSHKDFSecret, T_0 \parallel \TLSHKDFExpLabel \parallel 0\textrm{x}01 ) \\
  T_2 &= \HMACHash(\TLSHKDFSecret, T_1 \parallel \TLSHKDFExpLabel \parallel 0\textrm{x}02 ) \\
  &\vdots
\end{align*}

\noindent
Function \HKDFExpandLabel\ may input \TLSContext\ as the null ASCII character 0x00, denoted ``''.

\begin{tcolorbox}
Function \HKDFExpand\ is implemented by class \code{HKDF} (Listing~\ref{lst:HKDFB}) and
traffic keys are derived by class \code{SSLTrafficKeyDerivation} (Listing~\ref{lst:SSLTrafficKeyDerivation}). 
\ifImplNotes
\textcolor{red}{Note: Class \code{SSLTrafficKeyDerivation} shares similarities with 
\code{SSLSecretDerivation} and some refactoring may allow elimination of unnecessary code.}
\fi
\end{tcolorbox}

\begin{lstlisting}[
  float=tbp,
  widthgobble=0*0,
  linerange={
    137-139,
    150-150,
    %151-153
    154-159,
    162-184
  },
  label=lst:HKDFB,
  caption={[\code{HKDF} implements function \HKDFExpand]
    Class \code{HKDF} (omitted from Listing~\ref{lst:HKDF}) 
    defines method \code{expand} to implement function \HKDFExpand. 
    A buffer \code{kdfOutput} of length 
    $\HashLength\cdot\lceil \frac{\TLSHKDFLength}{\HashLength} \rceil$ is initialised
    (Lines~139 \& 154--155) and an HMAC is initialised with the input secret as a key (Line~150).
    The for-loop computes $T_1,T_2,...$ values as HMACs over messages that concatenate the 
    previous round's output (which is the empty string during the first round), label
    \code{info}, and the round number (Lines~167--170). Those values are stored in 
    buffer \code{kdfOutput} (Line~171), which is returned as a key of type 
    \code{javax.crypto.spec.SecretKeySpec} after truncating to length \code{outLen}
    (Line~183).
%
    \ifImplNotes\textcolor{red}{Lines~151-153 are omitted. As far as I can tell, info is never null}\fi
 %   
}]
    SecretKey expand(SecretKey pseudoRandKey, byte[] info, int outLen,
            String keyAlg) throws InvalidKeyException {
        byte[] kdfOutput;
        hmacObj.init(pseudoRandKey);
\end{lstlisting}

\begin{lstlisting}[
  float=tbp,
  %widthgobble=0*0,
  linerange={
    %43-43,   %%Outer class (SSLTrafficKeyDerivation)
    %47-47,
    %49-50,
    %52-56,
    %77-80,
    %121-122,
    %128-133,
    135-137,
    139-143,
    146-160,
%    162-176,
    177-177,
    179-202
  },
  label=lst:SSLTrafficKeyDerivation,
  caption={[\code{SSLTrafficKeyDerivation.T13TrafficKeyDerivation} derives traffic keys]
  Class \code{SSLTrafficKeyDerivation.T13TrafficKeyDerivation} derives traffic keys. 
  Method \code{deriveKey} is instantiated with a string that references a label and returns
  an HMAC computed by application of method \code{HKDR.expand} (Lines~153--155) to inputs including 
  \TLSHKDFLabelExt, which is computed (Lines~151--152) over the negotiated hash function's output length,
  the label prepended with $``\textrm{tls13\textvisiblespace}"$, and null ASCII character 0x00, using 
  static method \code{SSLTrafficKeyDerivation.T13TrafficKeyDerivation.createHkdfInfo}
  to handle concatenation and to introduce 0x00.
}]
/*
 * Copyright (c) 2018, Oracle and/or its affiliates. All rights reserved.
 * DO NOT ALTER OR REMOVE COPYRIGHT NOTICES OR THIS FILE HEADER.
 *
 * This code is free software; you can redistribute it and/or modify it
 * under the terms of the GNU General Public License version 2 only, as
 * published by the Free Software Foundation.  Oracle designates this
 * particular file as subject to the "Classpath" exception as provided
 * by Oracle in the LICENSE file that accompanied this code.
 *
 * This code is distributed in the hope that it will be useful, but WITHOUT
 * ANY WARRANTY; without even the implied warranty of MERCHANTABILITY or
 * FITNESS FOR A PARTICULAR PURPOSE.  See the GNU General Public License
 * version 2 for more details (a copy is included in the LICENSE file that
 * accompanied this code).
 *
 * You should have received a copy of the GNU General Public License version
 * 2 along with this work; if not, write to the Free Software Foundation,
 * Inc., 51 Franklin St, Fifth Floor, Boston, MA 02110-1301 USA.
 *
 * Please contact Oracle, 500 Oracle Parkway, Redwood Shores, CA 94065 USA
 * or visit www.oracle.com if you need additional information or have any
 * questions.
 */

package sun.security.ssl;

import java.io.IOException;
import java.nio.ByteBuffer;
import java.security.GeneralSecurityException;
import java.security.ProviderException;
import java.security.spec.AlgorithmParameterSpec;
import javax.crypto.KeyGenerator;
import javax.crypto.SecretKey;
import javax.crypto.spec.IvParameterSpec;
import javax.crypto.spec.SecretKeySpec;
import javax.net.ssl.SSLHandshakeException;
import sun.security.internal.spec.TlsKeyMaterialParameterSpec;
import sun.security.internal.spec.TlsKeyMaterialSpec;
import sun.security.ssl.CipherSuite.HashAlg;
import static sun.security.ssl.CipherSuite.HashAlg.H_NONE;

enum SSLTrafficKeyDerivation implements SSLKeyDerivationGenerator {
    SSL30       ("kdf_ssl30", new S30TrafficKeyDerivationGenerator()),
    TLS10       ("kdf_tls10", new T10TrafficKeyDerivationGenerator()),
    TLS12       ("kdf_tls12", new T12TrafficKeyDerivationGenerator()),
    TLS13       ("kdf_tls13", new T13TrafficKeyDerivationGenerator());

    final String name;
    final SSLKeyDerivationGenerator keyDerivationGenerator;

    SSLTrafficKeyDerivation(String name,
            SSLKeyDerivationGenerator keyDerivationGenerator) {
        this.name = name;
        this.keyDerivationGenerator = keyDerivationGenerator;
    }

    static SSLTrafficKeyDerivation valueOf(ProtocolVersion protocolVersion) {
        switch (protocolVersion) {
            case SSL30:
                return SSLTrafficKeyDerivation.SSL30;
            case TLS10:
            case TLS11:
            case DTLS10:
                return SSLTrafficKeyDerivation.TLS10;
            case TLS12:
            case DTLS12:
                return SSLTrafficKeyDerivation.TLS12;
            case TLS13:
                return SSLTrafficKeyDerivation.TLS13;
        }

        return null;
    }

    @Override
    public SSLKeyDerivation createKeyDerivation(HandshakeContext context,
            SecretKey secretKey) throws IOException {
        return keyDerivationGenerator.createKeyDerivation(context, secretKey);
    }

    private static final class S30TrafficKeyDerivationGenerator
            implements SSLKeyDerivationGenerator {
        private S30TrafficKeyDerivationGenerator() {
            // blank
        }

        @Override
        public SSLKeyDerivation createKeyDerivation(
            HandshakeContext context, SecretKey secretKey) throws IOException {
            return new LegacyTrafficKeyDerivation(context, secretKey);
        }
    }

    private static final class T10TrafficKeyDerivationGenerator
            implements SSLKeyDerivationGenerator {
        private T10TrafficKeyDerivationGenerator() {
            // blank
        }

        @Override
        public SSLKeyDerivation createKeyDerivation(
            HandshakeContext context, SecretKey secretKey) throws IOException {
            return new LegacyTrafficKeyDerivation(context, secretKey);
        }
    }

    private static final class T12TrafficKeyDerivationGenerator
            implements SSLKeyDerivationGenerator {
        private T12TrafficKeyDerivationGenerator() {
            // blank
        }

        @Override
        public SSLKeyDerivation createKeyDerivation(
            HandshakeContext context, SecretKey secretKey) throws IOException {
            return new LegacyTrafficKeyDerivation(context, secretKey);
        }
    }

    private static final class T13TrafficKeyDerivationGenerator
            implements SSLKeyDerivationGenerator {
        private T13TrafficKeyDerivationGenerator() {
            // blank
        }

        @Override
        public SSLKeyDerivation createKeyDerivation(
                HandshakeContext context,
                SecretKey secretKey) throws IOException {
            return new T13TrafficKeyDerivation(context, secretKey);
        }
    }

    static final class T13TrafficKeyDerivation implements SSLKeyDerivation {
        private final CipherSuite cs;
        private final SecretKey secret;

        T13TrafficKeyDerivation(
                HandshakeContext context, SecretKey secret) {
            this.secret = secret;
            this.cs = context.negotiatedCipherSuite;
        }

        @Override
        public SecretKey deriveKey(String algorithm,
                AlgorithmParameterSpec params) throws IOException {
            KeySchedule ks = KeySchedule.valueOf(algorithm);
            try {
                HKDF hkdf = new HKDF(cs.hashAlg.name);
                byte[] hkdfInfo =
                        createHkdfInfo(ks.label, ks.getKeyLength(cs));
                return hkdf.expand(secret, hkdfInfo,
                        ks.getKeyLength(cs),
                        ks.getAlgorithm(cs, algorithm));
            } catch (GeneralSecurityException gse) {
                throw (SSLHandshakeException)(new SSLHandshakeException(
                    "Could not generate secret").initCause(gse));
            }
        }

        private static byte[] createHkdfInfo(
                byte[] label, int length) throws IOException {
            byte[] info = new byte[4 + label.length];
            ByteBuffer m = ByteBuffer.wrap(info);
            try {
                Record.putInt16(m, length);
                Record.putBytes8(m, label);
                Record.putInt8(m, 0x00);    // zero-length context
            } catch (IOException ioe) {
                // unlikely
                throw new RuntimeException("Unexpected exception", ioe);
            }

            return info;
        }
    }

    private enum KeySchedule {
        // Note that we use enum name as the key/ name.
        TlsKey              ("key", false),
        TlsIv               ("iv",  true),
        TlsUpdateNplus1     ("traffic upd", false);

        private final byte[] label;
        private final boolean isIv;

        private KeySchedule(String label, boolean isIv) {
            this.label = ("tls13 " + label).getBytes();
            this.isIv = isIv;
        }

        int getKeyLength(CipherSuite cs) {
            if (this == KeySchedule.TlsUpdateNplus1)
                return cs.hashAlg.hashLength;
            return isIv ? cs.bulkCipher.ivSize : cs.bulkCipher.keySize;
        }

        String getAlgorithm(CipherSuite cs, String algorithm) {
            return isIv ? algorithm : cs.bulkCipher.algorithm;
        }
    }

    @SuppressWarnings("deprecation")
    static final class LegacyTrafficKeyDerivation implements SSLKeyDerivation {
        private final HandshakeContext context;
        private final SecretKey masterSecret;
        private final TlsKeyMaterialSpec keyMaterialSpec;

        LegacyTrafficKeyDerivation(
                HandshakeContext context, SecretKey masterSecret) {
            this.context = context;
            this.masterSecret = masterSecret;

            CipherSuite cipherSuite = context.negotiatedCipherSuite;
            ProtocolVersion protocolVersion = context.negotiatedProtocol;

            /*
             * For both the read and write sides of the protocol, we use the
             * master to generate MAC secrets and cipher keying material.  Block
             * ciphers need initialization vectors, which we also generate.
             *
             * First we figure out how much keying material is needed.
             */
            int hashSize = cipherSuite.macAlg.size;
            boolean is_exportable = cipherSuite.exportable;
            SSLCipher cipher = cipherSuite.bulkCipher;
            int expandedKeySize = is_exportable ? cipher.expandedKeySize : 0;

            // Which algs/params do we need to use?
            String keyMaterialAlg;
            HashAlg hashAlg;

            byte majorVersion = protocolVersion.major;
            byte minorVersion = protocolVersion.minor;
            if (protocolVersion.isDTLS) {
                // Use TLS version number for DTLS key calculation
                if (protocolVersion.id == ProtocolVersion.DTLS10.id) {
                    majorVersion = ProtocolVersion.TLS11.major;
                    minorVersion = ProtocolVersion.TLS11.minor;

                    keyMaterialAlg = "SunTlsKeyMaterial";
                    hashAlg = H_NONE;
                } else {    // DTLS 1.2+
                    majorVersion = ProtocolVersion.TLS12.major;
                    minorVersion = ProtocolVersion.TLS12.minor;

                    keyMaterialAlg = "SunTls12KeyMaterial";
                    hashAlg = cipherSuite.hashAlg;
                }
            } else {
                if (protocolVersion.id >= ProtocolVersion.TLS12.id) {
                    keyMaterialAlg = "SunTls12KeyMaterial";
                    hashAlg = cipherSuite.hashAlg;
                } else {
                    keyMaterialAlg = "SunTlsKeyMaterial";
                    hashAlg = H_NONE;
                }
            }

            // TLS v1.1+ and DTLS use an explicit IV in CBC cipher suites to
            // protect against the CBC attacks.  AEAD/GCM cipher suites in
            // TLS v1.2 or later use a fixed IV as the implicit part of the
            // partially implicit nonce technique described in RFC 5116.
            int ivSize = cipher.ivSize;
            if (cipher.cipherType == CipherType.AEAD_CIPHER) {
                ivSize = cipher.fixedIvSize;
            } else if (
                    cipher.cipherType == CipherType.BLOCK_CIPHER &&
                    protocolVersion.useTLS11PlusSpec()) {
                ivSize = 0;
            }

            TlsKeyMaterialParameterSpec spec = new TlsKeyMaterialParameterSpec(
                    masterSecret, (majorVersion & 0xFF), (minorVersion & 0xFF),
                    context.clientHelloRandom.randomBytes,
                    context.serverHelloRandom.randomBytes,
                    cipher.algorithm, cipher.keySize, expandedKeySize,
                    ivSize, hashSize,
                    hashAlg.name, hashAlg.hashLength, hashAlg.blockSize);

            try {
                KeyGenerator kg = JsseJce.getKeyGenerator(keyMaterialAlg);
                kg.init(spec);

                this.keyMaterialSpec = (TlsKeyMaterialSpec)kg.generateKey();
            } catch (GeneralSecurityException e) {
                throw new ProviderException(e);
            }
        }

        SecretKey getTrafficKey(String algorithm) {
            switch (algorithm) {
                case "clientMacKey":
                    return keyMaterialSpec.getClientMacKey();
                case "serverMacKey":
                    return keyMaterialSpec.getServerMacKey();
                case "clientWriteKey":
                    return keyMaterialSpec.getClientCipherKey();
                case "serverWriteKey":
                    return keyMaterialSpec.getServerCipherKey();
                case "clientWriteIv":
                    IvParameterSpec cliIvSpec = keyMaterialSpec.getClientIv();
                    return  (cliIvSpec == null) ? null :
                            new SecretKeySpec(cliIvSpec.getIV(), "TlsIv");
                case "serverWriteIv":
                    IvParameterSpec srvIvSpec = keyMaterialSpec.getServerIv();
                    return  (srvIvSpec == null) ? null :
                            new SecretKeySpec(srvIvSpec.getIV(), "TlsIv");
            }

            return null;
        }

        @Override
        public SecretKey deriveKey(String algorithm,
                AlgorithmParameterSpec params) throws IOException {
            return getTrafficKey(algorithm);
        }
    }
}

\end{lstlisting}

\noindent
Returning to key derivation, we derive the following traffic keys:
\begin{align*}
  \TLSwriteKey &= \HKDFExpandLabel(\TLSField{[secret]}, ``key", ``", key\_length) \\
  \TLSwriteIV  &= \HKDFExpandLabel(\TLSField{[secret]}, ``iv", ``", iv\_length)
\end{align*}
where \TLSField{[sender]} is either \TLSField{client} or \TLSField{server}, and
\TLSField{[secret]} is taken from the secrets listed in Table~\ref{table:secrets}.

\begin{tcolorbox}
Server- and client-side handshake-traffic key derivation is implemented by classes 
\code{Server\-Hello.T13\-Server\-Hello\-Producer} %(Listing~\ref{lst:T13ServerHelloProducerC})
and \code{Server\-Hello.T13\-Server\-Hello\-Consumer}, %(Listing~\ref{lst:T13ServerHelloConsumerB}), 
respectively.
The former class defines method \code{produce} to write a \ServerHello\ message to an 
output stream (Listings~\ref{lst:T13ServerHelloProducer} 
\&~\ref{lst:T13ServerHelloProducerB}), and that method derives handshake-traffic 
keys immediately after writing the \ServerHello\ message; the keys are used to encrypt subsequent 
outgoing handshake messages (including an \EncryptedExtensions\ message) and to decrypt subsequent
incoming handshake messages. Similarly, the latter class defines 
method \code{consume} to read a \ServerHello\ message from an input buffer 
(Listing~\ref{lst:T13ServerHelloConsumer}), and that method derives handshake-traffic 
keys immediately before reading an \EncryptedExtensions\ message (and prior to reading 
further extensions, including \Certificate\ and \CertificateVerify\ messages 
for (EC)DHE-only key exchange, and a \Finished\ message); the keys are used to decrypt subsequent
incoming handshake messages, including that \EncryptedExtensions\ message, and to encrypt subsequent
outgoing handshake messages.
The implementations 
are identical up to contexts (namely, \code{Server\-Handshake\-Context} and 
\code{Client\-Handshake\-Context}, that share parent \code{Handshake\-Context}), 
labels 
%hack for visible space:
{\code{s}\textvisiblespace{}\code{hs}\textvisiblespace{}\code{traffic}} and 
{\code{c}\textvisiblespace{}\code{hs}\textvisiblespace{}\code{traffic}}
(which are instantiated by enum \code{SSL\-Secret\-Derivation.Secret\-Schedule}
using strings \code{TlsServerHandshakeTrafficSecret} and 
\code{TlsClientHandshakeTrafficSecret}, respectively), treatment of null
in tricks to make the compiler happy (cf. \code{return null;} and \code{return;}
in catch-branches), $\alpha$-renaming of one variable, and whitespace
(and some obsolete, commented-out code). 
(Refactoring could eliminate unnecessary code.\footnote{%
  The OpenJDK team are aware of refactoring opportunities 
  (\url{https://mail.openjdk.java.net/pipermail/security-dev/2020-May/021928.html})
  and are tracking changes (\url{https://bugs.openjdk.java.net/browse/JDK-8245983}).
}) 
So, for brevity, we only present server-side handshake-traffic key derivation 
(Listings~\ref{lst:T13ServerHelloProducerC} \&~\ref{lst:T13ServerHelloProducerD}). 
%
%% The following has been moved to the previous section, where \TLSEarlySecret\
%% is introduced.
%
\begin{comment}
Derivation is dependent on \TLSEarlySecret\ and 
the negotiated (EC)DHE key (PSK-only key exchange is unsupported), 
which are computed by classes \code{DHKeyExchange.DHEKAGenerator.DHEKAKeyDerivation} and
\code{ECDHKeyExchange.ECDHEKAKeyDerivation}. Those classes
are identical up to constructor names, strings \code{"DiffieHellman"}
and \code{"ECDH"}, and whitespace. (Refactoring could elimate
unnecessary code.) So, for brevity, we only present the 
former class (Listing~\ref{lst:DHEKAKeyDerivation}).
\end{comment}

\end{tcolorbox}

\lstinputlisting[
  float=tbp,
  linerange={
    %587-587,
    588-589,
    591-592,
    593-598, %error handling for shc.handshakeKeyExchange == null 
    600-605,  
    613-615
  },
  label=lst:T13ServerHelloProducerC,
  caption={[\code{ServerHello.T13ServerHelloProducer} deriving keys]
  Class \code{ServerHello.T13ServerHelloProducer} (omitted from 
  Listing~\ref{lst:T13ServerHelloProducerB}) updates the transcript
  hash's digest to include all handshake protocol messages (Line~589), 
  derives an (EC)DHE key (Line~600), and establishes \TLSHandshakeSecret\
  (Lines~601--602). Variable \code{shc.handshakeKeyExchange} is assigned 
  by class \code{KeyShareExtension} (PSK-only key exchange is unsupported, 
    %%https://bugs.openjdk.java.net/browse/JDK-8049402
    %%https://bugs.openjdk.java.net/browse/JDK-8145252
    %%https://bugs.openjdk.java.net/browse/JDK-8209392
  hence, \code{ke} is not null) as an instance of class \code{SSLKeyExchange}
  parameterised with \code{SSLKeyExchange.T13KeyAgreement} (of type \code{SSLKeyAgreement})
  and \code{ke.createKeyDerivation(shc)} returns either 
    \code{ECDHKeyExchange.ecdheKAGenerator.createKeyDerivation(shc)} or
    \code{DHKeyExchange.kaGenerator.createKeyDerivation(shc)}, i.e., 
  an (EC)DHE key (Line~600). The class also initialises variables \code{kdg}
  (Line~604--605) and \code{kd} (Lines 614--615) which will be used to 
  derive traffic secrets and the corresponding traffic keys, respectively.
  The former is an instance of class \code{SSLTrafficKeyDerivation} that 
  overrides method \code{createKeyDerivation} such that it returns 
  an instance of class \code{SSLTrafficKeyDerivation.T13TrafficKeyDerivation}.
}]{listings/ServerHello.java}%

\begin{lstlisting}[
  float=tbp,
  linerange={
    617-636,
    640-664,
    668-675
  },
  label=lst:T13ServerHelloProducerD,
  caption={[\code{ServerHello.T13ServerHelloProducer} deriving keys (cont.)]
  Class \code{ServerHello.T13ServerHelloProducer} (continued from 
  Listing~\ref{lst:T13ServerHelloProducerC}) derives traffic secret
  \TLSclientHandshakeTrafficSecret\ (Lines~618--619), constructs an instance
  of \code{SSLTrafficKeyDerivation.T13TrafficKeyDerivation} from that secret
  (Lines~620--621), and uses that instance to derive the corresponding traffic 
  keys \TLSField{client\_write\_key} (Lines~622--623) and
  \TLSField{client\_write\_iv} (Lines~624--625), which will be used to decrypt (and read)
  incoming client traffic (Lines~626--643). Similarly, traffic secret   
  \TLSserverHandshakeTrafficSecret\ is derived (Lines~646--647), along with 
  traffic keys \TLSField{server\_write\_key} (Lines~650--651) and
  \TLSField{server\_write\_iv} (Lines~652--653), used to encrypt (and write) outgoing 
  traffic (Lines~654--672).
}]
            // update the handshake traffic read keys.
            SecretKey readSecret = kd.deriveKey(
                    "TlsClientHandshakeTrafficSecret", null);
            SSLKeyDerivation readKD =
                    kdg.createKeyDerivation(shc, readSecret);
            SecretKey readKey = readKD.deriveKey(
                    "TlsKey", null);
            SecretKey readIvSecret = readKD.deriveKey(
                    "TlsIv", null);
            IvParameterSpec readIv =
                    new IvParameterSpec(readIvSecret.getEncoded());
            SSLReadCipher readCipher;
            try {
                readCipher =
                    shc.negotiatedCipherSuite.bulkCipher.createReadCipher(
                        Authenticator.valueOf(shc.negotiatedProtocol),
                        shc.negotiatedProtocol, readKey, readIv,
                        shc.sslContext.getSecureRandom());
            } catch (GeneralSecurityException gse) {
                // unlikely
            }

            shc.baseReadSecret = readSecret;
            shc.conContext.inputRecord.changeReadCiphers(readCipher);

            // update the handshake traffic write secret.
            SecretKey writeSecret = kd.deriveKey(
                    "TlsServerHandshakeTrafficSecret", null);
            SSLKeyDerivation writeKD =
                    kdg.createKeyDerivation(shc, writeSecret);
            SecretKey writeKey = writeKD.deriveKey(
                    "TlsKey", null);
            SecretKey writeIvSecret = writeKD.deriveKey(
                    "TlsIv", null);
            IvParameterSpec writeIv =
                    new IvParameterSpec(writeIvSecret.getEncoded());
            SSLWriteCipher writeCipher;
            try {
                writeCipher =
                    shc.negotiatedCipherSuite.bulkCipher.createWriteCipher(
                        Authenticator.valueOf(shc.negotiatedProtocol),
                        shc.negotiatedProtocol, writeKey, writeIv,
                        shc.sslContext.getSecureRandom());
            } catch (GeneralSecurityException gse) {
                // unlikely
            }

            shc.baseWriteSecret = writeSecret;
            shc.conContext.outputRecord.changeWriteCiphers(
                    writeCipher, (clientHello.sessionId.length() != 0));

            // Update the context for master key derivation.
            shc.handshakeKeyDerivation = kd;
\end{lstlisting}%

Traffic secrets \TLSclientHandshakeTrafficSecret\ and \TLSserverHandshakeTrafficSecret\
are used to derive handshake-traffic keys that protect handshake extensions (\S\ref{sec:EE} 
\&~\ref{sec:handshakeAuth}). After those extensions are processed,
application-traffic keys to protect application data can be derived (\S\ref{sec:FIN}).

  \subsection{Server parameters: \EncryptedExtensions}\label{sec:EE}

%%Endpoints may include extensions beyond those already discussed to request extended
%%functionality. A client includes such extensions in \ClientHello\ messages, whereas 
%%a server includes such extensions in an \EncryptedExtensions\ message (which must 
%%follow immediately after a \ServerHello\ message)\sout{, typically in response to extended 
%%functionality requested by the client, but also to indicate additional functionality 
%%that does not require a response from the client}\textcolor{red}{not true for 
%%\EncryptedExtensions\ messages: "Implementations MUST NOT send extension responses 
%%if the remote endpoint did not send the corresponding extension requests, with the
%%exception of the "cookie" extension in the HelloRetryRequest." It follows that
%%\EncryptedExtensions\ messages only contain responses to requests in \ClientHello\ 
%%messages.

To request extended functionality, a client may include extensions -- beyond those 
already discussed -- in \ClientHello\ messages. Such functionality 
is not required to establish handshake-traffic keys, hence, those extensions 
can be encrypted, and a server responds to client requests by including extensions 
in \EncryptedExtensions\ and \Certificate\ messages.
(Appendix~\ref{sec:extensions} lists all extensions and formally states which extensions 
can be listed in the \TLSextensions\ field of \EncryptedExtensions\ and \Certificate\ 
messages, and of other handshake protocol messages.) The former message lists extensions 
which are not associated with individual certificates, and the latter lists those that are. 

An \EncryptedExtensions\ message (which must follow immediately after a \ServerHello\ 
message) comprises of the following field:

\begin{description}

\item \TLSextensions: A list of extensions responding to requests for extended 
  functionalities, i.e., functionalities not required to establish  handshake-traffic 
  keys (hence, can be encrypted with such keys), excluding extensions 
  associated with individual certificates.

\end{description}

\noindent
Each \EncryptedExtensions\ message is encrypted using the handshake-traffic key generated 
from traffic secret \TLSserverHandshakeTrafficSecret, as are subsequent 
handshake messages sent by the server.

\begin{tcolorbox}
\EncryptedExtensions\ messages are implemented, produced, and consumed by 
inner-classes of class \code{EncryptedExtensions}, namely, inner-classes \code{EncryptedExtensionsMessage},
\code{EncryptedExtensionsProducer}, and \code{EncryptedExtensionsConsumer}, respectively.
\ifPresentationNotes
\textcolor{red}{Perhaps explain how \code{writeCipher} encrypts traffic.}
\fi
\end{tcolorbox}

  \subsection{Authentication}\label{sec:handshakeAuth}

The handshake protocol concludes with unilateral authentication of the server.
(Client authentication is also possible, as discussed in Appendix~\ref{sec:CR}.)
For (EC)DHE-only key exchange, the server must send a \Certificate\ message
followed by a \CertificateVerify\ message (\S\ref{sec:CT}), immediately after an 
\EncryptedExtensions\ message (except when client authentication is requested).
Those messages are followed by a \Finished\ message (\S\ref{sec:FIN}). 
For PSK-based key exchange, the pre-shared key serves to authenticate the
handshake (without certificates), hence, \Certificate\ and \CertificateVerify\ 
messages are not sent, and the server only sends a \Finished\ message.\footnote{%
  RFC 8446 does not permit PSK-based key exchange with \Certificate\ and 
  \CertificateVerify\ messages from the server; (direct) certificate-based server 
  authentication is unsupported for PSK-based key exchange. (The specification notes 
  that future documents may support such authentication.) Certificate-based
  client authentication is compatible with PSK-based key exchange  
  (Appendix~\ref{sec:CR}).
}

\subsubsection{\Certificate\ and \CertificateVerify}\label{sec:CT}\label{sec:CV}

\ifPresentationNotes
\textcolor{red}{This section defines \Certificate\ and \CertificateVerify\ messages
  originating from servers. Some remarks are made with regards such messages originating from
  clients, especially when the design looks peculiar. E.g., the existence of field 
  \TLScertificateRequestContext\ makes little sense for \Certificate\ messages  
  originating from servers, because the field contains a zero-length identifier.
  To avoid (over) complicating the discourse, such details are kept to a minimum.
  Is it worth sign-posting that \Certificate\ and \CertificateVerify\ messages 
  originating from the client differ? I think not, that detail can be made 
  explicit in the appendix, if at all.}
\fi  

A \Certificate\ message contains a certificate (along with its certificate chain)
for authentication, and a \CertificateVerify\ message contains a 
signature (constructed with the private key corresponding to the public key in the certificate)
over a hash of the protocol's transcript, thereby, proving 
possession of the private key used for signing, hence, identifying the server.
%The former comprises of the following fields:

A \Certificate\ message comprises of the following fields:

\begin{description}

\item \TLScertificateRequestContext: A zero-length identifier. (A
  \Certificate\ message may also be sent in response to a 
  \CertificateRequest\ message during post-handshake authentication, 
  as discussed in Appendix~\ref{sec:CR}, in which case this field 
  echos the identifier used by the \CertificateRequest\ message.)

\item \TLScertificateList: A (non-empty) list of certificates and any associated
  extensions.   %(The list may be empty for \Certificate\ messages sent by the client.) 
  (Any extensions must respond to ones listed in the \ClientHello\ message.
  Moreover, an extension that applies to the entire chain should appear 
  in the first extension listed.)
  Certificates must be DER-encoded X.509v3 certificates, unless
  an alternative certificate type was negotiated (using extension 
  \TLSserverCertificateType). 
  The server's certificate 
  \ifPresentationNotes
  \textcolor{red}{\emph{the server's certificate} seems 
  a little ambiguous (a server may have many), but perhaps that's okay}
  \fi
  must appear first and every subsequent certificate should certify the 
  previous one (i.e., every subsequent certificate should contain a 
  signature -- using the private key corresponding to the certificate's
  public key -- over the previous certificate's public key), hence, the 
  list is a certificate chain. That first 
  \begin{comment}
  \sout{certificate must be signed using 
  an algorithm amongst those offered by the client 
  (\ClientHello.\TLSsignatureAlgorithms) and the}
  \textcolor{red}{I can't find any evidence to support that -- my mistake?}
  \end{comment}
  certificate's public key 
  should be compatible with \begin{comment}\sout{that algorithm}\end{comment}
  an algorithm amongst those offered, by the client, for \CertificateVerify\
  messages (i.e., advertised by \ClientHello.\TLSsignatureAlgorithms).
  \ifSpecNotes
  \textcolor{red}{
    The spec requires that the certificate's public key be compatible
    with the selected authentication algorithm from the client's 
    "signature\_algorithms" extension, but does not require the server
    to select such an algorithm.
  }
  \fi
  Any remaining certificates' public keys should be compatible with an algorithm 
  offered %by the client 
  for \Certificate\ messages (i.e., those advertised 
  by extension \TLSsignatureAlgorithmsCert\ if present and extension 
  \TLSsignatureAlgorithms\ otherwise).
  (When a certificate chain cannot be constructed from compatible algorithms, 
  the chain may rely on algorithms not offered by the client, 
  except for SHA-1, which must not be used, unless offered.)
  All certificates must (explicitly) permit signature verification (whenever
  certificates include a Key Usage extension).
  (Self-signed certificates or trust anchors may be signed 
  with any algorithm, trust anchor certificates 
  may be omitted when they are known to be in the client's 
  possession, and, for raw public keys, the list must contain 
  \ifSpecNotes 
  \textcolor{red}{
  The spec requires \emph{no more than one certificate}.
  The spec also requires a non-empty list for servers. 
  So, we can infer \emph{exactly one} here [for the case of servers].
  }
  \fi
  exactly one certificate.)

\end{description}

\noindent
A server's \Certificate\ message is consumed by the client, which aborts
with a \TLSdecodeError\ alert if the \Certificate\ message is empty
and with a \TLSbadCertificate\ alert if a certificate relies on MD5, moreover,
it is recommended that a client also aborts with a \TLSbadCertificate\ alert if
a certificate relies on SHA-1. The client may validate certificates using 
procedures beyond the scope of TLS. (The TLS 1.3 specification
cites RFC~5280 as a reference for validation procedures.)

A \CertificateVerify\ message comprises of the following fields:

\begin{description}

\begin{sloppypar}
\item \TLSalgorithm: A signing algorithm, which must be amongst those
  offered by the client (\ClientHello.\TLSsignatureAlgorithms), unless 
  unless a certificate chain cannot be constructed from compatible algorithms.
\end{sloppypar}
  
\item \TLSsignature: A signature, produced by the aforementioned algorithm, over the concatenation of: 
  0x20 repeated 64 times, string ``TLS 1.3, server CertificateVerify'',
  0x00, and the transcript hash (\S\ref{sec:hkdf:transcript}).

\end{description}

\noindent
A server's \Certificate\ message is consumed by the client, which aborts
with a  \TLSbadCertificate\ alert if the signature does not verify.

\begin{tcolorbox}
\Certificate\ and \CertificateVerify\ messages are implemented, produced, and consumed by 
inner-classes of class \code{CertificateMessage} (Listings~\ref{lst:CertificateMessage}--\ref{lst:CertificateMessageD}) 
and \code{CertificateVerify} (Listings~\ref{lst:CertificateVerify}--\ref{lst:CertificateVerifyD}), respectively.
\end{tcolorbox}

\begin{lstlisting}[
  float=tbp,
  linerange={
    732-734, %%CertificateEntry
    736-739,
    777-777,
    782-784, %%T13CertificateMessage
    786-798,
    913-913
  },
  label=lst:CertificateMessage,
  caption={[\code{CertificateMessage.T13CertificateMessage} defines \Certificate]
  Class \code{CertificateMessage.T13CertificateMessage} defines the two fields of a 
  \Certificate\ message (Lines~783--784) and a constructor to instantiate them
  (Lines~786--798), where the latter field is defined over a list of pairs, comprising a 
  certificate and any associated extensions (Lines~732--777).
  A further (omitted) constructor is defined to instantiate a \Certificate\ message 
  from an input buffer.
}]
    static final class CertificateEntry {
        final byte[] encoded;       // encoded cert or public key
        private final SSLExtensions extensions;
\end{lstlisting}%

\begin{lstlisting}[
  float=tbp,
  linerange={
    73-74,
    %732-739, %%CertificateEntry
    %777-777,
    %782-784, %%T13CertificateMessage
    %786-798,
    %800-806,
    %913-913,
    918-919, %%T13CertificateProducer
    926-927,
    %929-929,
    %934-935,
    %937-937,
    929-937,
    939-941,   %%onProduceCertificate
    943-943,
    955-956,
    963-971,
    974-975,
    %977-981,       %%stapling
    983-995,
    1001-1003,
    1005-1007,
    1126-1126
  },
  label=lst:CertificateMessageB,
  caption={[\code{CertificateMessage.T13CertificateProducer} produces \Certificate]
  Class \code{CertificateMessage.T13CertificateProducer} defines method \code{produce} to write 
  (to an output stream) a \Certificate\ message, originating from a client (Lines~931--932) 
  or server (Lines~934--935). 
  For the latter, a private key and authenticating certificates are wrapped inside an instance 
  of class \code{X509Authentication.X509Possession} (Lines~943--955), using method \code{choosePossession} 
  (Listing~\ref{lst:CertificateMessageC}); the server's active context is updated to include that 
  private key and associated certificates (Lines~964--967); a \Certificate\ message is constructed
  from the certificates (Lines~968--975); and the message is written to an output stream 
  (Lines~1002--1003).
}]
    static final HandshakeProducer t13HandshakeProducer =
        new T13CertificateProducer();
\end{lstlisting}%

\begin{lstlisting}[
  float=tbp,
  linerange={
    1009-1011,
    %1021-1021,
    1022-1022,
    1031-1035,
    1043-1044,
    1046-1047,
    1053-1054,
    1056-1057,
    1062-1063,
    1065-1066,
    1071-1072
  },
  label=lst:CertificateMessageC,
  caption={[\code{CertificateMessage.T13CertificateProducer} produces \Certificate\ (cont.)]
  Class \code{CertificateMessage.T13CertificateProducer} (omitted from Listing~\ref{lst:CertificateMessageB}) 
  defines method \code{choosePossession} to iterate over the client offered signature 
  algorithms for certificates (defined by extension \TLSsignatureAlgorithmsCert, 
  or \TLSsignatureAlgorithms\ if the former is absent), which class 
  \code{CertSignAlgsExtension.CHCertSignatureSchemesUpdate} 
  (respectively \code{SignatureAlgorithmsExtension.CHSignatureSchemesUpdate})
  assigns to variable \code{hc.peerRequestedCertSignSchemes}; disregard
  algorithms not offered for signing \CertificateVerify\ requests
  (Lines~1033--1044), unsupported algorithms (Lines~1046--1054), 
  or algorithms for which no suitable private key is available (1056--1063); and return a private key 
  for the first suitable algorithm (Line~1065), or null if no such key
  exists (Line~1071).
}]
        private static SSLPossession choosePossession(
                HandshakeContext hc,
                ClientHelloMessage clientHello) throws IOException {
\end{lstlisting}%

\begin{lstlisting}[
  float=tbp,
  linerange={
    71-72,
    %732-739, %%CertificateEntry
    %777-777,
    %782-784, %%T13CertificateMessage
    %786-798,
    %800-806,
    1131-1131, %%T13CertificateConsumer
    1138-1139,
    1141-1141,
    1144-1146,
    1151-1152,
    1157-1159,
    1186-1187,
    1194-1200,
    1202-1204,
    1207-1207,
    1209-1212,
    %%
    1369-1369
  },
  label=lst:CertificateMessageD,
  caption={[\code{CertificateMessage.T13CertificateConsumer} consumes \Certificate]
  Class \code{CertificateMessage.T13CertificateConsumer} 
  defines method \code{consume} to instantiate a \Certificate\ message from an 
  input buffer (Line~1145) and consume the message as originating from a server (Line~1151)
  or client (Lines~1157). For the former, certificates are checked (Lines~1203--1204) and
  the active context is updated (Lines~1209--1211).
}]
    static final SSLConsumer t13HandshakeConsumer =
        new T13CertificateConsumer();
\end{lstlisting}%

\begin{lstlisting}[
  float=tbp,
  linerange={
    793-795,
    813-813,
    823-823,
    825-826,
    844-844,
    854-854,
    857-858,
    860-861,
    863-870,
    878-910,
    1030-1030
  },
  label=lst:CertificateVerify,
  caption={[\code{CertificateVerify.T13CertificateVerifyMessage} defines \CertificateVerify]
  Class \code{CertificateVerify.T13CertificateVerifyMessage} defines the two fields of a 
  \CertificateVerify\ message (Lines~858 \&~861) and constructors to instantiate them from parameters
  (Lines~863--910) or an input buffer (Listing~\ref{lst:CertificateVerifyB}). The former instantiates 
  the first field with the 
  chosen signature algorithm (Lines~867--870); derives the string over which to compute the signature
  (Lines~878--890), using constant \code{serverSignHead} (Lines~764--823) for messages originating  
  from a server, and constant \code{clientSignHead} (Lines~825--854) for messages originating from
  a client, where bytes used to construct those contents are omitted for brevity; and instantiates
  the second field as a signature over that string (Lines~892--909).
}]
    static final class T13CertificateVerifyMessage extends HandshakeMessage {
        private static final byte[] serverSignHead = new byte[] {
            // repeated 0x20 for 64 times
            // "TLS 1.3, server CertificateVerify" + 0x00
        };
        private static final byte[] clientSignHead = new byte[] {
            // repeated 0x20 for 64 times
            // "TLS 1.3, client CertificateVerify" + 0x00
        };
        // the signature algorithm
        private final SignatureScheme signatureScheme;
        // signature bytes
        private final byte[] signature;
        T13CertificateVerifyMessage(HandshakeContext context,
                X509Possession x509Possession) throws IOException {
            super(context);

            this.signatureScheme = SignatureScheme.getPreferableAlgorithm(
                    context.peerRequestedSignatureSchemes,
                    x509Possession.popPrivateKey,
                    context.negotiatedProtocol);
            byte[] hashValue = context.handshakeHash.digest();
            byte[] contentCovered;
            if (context.sslConfig.isClientMode) {
                contentCovered = Arrays.copyOf(clientSignHead,
                        clientSignHead.length + hashValue.length);
                System.arraycopy(hashValue, 0, contentCovered,
                        clientSignHead.length, hashValue.length);
            } else {
                contentCovered = Arrays.copyOf(serverSignHead,
                        serverSignHead.length + hashValue.length);
                System.arraycopy(hashValue, 0, contentCovered,
                        serverSignHead.length, hashValue.length);
            }

            byte[] temproary = null;
            try {
                Signature signer =
                    signatureScheme.getSignature(x509Possession.popPrivateKey);
                signer.update(contentCovered);
                temproary = signer.sign();
            } catch (NoSuchAlgorithmException |
                    InvalidAlgorithmParameterException nsae) {
                context.conContext.fatal(Alert.INTERNAL_ERROR,
                        "Unsupported signature algorithm (" +
                        signatureScheme.name +
                        ") used in CertificateVerify handshake message", nsae);
            } catch (InvalidKeyException | SignatureException ikse) {
                context.conContext.fatal(Alert.HANDSHAKE_FAILURE,
                        "Cannot produce CertificateVerify signature", ikse);
            }

            this.signature = temproary;
        }
    }
\end{lstlisting}%

\begin{lstlisting}[
  float=tbp,
  linerange={
    912-914,
    925-927,
    941-948,
    957-991
  },
  label=lst:CertificateVerifyB,
  caption={[\code{CertificateVerify.T13CertificateVerifyMessage} defines \CertificateVerify\ (cont.)]
  Class \code{CertificateVerify.T13CertificateVerifyMessage} (omitted from Listing~\ref{lst:CertificateVerify})
  defines a constructor which instantiates a \CertificateVerify\ message from an input buffer, parametrising 
  the first field with the chosen signature algorithm (Lines~926--927) and the second with the signature
  (Line~957), if the signature verifies (Lines~974--980) with respect to the expected string
  (Lines~959--971).
}]
        T13CertificateVerifyMessage(HandshakeContext context,
                ByteBuffer m) throws IOException {
             super(context);
            // SignatureAndHashAlgorithm algorithm
            int ssid = Record.getInt16(m);
            this.signatureScheme = SignatureScheme.valueOf(ssid);
            // read and verify the signature
            X509Credentials x509Credentials = null;
            for (SSLCredentials cd : context.handshakeCredentials) {
                if (cd instanceof X509Credentials) {
                    x509Credentials = (X509Credentials)cd;
                    break;
                }
            }
            this.signature = Record.getBytes16(m);

            byte[] hashValue = context.handshakeHash.digest();
            byte[] contentCovered;
            if (context.sslConfig.isClientMode) {
                contentCovered = Arrays.copyOf(serverSignHead,
                        serverSignHead.length + hashValue.length);
                System.arraycopy(hashValue, 0, contentCovered,
                        serverSignHead.length, hashValue.length);
            } else {
                contentCovered = Arrays.copyOf(clientSignHead,
                        clientSignHead.length + hashValue.length);
                System.arraycopy(hashValue, 0, contentCovered,
                        clientSignHead.length, hashValue.length);
            }

            try {
                Signature signer =
                    signatureScheme.getSignature(x509Credentials.popPublicKey);
                signer.update(contentCovered);
                if (!signer.verify(signature)) {
                    context.conContext.fatal(Alert.HANDSHAKE_FAILURE,
                        "Invalid CertificateVerify signature");
                }
            } catch (NoSuchAlgorithmException |
                    InvalidAlgorithmParameterException nsae) {
                context.conContext.fatal(Alert.INTERNAL_ERROR,
                        "Unsupported signature algorithm (" +
                        signatureScheme.name +
                        ") used in CertificateVerify handshake message", nsae);
            } catch (InvalidKeyException | SignatureException ikse) {
                context.conContext.fatal(Alert.HANDSHAKE_FAILURE,
                        "Cannot verify CertificateVerify signature", ikse);
            }
        }
\end{lstlisting}%

\begin{lstlisting}[
  float=tbp,
  linerange={
    60-61,
    1035-1036,
    1043-1046,
    1048-1053,
    1066-1073,
    1075-1078,  
    1084-1086,
    1088-1090,
    1108-1108
  },
  label=lst:CertificateVerifyC,
  caption={[\code{CertificateVerify.T13CertificateVerifyProducer} produces \CertificateVerify]
  Class \code{CertificateVerify.T13CertificateVerifyProducer} defines method \code{produce} to
  write (to an output stream) a \CertificateVerify\ message, originating from a client (Lines~1067--1068) or server (Lines~1070--1071).  
  \ifImplNotes
  \textcolor{red}{The implementations of method \code{onProduceCertificateVerify} parameterised 
  on \code{ServerHandshakeContext} and \code{ClientHandshakeContext} are identical up to 
  variables \code{shc} and \code{chc}, and string \code{"server"} and \code{"client"}, refactoring
  could eliminate unnecessary code.}
  \textcolor{red}{Reported to security-dev@openjdk.java.net on 27 May 2020}
  \fi
  For the latter, a \CertificateVerify\ message is constructed (Lines~1077-1078)
  and written to an output stream (Lines~1085-1086).
}]
    static final HandshakeProducer t13HandshakeProducer =
        new T13CertificateVerifyProducer();
    private static final
            class T13CertificateVerifyProducer implements HandshakeProducer {
        public byte[] produce(ConnectionContext context,
                HandshakeMessage message) throws IOException {
            // The producing happens in handshake context only.
            HandshakeContext hc = (HandshakeContext)context;
            X509Possession x509Possession = null;
            for (SSLPossession possession : hc.handshakePossessions) {
                if (possession instanceof X509Possession) {
                    x509Possession = (X509Possession)possession;
                    break;
                }
            if (hc.sslConfig.isClientMode) {
                return onProduceCertificateVerify(
                        (ClientHandshakeContext)context, x509Possession);
            } else {
                return onProduceCertificateVerify(
                        (ServerHandshakeContext)context, x509Possession);
            }
        }
        private byte[] onProduceCertificateVerify(ServerHandshakeContext shc,
                X509Possession x509Possession) throws IOException {
            T13CertificateVerifyMessage cvm =
                    new T13CertificateVerifyMessage(shc, x509Possession);
            // Output the handshake message.
            cvm.write(shc.handshakeOutput);
            shc.handshakeOutput.flush();
            // The handshake message has been delivered.
            return null;
        }
    }
\end{lstlisting}%

\begin{lstlisting}[
  float=tbp,
  linerange={
    58-59, 
    1113-1114,
    1121-1122,
    1124-1126,
    1141-1142    
  },
  label=lst:CertificateVerifyD,
  caption={[\code{CertificateVerify.T13CertificateVerifyConsumer} consumes \CertificateVerify]
  Class \code{CertificateVerify.T13CertificateVerifyConsumer} defines method \code{consume} to 
  instantiate a \CertificateVerify\ message from an input buffer (Line~1125--1126), checking 
  validity of the message's signature as a side effect.
}]
    static final SSLConsumer t13HandshakeConsumer =
        new T13CertificateVerifyConsumer();
    private static final
            class T13CertificateVerifyConsumer implements SSLConsumer {
        public void consume(ConnectionContext context,
                ByteBuffer message) throws IOException {
            HandshakeContext hc = (HandshakeContext)context;
            T13CertificateVerifyMessage cvm =
                    new T13CertificateVerifyMessage(hc, message);
        }
    }
\end{lstlisting}%

%%%%%%%%%%%%%%%%%%%%%%%%%%%%%%%%%%%%%%%%%%%
%%%%%%%%%%%%%%%%%%%%%%%%%%%%%%%%%%%%%%%%%%%
%%%%%%%%%%%%%%%%%%%%%%%%%%%%%%%%%%%%%%%%%%%

\subsubsection{\Finished}\label{sec:FIN}

The handshake protocol concludes with a \Finished\ message, which provides key confirmation, 
binds the server's identity to the exchanged keys (and the client's identity, if client 
authentication is used), and, for PSK-based key exchange, authenticates the handshake. 
A \Finished\ message comprises of the following field:

\begin{description}
\item \TLSverifyData: An HMAC over the entire handshake.
\end{description}

\noindent
The HMAC is computed as  
\[
  \HMACHash(\TLSfinishedKey,\TranscriptHash(\TLSHandshakeMessage))
\]
where \TLSHandshakeMessage\ is a concatenation of the protocol's messages (\S\ref{sec:hkdf:transcript}),
\begin{multline*}
  \TLSfinishedKey =   \HKDFExpandLabel(\TLShandshakeTrafficSecret,\\ ``finished", ``", \HashLength),
\end{multline*}
%
%traffic secret $S$ is \TLShandshakeTrafficSecret\ when the \Finished\ message concludes an initial 
%handshake and \TLSapplicationTrafficSecretN\ when concluding post-handshake authentication, and 
%\TLSField{[sender]} is either \TLSField{client} or \TLSField{server}.
%\ifSpecNotes
%\textcolor{red}{
%  Q: In the final instance, shouldn't \TLSapplicationTrafficSecret[client]{N}
%  be \TLSapplicationTrafficSecretN? A: No, it should not. During post-handshake authentication, only the client sends a finished message (as far
%  as I can tell).
%}
%\fi
%and traffic secret $S$ is \TLSserverHandshakeTrafficSecret\ when the \Finished\ message originates from
%a server to conclude an initial handshake, \TLSclientHandshakeTrafficSecret\ when originating 
%from a client to conclude an initial handshake, and \TLSapplicationTrafficSecret[client]{N} when 
%concluding post-handshake authentication. 
%(Post-handshake authentication is only concerned with updating the client's application-traffic key, 
%for the purposes of blinding the client's identity to that key. Hence, secret \TLSfinishedKey\ is not 
%concerned with traffic secret \TLSapplicationTrafficSecret[server]{N}. Beyond traffic keys, a key 
%established by a \NewSessionTicket\ message, sent after post-handshake authentication, will also be 
%bound to the client's identity.)
%
%% Defer post-handshake aspects to the appendix.
%
and \TLSField{[sender]} is \TLSField{server} when the \Finished\ message originates from
a server to conclude a handshake and \TLSField{client} when originating from a client. %to conclude a handshake. 

A \Finished\ message is first sent by the server (immediately after a \CertificateVerify\ message 
for (EC)DHE-only key exchange and immediately after an \EncryptedExtensions\ message for PSK-based 
key exchange). That message is consumed by the client, which recomputes the HMAC (using secret
\TLSserverHandshakeTrafficSecret) and checks that it matches the \Finished\ message's HMAC 
(\Finished.\TLSverifyData), terminating the connection with a \TLSdecryptError\ alert if the check 
fails. A client that successfully consumes a server's \Finished\ message responds with its own \Finished\
message, which is similarly consumed by the server (albeit using secret \TLSclientHandshakeTrafficSecret).
(That message is preceded by client generated \Certificate\ and \CertificateVerify\ messages, if 
client authentication is used.) Once endpoints have successfully consumed \Finished\ messages, 
(encrypted) application data may be exchanged. Moreover, a server may send (encrypted) application 
data immediately after sending its \Finished\ message (i.e., without consuming a \Finished\ message), 
albeit, since \ClientHello\ messages may be replayed, any such data is sent without assurance of the 
client's liveness (nor identity). 

\begin{tcolorbox}
\Finished\ messages are implemented, produced, and consumed by inner-classes of class \code{Finished} 
(Listings~\ref{lst:Finished}--\ref{lst:T13FinishedConsumer:ServerSide}).
    \ifImplNotes
\textcolor{blue}{Classes \code{Finished.T13FinishedProducer} and \code{Finished.T13FinishedConsumer}
  seem to contain some obsolete secure renegotiation code.} 
    \fi
\ifPresentationNotes
\textcolor{red}{Perhaps separate the listings into production/consumption, and establishing keys, as we 
did earlier. Or perhaps not, the reader should be able to consume it all (no pun intended) at this stage.}
\fi
\end{tcolorbox}

\lstinputlisting[
  float=tbp,
  linerange={
    69-70,
    72-73,
    75-76,
    78-84,
    86-87,
    89-98,
    106-107,
    109-114,
    117-123, 
    156-156
  },
  label=lst:Finished,
  caption={[\code{Finished.FinishedMessage} defines \Finished]
  Class \code{Finished.FinishedMessage} defines the one field of a \Finished\ message (Line~70)
  and two constructors to instantiate it. The first constructor parameterises the field with an HMAC
  it constructs (Lines 72--87) and the second parses an HMAC from an input buffer (Lines 92--107),
  recomputes the expected HMAC itself (Lines~109--118), and checks that the HMACs match
  (Lines~119--122). The HMACs are (indirectly) computed using method 
  \code{T13VerifyDataGenerator.createVerifyData} (Listing~\ref{lst:FinishedB}).
}]{listings/Finished.java}%

\lstinputlisting[
  float=tbp,
  linerange={
    326-329,
    332-333,
    335-358
  },
  label=lst:FinishedB,
  caption={[\code{Finished.T13VerifyDataGenerator} defines \Finished\ (cont.)]
  Class \code{Finished.T13VerifyDataGenerator} defines method \code{createVerifyData}
  to compute HMACs for \Finished\ messages. 
  \ifImplNotes
  \textcolor{red}{Comment on Line 325 doesn't match the class}
  \fi
  That method computes variable \code{finishedSecret} by indirect application
  of method \code{HKDF.expand} to inputs including secret \code{context.baseReadSecret} or
  \code{context.baseWriteSecret}, and \TLSHKDFLabelExt, which is computed over the 
  negotiated hash function's output length, 
  label $``\textrm{tls13\textvisiblespace{}finished}"$, and null ASCII character 0x00,
  using class \code{SSLBasicKeyDerivation} to apply method \code{HKDF.expand}.
  (Class \code{SSLBasicKeyDerivation} uses method \code{createHkdfInfo}
  to handle concatenation and is reliant on method \code{Record.putBytes8} to introduce 
  0x00. This differs from a similar application of method \code{HKDF.expand} by 
  class \code{SSLTrafficKeyDerivation.T13TrafficKeyDerivation}, which introduces 
  0x00 itself.)
  \ifImplNotes
  \textcolor{red}{Classes \code{SSLBasicKeyDerivation} and 
  \code{SSLTrafficKeyDerivation.T13TrafficKeyDerivation} are almost identical, 
  and some refactoring could eliminate one of the classes.}
  \fi
}]{listings/Finished.java}%

\begin{comment}
\lstinputlisting[
  float=tbp,
  widthgobble=0*0,
  linerange={
    35-38,
    40-45,
    48-58,
    81-81
  },
  label=lst:SSLBasicKeyDerivation,
  caption={[\code{SSLBasicKeyDerivation} defines \Finished\ (cont.)]
  Class \code{SSLBasicKeyDerivation} defines three fields (Lines~36--38),
  a constructor to instantiate them (Lines~40-45), and method \code{deriveKey}
  to return an HMAC computed by application of method \code{HKDR.expand}
  (Lines~52--53) to inputs including \TLSHKDFLabelExt, which is computed 
  (Line~44) over some label, context, and length, using static method 
  \code{SSLBasicKeyDerivation.createHkdfInfo} to handle concatenation.
    \textcolor{red}{Drop this figure, pushing details into the previous figure}
}]{listings/SSLBasicKeyDerivation.java}%
\end{comment}

\lstinputlisting[
  float=tbp,
  widthgobble=0*0,
  linerange={
    63-64,
    630-631,
    638-639,  
    641-649,
    738-741,
    743-743,
    749-751,
    753-754,
    762-763,
    772-783
  },
  label=lst:T13FinishedProducer,
  caption={[\code{Finished.T13FinishedProducer} produces server-side \Finished]
  Class \code{Finished.T13FinishedProducer} defines method \code{produce}
  to write (to an output stream) a \Finished\ message, originating from a 
  client or a server, and to establish \TLSMasterSecret\ when the message 
  originates from such a server. For messages originating 
  from servers, processing  
  proceeds with method \code{onProduceFinished}, parameterised by the 
  server's active context. That method updates the transcript hash's 
  digest to include all handshake protocol messages (Line~741), 
  instantiates and outputs a \Finished\ message (Lines~743--751), and 
  establishes \TLSMasterSecret\ (Lines~782--783). Variable \code{shc.handshakeKeyDerivation} 
  (Line~754) is assigned by class \code{ServerHello.T13ServerHelloProducer} (Listing 27)
  as an instance of class \code{SSLSecretDerivation}, parameterised by 
  \TLSHandshakeSecret, hence, the salt necessary to establish 
  \TLSMasterSecret\ is correctly derived (Line~774), as-is the necessary 
  \HashLength-length string of zeros (Lines~780--781).
}]{listings/Finished.java}%

\lstinputlisting[
  float=tbp,
  widthgobble=0*0,
  linerange={
    785-786,
    788-811,
    814-815,
    824-831
  },
  label=lst:T13FinishedProducerB,
  caption={[\code{Finished.T13FinishedProducer} produces server-side \Finished\ (cont.)]
  Class \code{Finished.T13FinishedProducer} defines method \code{onProduceFinished}
  (continued from Listing~\ref{lst:T13FinishedProducer}) to derive traffic   
  secret \TLSserverApplicationTrafficSecret\ from an instance of  
  class \code{SSLSecretDerivation}, parameterised by \TLSMasterSecret\ (Lines~785--790); 
  constructs an instance of \code{SSLTrafficKeyDerivation.T13TrafficKeyDerivation}
  from that secret (Lines~791--792); and uses that instance to derive the corresponding 
  traffic keys \TLSField{server\_write\_key} (Lines~793--794) and
  \TLSField{server\_write\_iv} (Lines~795--796), used to encrypt (and write) outgoing 
  traffic (Lines~797--807). Moreover, the method prepares the server's active context
  for the client's response (Lines~825--826).
}]{listings/Finished.java}%

\lstinputlisting[
  float=tbp,
  widthgobble=0*0,
  linerange={
    61-62,
    836-836,
    843-844,
    846-854,
    856-858,
    874-884,
    892-893,
    909-920
  },
  label=lst:T13FinishedConsumer,
  caption={[\code{Finished.T13FinishedConsumer} consumes server-generated \Finished]
  Class \code{Finished.T13FinishedConsumer} defines method \code{consume}
  to read (from an input buffer) a \Finished\ message, originating from a 
  client or a server, and to establish \TLSMasterSecret\ when the message 
  originates from such a server. For messages 
  originating from servers, processing proceeds with method \code{onConsumeFinished}, 
  parameterised by the client's active context. That method instantiates 
  a \Finished\ message from the input buffer (Line~858), updates the transcript hash's 
  digest to include all handshake protocol messages (Line~883), and 
  establishes \TLSMasterSecret\ (Lines~919--920). (Computations are similar to 
  Listing~\ref{lst:T13FinishedProducer} and refactoring could eliminate unnecessary 
  code.) 
}]{listings/Finished.java}%

\lstinputlisting[
  float=tbp,
  widthgobble=0*0,
  linerange={
    922-947,
    950-951,
    953-963,
    965-972
  },
  label=lst:T13FinishedConsumerB,
  caption={[\code{Finished.T13FinishedConsumer} consumes server-generated \Finished\ (cont.)]
  Class \code{Finished.T13FinishedConsumer} defines method \code{consume}
  (continued from Listing~\ref{lst:T13FinishedConsumer}) to derive traffic   
  secret \TLSserverApplicationTrafficSecret\ (Lines~922--927) and corresponding 
  traffic keys \TLSField{server\_write\_key} (Lines~930--931) and
  \TLSField{server\_write\_iv} (Lines~932--933), used to decrypt (and read) incoming 
  traffic (Lines~934--943). (Computations are similar to Listing~\ref{lst:T13FinishedProducerB} 
  and refactoring could eliminate unnecessary code.) Moreover, the method updates the 
  client's active context to include a producer for \Finished\ messages (Lines~956--957);
  constructs an array of producers clients might use during the remainder of the 
  handshake protocol, namely, produces for messages \Certificate, \CertificateVerify, 
  and \Finished, in the order that they might be used (Lines~958--963); and uses 
  those producers to produce messages when the active context includes the producer
  (Lines~965--971). Since a \Finished\ message producer is included, a \Finished\
  message is always produced, using class \code{Finished.T13FinishedProducer}
  (Listing~\ref{lst:T13FinishedProducer} \&~\ref{lst:T13FinishedProducer:ClientSide}).
}]{listings/Finished.java}%

\lstinputlisting[
  float=tbp,
  widthgobble=0*0,
  linerange={
    651-654,
    656-656,
    662-664,
    672-673,
    681-682,
    691-712,
    714-714,
    717-718,
    %727-736
    734-736
  },
  label=lst:T13FinishedProducer:ClientSide,
  caption={[\code{Finished.T13FinishedProducer} produces client-side \Finished\ (cont.)]
  Class \code{Finished.T13FinishedProducer} defines method \code{onProduceFinished} 
  parameterised by a client's active context (omitted from Listing~\ref{lst:T13FinishedProducer})
  to write (to an output stream) a \Finished\ message originating from a 
  client, and to derive traffic secret \TLSclientApplicationTrafficSecret\ (Lines~692--693) 
  and corresponding traffic keys \TLSField{client\_write\_key} (Lines~698--699) and
  \TLSField{client\_write\_iv} (Lines~700--701), used to encrypt (and write) outgoing
  traffic (Lines~702--712). %Moreover, the method \textcolor{red}{...XXX...}
}]{listings/Finished.java}%

\begin{lstlisting}[
  float=tbp,
  widthgobble=0*0,
  linerange={
    974-976,
    991-995,
    1003-1004,
    %1013-1019,
    1020-1040,
    %1041-1048,
    1049-1049,
    1052-1053,
    %1073-1073,
    1075-1075
  },
  label=lst:T13FinishedConsumer:ServerSide,
  caption={[\code{Finished.T13FinishedConsumer} consumes client-generated \Finished\ (cont.)]
  Class \code{Finished.T13FinishedConsumer} defines method \code{onConsumeFinished} 
  parameterised by a server's active context (omitted from Listing~\ref{lst:T13FinishedConsumer})
  to read (from an input buffer) a \Finished\ message originating from a 
  client, and to derive traffic secret \TLSclientApplicationTrafficSecret\ (Lines~1022--1023) 
  and corresponding traffic keys \TLSField{client\_write\_key} (Lines~1027--1028) and
  \TLSField{client\_write\_iv} (Lines~1029--1030), used to decrypt (and read) incoming
  traffic (Lines~1031--1040). (Computations are similar to Listing~\ref{lst:T13FinishedProducer:ClientSide} 
  and refactoring could eliminate unnecessary code.)
  \ifImplNotes
  \textcolor{blue}{Not only computation similar to Listing~\ref{lst:T13FinishedProducer:ClientSide}, 
    but computation is (unsurprisingly) similar between other traffic key derivations. This could be 
    resolved by refactoring.}
  \textcolor{red}{Reported to security-dev@openjdk.java.net on 27 May 2020}
  \fi
}]
        private void onConsumeFinished(ServerHandshakeContext shc,
                ByteBuffer message) throws IOException {
            FinishedMessage fm = new FinishedMessage(shc, message);
            //
            // update
            //
            // Change client/server application traffic secrets.
            SSLKeyDerivation kd = shc.handshakeKeyDerivation;
            SSLTrafficKeyDerivation kdg =
                    SSLTrafficKeyDerivation.valueOf(shc.negotiatedProtocol);
\end{lstlisting}%

Traffic secrets \TLSserverApplicationTrafficSecret\ and \TLSclientApplicationTrafficSecret\
are used to derive application-traffic keys to protect application data. 

\begin{comment}
\sout{Both secrets are 
computed from the protocol's transcript, but at different stages of the transcripts
evolution. Indeed, the former traffic secret is computed from the transcript after the server
generates its \Finished\ message (and when the transcript concludes with that server-generated 
\Finished\ message), whereas the latter traffic secret is computed after the client generates 
its \Finished\ message (and when the transcript concludes with that client-generated \Finished\
message).} \textcolor{red}{That's false according to the spec: Figure~\ref{fig:keyDerivation} 
shows that both secrets are computed from transcripts ending with a server-generated \Finished\ 
message. But, the code suggests the crossed-out text is true. Truthify.}
%%
%% I have confirmed that both secrets are computed from transcripts ending with a 
%% server-generated \Finished\ message. I haven't dug into the code too deeply. My 
%% guess is that 

\end{comment}

\subsubsection*{Application data}

TLS protects application-layer communication independently of specific applications.
%How applications should use TLS is unspecified. 
Independence %(between applications and TLS) 
is readily apparent from the specification: There 
is no mention of interaction between applications and TLS. Designers and implementors
must decide for themselves how to use TLS within their applications. For instance,
when to initiate a handshake and how to validate certificates.

  \subsection{Early data}\label{sec:handshakeEarlyData}

For PSK-based key exchange, clients may exceptionally start sending encrypted application data 
immediately after \ClientHello\ messages (before receiving a \ServerHello\ message),\footnote{%
  RFC 8446 only permits clients to send early data when the pre-shared key is associated with 
  data permitting them to do so. For resumption PSKs, permission is granted by inclusion of 
  extension \TLSearlyData\ in \NewSessionTicket\ messages (\S\ref{sec:NST}).
}
enabling 
a zero round-trip time (0-RTT), at the cost of forward secrecy 
(since application data is solely encrypted by the 
pre-shared key, %rather than a combination of that key and Diffie-Hellman
%  key shares, hence, forward secret cannot be obtained
which does not afford forward secrecy, as per PSK-only key exchange) and replay protection
(since such protection is derived from the server's nonce, which is constructed
after encrypted application data is sent).\footnote{
  RFC 8446 discusses anti-replay defences and notes that 
  single-use PSKs enjoy forward secrecy.
}
Such early data requires the
\ClientHello\ message to include extensions \TLSearlyData\ and \TLSpsk,
and 
application data must be encrypted using 
%the pre-shared key identified by the first entry in the client's list of pre-shared key identifier .
the client's first identified pre-shared key. (Extension \TLSearlyData\
is not associated with data, encrypted application data is sent separately.)

To consume early data, a server must select the client's first pre-shared key
identifier and an offered cipher suite associated with that identifier. The
server must check the identifier is associated with the server-selected 
protocol version and (if extension \TLSapplicationLayerProtocolNegotiation\ is present) 
application protocol. (These checks are a superset of those for PSK-based key 
exchange without early data.) Additionally, for resumption PSKs, the server 
must check that the PSK is not beyond its lifetime. 
If checks succeed (and the server is willing to consume early data), 
then the server will include a corresponding \TLSearlyData\ extension
in their \EncryptedExtensions\ message.
(When consuming that extension, the client must check that the server selected
the client's first pre-shared key identifier, aborting with an \TLSillegalParameter\
alert, if the check fails.)
Otherwise, no such extension will 
be sent and no early data will be consumed (extension \TLSearlyData\ is
ignored), and the server will proceed in one of the following two ways (which 
must also be followed by servers not supporting early data): Respond 
with a \ServerHello\ message, excluding extension \TLSearlyData,
or respond with a \HelloRetryRequest\ message, forcing the client to send 
a second \ClientHello\ message without extension \TLSearlyData.
In both cases, the server must skip past early data. For the former, 
given that all messages will be encrypted, the server must decrypt 
messages with the handshake traffic key, discard messages when 
decryption fails, and treat the first successfully decrypted
message as the client's next handshake message, thereafter 
proceeding as if no early data were sent. 
For the latter, the second \ClientHello\ message will be unencrypted
and the server can discard all encrypted messages (identified by
record type \TLSapplicationData\ (0x23), rather than type \TLShandshake\
(0x22), as introduced in Section~\ref{sec:record}), before proceeding as 
if no early data were sent when the second \ClientHello\ message
is identified.
(When the pre-shared key is associated with a maximum amount of early 
data, the server should abort with an \TLSunexpectedMessage\ alert if 
the maximum is exceeded when skipping past early data.)

\begin{tcolorbox}
Early data is not supported by JDK~11 (\url{https://openjdk.java.net/jeps/332}),
nor subsequent versions: When extension \TLSearlyData\ is included in message
\ClientHello, that extension will be processed (Line~1119, Listing~\ref{lst:T13ClientHelloConsumer})
and runtime exception \code{UnsupportedOperationException} will be thrown 
(omitted from Listing~\ref{lst:SSLExtension}).
\ifImplNotes
\textcolor{red}{
Since JDK throws a runtime exception rather than skipping early data, 
is JDK non-compliant with the spec, or can runtime exceptions be used
to avoid non-compliance?
}
\fi
\end{tcolorbox}

\subsubsection*{Data associated with pre-shared keys}

An external PSK (established independently of TLS) must minimally be associated 
with a hash function and an identity. (The hash function may default to SHA-256, 
if no function is explicitly associated.)
Such a PSK grants freedom over AEAD algorithms, whilst 
fixing the hash function. By comparison, a resumption PSK (established by 
\NewSessionTicket\ messages) is associated with values negotiated in the 
connection that provisioned the PSK, which fixes a cipher suite, hence, 
an AEAD algorithm and a hash function.\footnote{%
  Although resumption PSKs are associated with cipher suites, they need 
  not be used with defined AEAD algorithms, except for compatibility 
  with early data.}
(It follows that a connection established using a resumption PSK will 
inherit security from the connection in which the resumption PSK was 
established.)
Resumption PSKs are compatible with 
early data by default (assuming suitable provisioning with extension 
\TLSapplicationLayerProtocolNegotiation, if relevant), whereas (minimally 
associated) external PSKs are not. They must be associated with a cipher suite 
(rather than just a hash function), a protocol version, and (if relevant)  an 
application protocol, for compatibility with early data.

\subsubsection{\EndOfEarlyData}

A client can transmit early data until they receive a server's \Finished\
message. After which, the client transmits an \EndOfEarlyData\ message
(encrypted using a key derived from secret \TLSclientEarlyTrafficSecret), 
if the server's \EncryptedExtensions\ message included extension
\TLSearlyData. Otherwise, early data has not and will not be consumed 
by the server, and no \EndOfEarlyData\ message is sent. The \EndOfEarlyData\ 
message indicates that all early data has been transmitted and subsequent
handshake messages will be encrypted with the client's handshake-traffic key.
(Servers must not send \EndOfEarlyData\ messages and clients receiving 
such messages must abort with an \TLSunexpectedMessage\ alert.)

  \subsection{Further features}
  \subsubsection{\NewSessionTicket}\label{sec:NST}

After receiving a client's \Finished\ message, a server can initiate establishment 
of a new pre-shared key, which will be derived from the resumption master secret
\TLSresumptionMasterSecret\ (Figure~\ref{fig:keyDerivation}). 
Such a pre-shared key may be used to establish subsequent channels. Establishment 
is initiated with a \NewSessionTicket\ message, comprising the following fields:

\begin{description}
\item \TLSticketLifetime: A 32-bit unsigned integer indicating the lifetime 
      in seconds %(from the time of issuance) 
      of the pre-shared key, which must not exceed seven days (604800 seconds).

\item \TLSticketAgeAdd: A 32-bit nonce to obscure the lifetime.

\item \TLSticketNonce: A nonce for key derivation.

\item \TLSticket: A key identifier. 

\item \TLSextensions: A list of extensions, currently limited to extension \TLSearlyData, 
  which indicates that early data is permitted and defines a maximum amount 
  of early data.

\end{description}

\begin{sloppypar}
\noindent
The associated pre-shared key is computed as follows:
\[
  \HKDFExpandLabel(\TLSresumptionMasterSecret, ``resumption", \TLSticketNonce, \HashLength),
\]
Since the pre-shared key is computed from nonce \TLSticketNonce, it follows 
that each \NewSessionTicket\ message creates a distinct pre-shared key.
\end{sloppypar}

A \NewSessionTicket\ is consumed by the client, which derives and stores 
the pre-shared key along with associated data (including the negotiated 
hash function). That data may be stored by client and used in 
extension \TLSpsk\ of subsequent \ClientHello\ messages.
Data must not be used 
\ifSpecNotes
\textcolor{red}{the spec refers to storage rather than use}
\fi
longer than seven days or beyond its lifetime (specified by \TLSticketLifetime), 
whichever is shorter, and endpoints may retire data early.

   The sole extension currently defined for NewSessionTicket is
   "early_data", indicating that the ticket may be used to send 0-RTT
   data (Section 4.2.10).  It contains the following value:

   max_early_data_size:  The maximum amount of 0-RTT data that the
      client is allowed to send when using this ticket, in bytes.  Only
      Application Data payload (i.e., plaintext but not padding or the
      inner content type byte) is counted.  A server receiving more than
      max_early_data_size bytes of 0-RTT data SHOULD terminate the
      connection with an "unexpected_message" alert.  Note that servers
      that reject early data due to lack of cryptographic material will
      be unable to differentiate padding from content, so clients
      SHOULD NOT depend on being able to send large quantities of
      padding in early data records.

\begin{tcolorbox}
\NewSessionTicket\ messages are implemented, produced, and consumed by inner-classes of 
class \code{NewSessionTicket}. Those classes define the five fields of a \NewSessionTicket\ 
message and constructors to instantiate them from parameters or an input buffer, moreover,
they define methods to write such a message to an output stream and to read such a message 
from an input buffer.
\end{tcolorbox}

\subsubsection*{Extension \TLSpsk}

The pre-shared key identifiers listed by extension \TLSpsk\ (\S\ref{sec:CH}) may include 
identifiers established by \NewSessionTicket\ messages, identifiers established %out-of-band,
externally,
or both. Each identifier is coupled with an obfuscated age, which is derived by addition of 
\NewSessionTicket.\TLSticketLifetime\ and \NewSessionTicket.\TLSticketAgeAdd\ (modulo $2^{32}$)
for identifiers established by \NewSessionTicket\ messages, and $0$ for identifiers established
%out-of-band. 
externally.
Moreover, extension \TLSpsk\ associates each identifier with a PSK binder, 
which binds the pre-shared key with the current handshake, and to the session in which the 
pre-shared key was generated for pre-shared keys established by \NewSessionTicket\ messages.
The PSK binder is computed as an HMAC over a partial transcript, which excludes binders, 
namely,
\[
  \HMACHash(\TLSbinderKey,\TranscriptHash(\Truncate(CH)))
\]
for transcripts which include only a single \ClientHello\ message $CH$, and
\[
  \HMACHash(\TLSbinderKey,\TranscriptHash(CH,HRR,\Truncate(CH')))
\]
for transcripts that include an initial \ClientHello\ message $CH$, followed 
by a \HelloRetryRequest\ message $HRR$ and a subsequent \ClientHello\ message $CH'$,
where function \Truncate\ removes binders. 
When consuming a \ClientHello\ message that includes extension \TLSpsk, a server recomputes 
the HMAC for their selected pre-shared key and checks that it matches the corresponding binder 
listed by the extension, aborting if the check fails or the binder is not present.

\begin{tcolorbox}
Extension \TLSpsk\ is implemented, produced, and consumed by inner-classes of 
class \code{PreSharedKeyExtension}. Those classes define fields of extension \TLSpsk\ 
 and constructors to instantiate them from parameters or an input buffer, moreover,
they define methods to write such an extension to an output stream and to read such an extension 
from an input buffer, the latter is reliant on static methods \code{checkBinder}, 
\code{computeBinder} and \code{deriveBinderKey} to recompute HMACs for pre-shared keys 
and to check whether they match the corresponding binder listed by the extension.
\end{tcolorbox}

  \subsubsection{\KeyUpdate}

After sending a \Finished\ message, an endpoint may send a \KeyUpdate\ message
to notify their peer that they are updating their cryptographic key. The message 
comprises of the following field:

\begin{description}

\item \TLSrequestUpdate: A bit indicating whether the recipient should 
  respond with their own \KeyUpdate\ message and update their own cryptographic
  key.

\end{description}

\noindent
After sending a \KeyUpdate\ message, the sender must update their application-traffic 
secret and corresponding application-traffic keys (\S\ref{sec:secrets}--\ref{sec:hkdf:trafficKeys}).

A peer that receives a \KeyUpdate\ message prior to receiving a \Finished\ 
message aborts with an \TLSunexpectedMessage\ alert, moreover, the peer 
aborts with an \TLSillegalParameter\ alert if field \TLSrequestUpdate\ does not
contain a bit. Otherwise, the peer updates its receiving keys
(\S\ref{sec:secrets}--\ref{sec:hkdf:trafficKeys}). Moreover, when the sender
requests that the peer updates their sending keys, the peer must send a
\KeyUpdate\ message of its own (without requesting that the sender update 
its cryptographic key), prior to sending any further application data.

\begin{tcolorbox}
\KeyUpdate\ messages are implemented, produced, and consumed by inner-classes of 
class \code{KeyUpdate}. Those classes define the one field of a \NewSessionTicket\ 
message and constructors to instantiate it from parameters or an input buffer, moreover,
they define methods to write such a message to an output stream and to read such a message 
from an input buffer, those methods also update application-traffic secrets and 
corresponding application-traffic keys, and the latter produces a \KeyUpdate\ message
of the receiver when requested by the sender.
\end{tcolorbox}

\section{Record protocol}\label{sec:record}

Handshake messages are encapsulated into one or more \TLSPlaintext\ records (\S\ref{sec:TLSPlaintext}), 
which, for \ClientHello, \ServerHello\ and \HelloRetryRequest\ messages, are 
immediately written to the transport layer, otherwise, \TLSPlaintext\ records are 
translated to \TLSCiphertext\ records (\S\ref{sec:TLSCiphertext}), which add protection 
prior to writing to the transport layer. (Alerts are similarly encapsulated and when 
appropriate protected. Application data is always encapsulated and protected.)

\subsection{\TLSPlaintext}\label{sec:TLSPlaintext}

Handshake messages are fragmented and each fragment is encapsulated into a 
\TLSPlaintext\ record, comprising the following fields:

\begin{description}

\item \TLStype: Constant 0x22 (\TLShandshake). (Other constants are used for records 
  encapsulating data other than handshake messages, e.g., alerts and application 
  data.)

\item \TLSlegacyRecordVersion: Constant 0x0303, except for an initial \ClientHello\
  message, which may use constant 0x0301.

\item \TLSlength: The byte length of the following field (namely, \TLSfragment), which must not exceed
  $2^{14}$ bytes.

\item \TLSfragment: A handshake message fragment.

\end{description}

\noindent
An endpoint that receives a \TLSPlaintext\ record with field \TLSlength\ set 
greater than $2^{14}$ must abort with a \TLSrecordOverflow\ alert.

\subsection{\TLSCiphertext}\label{sec:TLSCiphertext}

For protection, a \TLSPlaintext\ record is transformed into 
a \TLSCiphertext\ record, comprising of the following fields:

\begin{description}

\item \TLSopaqueType: Constant 0x23. 

\item \TLSlegacyRecordVersion: Constant 0x0303.

\item \TLSlength: The byte length of the following field (namely, \TLSencryptedRecord), which must not 
  exceed $2^{14} + 256$ bytes.

\item \TLSencryptedRecord:  Encrypted data.

\end{description}

\noindent
Encrypted data is computed, using the negotiated AEAD algorithm,
as 
\[
  %\TLSAEADEncrypted =
    \AEADEnc(\textit{write\_key}, \textit{nonce}, \textit{additional_data}, \textit{plaintext}),
\]
where \textit{write\_key} is either \TLSclientWriteKey\ or \TLSserverWriteKey;
\textit{nonce} is derived from a sequence number XORed with \TLSclientWriteIV\ or 
\TLSserverWriteIV, respectively; \textit{additional_data} is the \TLSCiphertext\ 
record header, i.e., 
$\textit{additional_data} = 
            \TLSCiphertext.\TLSopaqueType \parallel 
            \TLSCiphertext.\TLSlegacyRecordVersion \parallel 
            \TLSCiphertext.\TLSlength$;
and \textit{plaintext} comprises of \TLSPlaintext.\TLSfragment\ appended with type
\TLSPlaintext.\TLStype\ and field \TLSzeros, which contains an arbitrary-length run 
of zero-valued bytes and is used to pad a TLS record (the resulting plaintext is known as 
record \TLSInnerPlaintext).

An endpoint that receives a \TLSCiphertext\ record with field \TLSlength\ set 
greater than $2^{14} + 256$ must abort with a \TLSrecordOverflow\ alert. Otherwise,
the endpoint computes 
\[
  \AEADDec(\textit{write\_key}, \textit{nonce},  \textit{additional_data}, %\TLSAEADEncrypted),
     \TLSCiphertext.\TLSencryptedRecord),
\]
which outputs a plaintext or terminates with an error. The endpoint aborts
with a \TLSbadRecordMac\ alert in the event of such an error.

\paragraph{Per-record nonce.}

The nonce used by the negotiated AEAD algorithm is derived from a 64-bit
sequence number, which is initialised as 0, incremented by one after 
reading or writing a record, and reset to 0 whenever the key is changed. 
That sequence number is XORed with \TLSclientWriteIV\ 
or \TLSserverWriteIV\ to derive the nonce.

\begin{tcolorbox}
Outgoing records are produced by class \code{SSLSocketOutputRecord} 
(Listing~\ref{lst:SSLSocketOutputRecord}) and parent \code{OutputRecord} 
(Listing~\ref{lst:OutputRecord}), using enum \code{SSLCipher} 
(Listing~\ref{lst:SSLCipher}) to protect outgoing records.
(Alternatively, outgoing records are constructed by class 
\code{SSLEngineOutputRecord}, which shares the same parent.) 
Incoming records are consumed by class \code{SSLSocketInputRecord} 
(or \code{SSLEngineInputRecord}) and parent \code{InputRecord}, which 
uses enum \code{SSLCipher} for record protection.
\end{tcolorbox}

\lstinputlisting[
  float=tbp,
  widthgobble=0*0,
  linerange={
    38-38,
    39-39,
    94-95,
    147-163,
    165-166,
    168-172,
    181-182,
    184-186,
    193-194,
    196-199,
    589-589
  },
  label=lst:SSLSocketOutputRecord,
  caption={[\code{SSLSocketOutputRecord.encodeHandshake} fragments outgoing handshake messages]
  Class \code{SSLSocketOutputRecord} defines method \code{encodeHandshake} to fragment 
  outgoing handshake messages and write fragments to (its parent's) buffer \code{buf} (Lines~159 \&~169), 
  using  method \code{ByteArrayOutputStream.write}, 
  which (if full) is processed by parent \code{OutputRecord} (Line~182) and delivered
  (Lines~185--186). \ifPresentationNotes\textcolor{red}{What special case is dealt with on Line~159? TO DO: 
  Give further explanation of the for-loop}\fi
  The class is also responsible for adding the encapsulated message to the transcript hash, 
  if appropriate (Lines~147--150).
}]{listings/SSLSocketOutputRecord.java}%

\begin{lstlisting}[
  float=tbp,
  widthgobble=0*0,
  linerange={
    42-43,
    %44-44,     %%Constructor
    %48-49,
    %66-66,
    %82-85,
    %87-87,
    %90-90,
    %248-256,   %%Used by SSLEngineOutputRecord
    %270-272,
    %279-279,
    %322-359,
    %
    %385-388,
    389-390,
    392-392,
    396-396,
    398-400,
    402-445  
  },
  label=lst:OutputRecord,
  caption={[\code{OutputRecord.t13Encrypt} produces records \TLSPlaintext\ or \TLSCiphertext]
  Class \code{OutputRecord} defines method \code{t13Encrypt} which appends constant 0x22
  (defined by variable \code{ContentType.HANDSHAKE.id}) and padding (defined by constant
  \code{zeros}) to buffer \code{buf} if outgoing data should be encrypted (Lines~404--408), 
  i.e., when producing record \TLSCiphertext, as opposed to \TLSPlaintext; encrypts the data 
  in that buffer (Lines~432--433), using a null cipher (\code{SSLCipher.NullReadCipherGenerator})
  %%%
  %%% Method SSLCipher.NullReadCipherGenerator.encrypt includes the following:
  %%%
  %%%   MAC signer = (MAC)authenticator;
  %%%   if (signer.macAlg().size != 0) {
  %%%      addMac(signer, bb, contentType);
  %%%   } else {
  %%%      authenticator.increaseSequenceNumber();
  %%%   }
  %%%
  %%% which suggests a MAC can be added. However, CipherSuite.MacAlg defines M_NULL
  %%% such that M_NULL.size = 0. So I suspect M_NULL is in use and no MAC is added.
  %%%
  if data should not be encrypted and a cipher in Galois/Counter Mode 
  (\code{SSLCipher.T13GcmWriteCipherGenerator}) otherwise; and 
  adds the header fields for record \TLSPlaintext\ or \TLSCiphertext\ (Lines~438--442), which 
  only differ on the first byte, in particular, the former uses constant 0x22 (which is input 
  by child \code{SSLSocketOutputRecord}, in the context of Listing~\ref{lst:SSLSocketOutputRecord}),
  whereas the latter uses constant 0x23 (Line~427).
}]
abstract class OutputRecord
        extends ByteArrayOutputStream implements Record, Closeable {
\end{lstlisting}%

\lstinputlisting[
  float=tbp,
  widthgobble=0*0,
  linerange={
    1945-1946,
    1956-1973,
    1987-1987,
    1990-2005,
    2006-2006,
    2009-2009,
    2011-2018,
    %2025-2033,
    2035-2039,
    2043-2043,
    2054-2055,
    2082-2083
  },
  label=lst:SSLCipher,
  caption={[\code{SSLCipher.T13GcmWriteCipherGenerator} encrypts data in Galois/Counter Mode]
  Class \code{SSLCipher.T13GcmWriteCipherGenerator} defines method \code{encrypt} which 
  XORs the sequence number and initialisation vector (Lines~1992--1997); initialises a 
  cipher (Line~2003), using algorithm parameters that define the bit length of the 
  authentication tag and the initialisation vector (Lines~2000-2001); and encrypts the
  input data (Line~2036), appending the authentication tag (Lines~2013--2016), which 
  increments the sequence number as a side effect. (Method \code{Authenticator.TLS13Authenticator.acquireAuthenticationBytes} increments the 
  sequence number using method \code{Authenticator.increaseSequenceNumber}.)
}]{listings/SSLCipher.java}%

\section{Java Secure Socket Extension (JSSE)}\label{sec:JSSE}

Java programmers need not concern themselves with the intricacies of TLS: They can 
use the Java Secure Socket Extension (JSSE), which provides an abstract, high-level
API to establish a TLS channel. Doing otherwise is outright dangerous! TLS 1.3 
was developed over four years by a team of almost one hundred security experts 
from more than forty institutions, including tech behemoths Amazon, Apple, Google, 
IBM, and Microsoft. \ifPresentationNotes \textcolor{red}{Part of this is probably 
better placed elsewhere.} \fi
Their work involved iterating over the subtle 
details to ensure that security objectives were achieved. JSSE provides OpenJDK's 
implementation of TLS; using it simplifies development and reduces risk.
We present toy applications that demonstrate the use of JSSE (Section~\ref{sec:monkeys})
and then delve into JSSE itself (Section~\ref{sec:SunJSSE}).

\subsection{Examples for code monkeys: Toy client and server}\label{sec:monkeys}

\ifPresentationNotes
\marginpar{``channel'' (throughout) is perhaps too academic}
\fi

JSSE trivialises the development of toy applications. For instance, the following code 
snippet establishes a TLS socket:\footnote{%
  Prepending the snippet with 
    \code{System.setProperty("javax.net.debug", "ssl handshake verbose")}
  prints additional information, which can be useful.}

\lstinputlisting[
  numbers=none,
  frame=none,
  nolol=true,
  linerange={
    16-22
  }
]{code/JSSEClient.java}

\noindent 
The established TLS socket protects communication, for example, the following HTTP
request and response is protected:

\lstinputlisting[
  numbers=none,
  frame=none,
  nolol=true,
  linerange={
    24-39
  }
]{code/JSSEClient.java}

\noindent
JSSE uses a ``provider''-based architecture, whereby services (e.g., \code{SSLSocket} and \code{SSLSocketFactory})
and implementations (e.g., \code{SSLSocketImpl} and \code{SSLSocketFactoryImpl}) are defined independently, 
and are (typically) instantiated by factory methods (e.g., \code{SSLSocketFactory.getDefault()}). Hence, 
programmers need not concern themselves with the inner-workings of implementations, such as those provided 
by \emph{SunJSSE} (we will nonetheless take a brief look in Section~\ref{sec:SunJSSE}). 
Let us now consider a toy server application, to compliment our (above) toy client.

Our client uses the default client-side context, whereas our server cannot, because 
server-side authentication is mandatory and a certification must be initialised, 
hence, we start by initialising a suitable context:

\lstinputlisting[
  numbers=none,
  frame=none,
  nolol=true,
  linerange={
    23-31
  }
]{code/JSSEServer.java}

\noindent
That context can be used to establish a TLS socket:

\lstinputlisting[
  numbers=none,
  frame=none,
  nolol=true,
  linerange={
    33-41
  }
]{code/JSSEServer.java}

\noindent
Communication over the TLS socket is protected, for example, any incoming character is
protected as is any subsequent response:

\lstinputlisting[
  numbers=none,
  frame=none,
  nolol=true,
  linerange={
    43-51
  }
]{code/JSSEServer.java}

\noindent 
Our client and server can communicate by assigning \code{InetAddress.getLocalHost()} 
to variable \code{host} and \code{8443} to variable \code{port},\footnote{Unix systems
  protect ports under 1024, hence, we use port 8443, rather than port 443.} 
rather than \code{example.com} and \code{443}, respectively. The key 
store necessary for this example can be constructed using \code{keytool} as follows:

\begin{verbatim}
bas $ keytool -genkey -keyalg RSA -keystore store
Enter keystore password:  
Re-enter new password: 
What is your first and last name?
  [Unknown]:  127.0.1.1
What is the name of your organizational unit?
  [Unknown]:   
What is the name of your organization?
  [Unknown]:  
What is the name of your City or Locality?
  [Unknown]:  
What is the name of your State or Province?
  [Unknown]:  
What is the two-letter country code for this unit?
  [Unknown]:  
Is CN=127.0.1.1, OU=Unknown, O=Unknown, L=Unknown, ST=Unknown, C=Unknown correct?
  [no]:  yes

bas $
\end{verbatim}

\noindent
where \code{InetAddress.getLocalHost()} is 127.0.1.1 and filename \code{store} can 
be replaced with alternatives. Since the above key store is 
self-signed, it must be added to the Java virtual machine's trust store, which can 
be achieved by prepending client code with the following:
\code{System.setProperty( "javax.net.ssl.trustStore", "store" ); System.setProperty( "javax.net.ssl.trustStorePassword", <<pwd>> )}.

\subsection{SunJSSE provider for architects, researchers, and the curious}\label{sec:SunJSSE}

Our toy client uses statement \code{SSLSocketFactory.getDefault().createSocket(host, port)} 
to instantiate an instance of 
\code{SSLSocketFactoryImpl} parameterised with an initial context, and uses that 
context along with variables \code{host} and \code{port} to instantiate and return an 
instance of \code{SSLSocketImpl}. Method \code{SSLSocketImpl.startHandshake()} proceeds as follows:

\lstinputlisting[
  frame=none,
  nolol=true,
  linerange={
    377-377,
    395-398,
    401-403,
    411-411
  }
]{listings/SSLSocketImpl.java}

\noindent
Line~395 indirectly calls method \code{ClientHello.kickstartProducer.produce} (\S\ref{sec:CH}) and processes 
responses (Line~401--403) as follows:

\lstinputlisting[
  frame=none,
  nolol=true,
  linerange={
    1060-1061,
    1063-1067,
    1077-1077,
    1079-1080,
    1148-1149,
    1152-1153,
    1170-1171
  }
]{listings/SSLSocketImpl.java}

\noindent
Hence, until a connection is negotiated (Lines 1064--1067), responses are processed by 
method \code{SSLTransport.decode} (Line~1152--1153) as follows:

\lstinputlisting[
  frame=none,
  nolol=true,
  linerange={
    101-105,
    107-108,
    145-149,
    162-165,
    %167-191,
    %197-199,
    200-204
 }
]{listings/SSLTransport.java}

\noindent
Line~108 parses a handshake (or an alert) record header and calls method 
\code{SSLSocketInputRecord.decodeInputRecord}, providing the header as 
input. That method parses and decodes the complete record, which is then 
processed by method \code{TransportContext.dispatch} (Line~164).
Method \code{SSLSocketInputRecord.decodeInputRecord} proceeds as follows:

\lstinputlisting[
  frame=none,
  nolol=true,
  linerange={
    204-210,
    230-230,
    232-246,
    257-257,
    259-259,
    261-264,
    280-280,
    282-344,
    346-353
 }
]{listings/SSLSocketInputRecord.java}

\noindent
Finally, method \code{TransportContext.dispatch} proceeds as follows:

\lstinputlisting[
  frame=none,
  nolol=true,
  linerange={
    143-144,
    149-149,
    156-179,
    191-192
 }
]{listings/TransportContext.java}

\noindent
Line~178 indirectly calls \code{HandshakeContext.dispatch} on variables
\code{handshakeType} and \code{plaintext.fragment}, which calls the 
relevant consumer and updates the handshake hash.

\ifPresentationNotes
\section{Security}

\textcolor{red}{Informally explain TLS security properties (as per the spec) and 
  provide informal reasoning as to how these properties are satisfied.}
\fi

\ifPresentationNotes
\section{Cryptography}

\textcolor{red}{Perhaps include some details on the underlying cryptography, especially in terms
  of establishing a shared secret from key shares (\S\ref{sec:dheKey}), perhaps also 
  on AEAD algorithms (\S\ref{sec:aead}).}

\subsection{(EC)DHE (shared secret) key derivation}\label{sec:dheKey}

\textcolor{red}{\code{KeyShareExtension} uses \code{SSLKeyExchange.valueOf} to 
  construct an instance of \code{SSLKeyExchange} that parametrises field 
  \code{SSLKeyExchange.keyAgreement} with an instance of \code{T13KeyAgreement}.
  Hence, calling method \code{SSLKeyExchange.createKeyDerivation} calls
  \code{T13KeyAgreement.createKeyDerivation}, which is used as part of 
  traffic key generation.}

\subsection{AEAD algorithms}\label{sec:aead}

...

Discussion on the advantages of AEAD: \url{https://crypto.stackexchange.com/questions/27243/what-is-the-advantage-of-aead-ciphers}

CBC + HMAC achieves AEAD: \url{https://tools.ietf.org/html/draft-mcgrew-aead-aes-cbc-hmac-sha2-01}

Tutorial: \url{https://www.youtube.com/watch?v=g_eY7JXOc8U}

\fi

%\section{Outlook}
%
%...

\appendix
\section{Extensions}
\label{sec:extensions}

Extensions listed by an endpoint are generally 
followed by a corresponding extension from their peer. 
Corresponding extensions must not be sent without solicitation, and  
endpoints must abort with an \TLSunsupportedExtension\ alert upon 
receipt of such unsolicited extensions. For instance, a \ClientHello\ message listing 
extension \TLSsupportedGroups\ is followed by a \ServerHello\ message listing 
the same extension, whereas a \ServerHello\ message must not list that 
extension in response to a \ClientHello\ message that does not and a client
should abort in such cases. 

Table~\ref{table:extensions} formally specifies which extensions can be listed 
in the \TLSextensions\ field of handshake protocol messages. Endpoints 
must abort with an \TLSillegalParameter\ alert if an extension is received
in a handshake protocol message for which it is not specified. Support for 
the following extensions is mandatory (unless an implementation explicitly opts out): 
\TLScookie,
\TLSkeyShare,
\TLSserverName,
\TLSsignatureAlgorithms,
\TLSsignatureAlgorithmsCert,
\TLSsupportedGroups, and
\TLSsupportedVersions.
A client requesting a non-mandatory extension may abort if the extension is
not supported by the server. A server may require \ClientHello\ messages
to include extension \TLSserverName\ and should abort with an \TLSmissingExtension\ 
alert if the extension is missing.

\begin{table}[H]
\caption{Extensions and the handshake protocol messages in which they
  may appear, where such messages are abbreviated as follows:
  CH (\ClientHello), SH (\ServerHello), EE (\EncryptedExtensions), 
  CT (\Certificate), CR (\CertificateRequest), NST (\NewSessionTicket), 
  and HRR (\HelloRetryRequest).}
\label{table:extensions}
\centering
\begin{tabular}{l|l|l}
Extension                                 &RFC & Handshake message \\ \hline
\TLSapplicationLayerProtocolNegotiation   &7301&      CH, EE    \\
\TLScertificateAuthorities                &8446&      CH, CR    \\
\TLSclientCertificateType                 &7250&      CH, EE    \\
\TLScookie                                &8446&     CH, HRR    \\
\TLSearlyData                             &8446& CH, EE, NST    \\
\TLSheartbeat                             &6520&      CH, EE    \\
\TLSkeyShare                              &8446& CH, SH, HRR    \\
\TLSmaxFragmentLength                     &6066&      CH, EE    \\
\TLSoidFilters                            &8446&          CR    \\
\TLSpadding                               &7685&          CH    \\
\TLSpostHandshakeAuth                     &8446&          CH    \\
\TLSpsk                                   &8446&      CH, SH    \\
\TLSpskModes                              &8446&          CH    \\
\TLSserverCertificateType                 &7250&      CH, EE    \\
\TLSserverName                            &6066&      CH, EE    \\
\TLSsignatureAlgorithms                   &8446&      CH, CR    \\
\TLSsignatureAlgorithmsCert               &8446&      CH, CR    \\
\TLSsignedCertificateTimestamp            &6962&  CH, CR, CT    \\
\TLSstatusRequest                         &6066&  CH, CR, CT    \\
\TLSsupportedGroups                       &7919&      CH, EE    \\
\TLSsupportedVersions                     &8446& CH, SH, HRR    \\
\TLSuseSrtp                               &5764&      CH, EE    
\end{tabular}
\end{table}

When designing new extensions, the following considerations should 
be taken into account:
First, a server that does not support a client-requested extension 
\ifSpecNotes
  \textcolor{red}{the spec states client requested feature, but my 
  narrowing seems correct}
\fi
  should indicate that the extension is unsupported by inclusion 
  of a suitable extension in their response, rather than aborting.
  By comparison, a server should abort when a client-supplied
  extension is erroneous. 
Secondly, prior to authentication, active attackers can remove and 
  inject messages, hence, they can modify handshake messages. 
  Since an HMAC is computed over the entire handshake, such 
  modifications can typically be detected and endpoints can 
  abort. However, to quote RFC 8446, ``extreme care is needed 
  when the extension changes the meaning of messages sent in the 
  handshake phase.'' Thus, extensions should be designed to 
  prevent an active adversary from unduly influencing parameter 
  negotiation, i.e., endpoints should negotiate their preferred
  parameters, even in the presence of an adversary.
In addition, any interactions with early data must be defined.

\begin{tcolorbox}
Extensions are enumerated and instantiated by enum \code{SSLExtension} (Listing~\ref{lst:SSLExtension}), 
and class \code{SSLExtensions} (Listings~\ref{lst:SSLExtensions}--\ref{lst:SSLExtensionsB}) represents
a list of extensions.
\end{tcolorbox}

\begin{lstlisting}[
  float=tbp,
  widthgobble=0*0,
  linerange={
    38-38,%class def
    489-505,%constructor
    529-530,532-532,537-537,%produce
    539-540,542-542,547-547,%consumeOnLoad
    549-550,552-552,557-557,%consumeOnTrade
    685-685%closing brace
  },
  label=lst:SSLExtension,
  caption={[\code{SSLExtension} enumerates and instantiates extensions]
  \code{SSLExtension} enumerates and instantiates extensions. 
  Each instantiation defines a hexadecimal value (Line~495) and a name (Line~497). 
  Moreover, they define variable \code{networkProducer} of (interface) type 
  \code{HandshakeProducer} which is instantiated by a constant 
  \code{ThisNameExtension.messageNetworkProducer}, where \code{ThisName} corresponds 
  to extension \TLSField{this\_name} and \code{message} is an abbreviation of
  the message type, e.g., \code{ch} abbreviates \ClientHello. For instance, constants 
  \code{SupportedVersionsExtension.chNetworkProducer} 
  and \code{PreSharedKeyExtension.chNetworkProducer} are used for extensions
  \TLSsupportedVersions\ and \TLSpsk, respectively, for \ClientHello\ messages. 
  Variable \code{networkProducer} is used by method \code{produce} to instantiate extensions
  (Lines~529--537).
  Variables \code{onLoadConsumer} and \code{onTradeConsumer} of (interface) type 
  \code{ExtensionConsumer} and \code{HandshakeConsumer}, respectively, are defined 
  similarly. The former is used by method \code{consumeOnLoad} to consume extensions 
  (Lines~539--547) and the latter is used by method \code{consumeOnTrade} to 
  update the active context to include extensions \ifImplNotes\textcolor{red}{accurate?}\fi
  (Lines~549--557).
  Hence, enum \code{SSLExtension} is reliant
  on classes implementing interfaces \code{HandshakeConsumer}, \code{HandshakeProducer}, and 
  \code{ExtensionConsumer}, e.g., inner-classes of class \code{PreSharedKeyExtension}.
}]
enum SSLExtension implements SSLStringizer {
\end{lstlisting}

\lstinputlisting[
  float=tbp,
  widthgobble=0*0,
  linerange={
    39-41,
    47-49,
    51-51,
    53-61,
    77-82,
    91-93,
    98-100,
    102-104,
    114-119,
    121-123,
    207-208,
    209-209,
    228-230,
    233-234,
    238-240
  },
  label=lst:SSLExtensions,
  caption={[\code{SSLExtensions} produces and consumes extensions]
  Class \code{SSLExtensions} defines a map of extensions and their 
  associated data (Line~41). That map can be instantiated by method \code{produce} 
  (Lines~207--240) or during construction from an input stream (Lines~53-123).
}]{listings/SSLExtensions.java}

\lstinputlisting[
  float=tbp,
  widthgobble=0*0,
  linerange={
    132-134,
    163-164,
    169-170,
    175-177,
    197-197,
    201-202,    
    293-307,
    362-362
  },
  label=lst:SSLExtensionsB,
  caption={[\code{SSLExtensions}  produces and consumes extensions (cont.)]
  Class \code{SSLExtensions} (continued from 
  Listing~\ref{lst:SSLExtensions}) defines method \code{consumeOnLoad} to 
  consume received extensions (Lines~132--170), using method 
  \code{SSLExtension.consumeOnLoad} (Listing~\ref{lst:SSLExtension});
  \code{consumeOnTrade} to update the active context to include extensions
  \ifImplNotes\textcolor{red}{accurate?}\fi
  (Lines~175--202), using method \code{SSLExtension.consumeOnTrade} (Listing~\ref{lst:SSLExtension});
  and method \code{send} to write extensions and associated data to an output stream 
  (Lines~293--307).
}]{listings/SSLExtensions.java}

\section{Alert protocol}
\label{sec:alerts}

TLS defines closure and error alerts, comprising a description field and a legacy 
severity-level field (which, in TLS 1.3, can be inferred from the alert type). 
Closure alerts indicate orderly termination of the established channel (in one direction 
only), and are necessary to avoid truncation attacks. Such closure alerts notify the 
receiver that the sender will not send any more messages on the channel and any data sent
after the alert must be ignored. (The channel must be closed in one direction only 
to avoid truncating messages in the other direction. This requirement
differs from prior versions of TLS, which required the receiver to discard pending 
messages and immediately send a closure alert of their own, thereby truncating the
pending messages.)
Error alerts indicate abortive closure and should be sent whenever an error 
is encountered. Upon transmission or receipt of such an error alert, the established 
channel must be closed immediately, without sending or receiving any further data.
(The listings in this manuscript omit most error alert handling and processing for brevity.)
All alerts are encrypted (by the record protocol) after message \ServerHello\ has been 
successfully consumed.

\section{Client authentication: \CertificateRequest}\label{sec:CR}

A server may request client authentication by sending a \CertificateRequest\ 
message, comprising the following fields:

\begin{description}

\item \TLScertificateRequestContext: A zero-length identifier. (A 
  \CertificateRequest\ message may also be sent to initiate post-handshake 
  authentication, as explained below, in which case a nonce may be used as
  an identifier.)

\item \TLSextensions: A list of extensions describing authentication
  properties. The list must contain at least extension \TLSsignatureAlgorithms.
  (Table~\ref{table:extensions}, Appendix~\ref{sec:extensions}, lists other
  permissible extensions.)

\end{description}

\begin{sloppypar}
\noindent
A client %consuming a \CertificateRequest\ message
may decline to authenticate by responding with a \Certificate\ message
that does not contain a certificate, followed by a \Finished\ message. 
(The server may continue %\sout{the handshake} 
without client authentication or abort with a \TLScertificateRequired\ alert.)
\ifSpecNotes
\textcolor{red}{
  The spec only defines the bracketed remark for handshakes. Presumably 
  it applies to post-handshake authentication too.
}
\fi
 Alternatively,
a client may authenticate by responding with \Certificate\ 
and \CertificateVerify\ messages 
(such that $\CertificateRequest.\TLScertificateRequestContext = \Certificate.\TLScertificateRequestContext$),
followed by a \Finished\ message.
The \CertificateVerify\ message includes a signature over string ``TLS 1.3, client CertificateVerify'',
rather than ``TLS 1.3, server CertificateVerify'', to distinguish client-
and server-generated \CertificateVerify\ messages, and to help defend 
against potential cross-protocol attacks. The signature
algorithm must be one of those listed in field
\TLSsupportedSignatureAlgorithms\ of extension \TLSsignatureAlgorithms\
in the \CertificateRequest\ message.
(The server may abort %\sout{the handshake} 
if the client's certificate chain is 
unacceptable, e.g., when the chain contains a signature from an unknown or 
untrusted certificate authority. Alternatively, the server may proceed, 
considering the client unauthenticated.)
\ifSpecNotes
\textcolor{red}{
  The spec only defines the bracketed remark for handshakes. Presumably 
  it applies to post-handshake authentication too.
}
\fi
Any extensions listed by the \Certificate\ message must respond to ones 
listed in the \CertificateRequest\ message.
\end{sloppypar}

For (EC)DHE-only key exchange, client authentication is possible during 
a handshake: a server includes a \CertificateRequest\ message immediately 
after their \EncryptedExtensions\ message (and before \Certificate, 
\CertificateVerify, and \Finished\ messages), and a client responds 
with \Certificate, (optionally) \CertificateVerify, and \Finished\ messages. 
For PSK-based key exchange, a server must only request client authentication
if their peer's \ClientHello\ message included extension \TLSpostHandshakeAuth. 
Such a request can be made by sending a \CertificateRequest\ message (with a 
non-zero length identifier) after the handshake protocol completes. A client 
responds with \Certificate, (optionally) \CertificateVerify, and \Finished\ 
messages, computing the HMAC with 
\begin{multline*}
  \TLSfinishedKey =   \HKDFExpandLabel( \TLSapplicationTrafficSecret[client]{N},\\
       ``finished", ``", \HashLength)
\end{multline*}
(Post-handshake authentication is only concerned with updating the client's 
application-traffic key, for the purposes of blinding the client's identity 
to that key. Hence, secret \TLSfinishedKey\ is not concerned with traffic 
secret \TLSapplicationTrafficSecret[server]{N}. Beyond traffic keys, a key 
established by a \NewSessionTicket\ message, sent after post-handshake 
authentication, will also be bound to the client's identity.)
A client receiving an unsolicited post-handshake authentication request 
(i.e., message \ClientHello\ did not include extension \TLSpostHandshakeAuth) 
must abort with an \TLSunexpectedMessage\ alert.

%\bibliographystyle{alpha}
%\bibliography{main-tls}

\end{document}